\documentclass[11pt]{article}
\usepackage{mathrsfs}
 \usepackage[utf8]{inputenc}
 \usepackage{epsfig,graphicx,color}
 \usepackage{subfig}
 \usepackage{latexsym}                 
 \usepackage{amsmath,amsfonts,verbatim,xkeyval,bm, amsthm, upgreek, amscd, wasysym, amssymb, amsbsy}
 \usepackage{cases}
\usepackage{bm}
 \usepackage[english]{babel}
 \usepackage{listings}
 \usepackage{multicol}
 \usepackage[colorlinks]{hyperref}  
 \hypersetup{hidelinks}  
 \usepackage{tikz-cd}
 \usepackage{booktabs}
\usepackage[numbers, sort&compress]{natbib}
 \usepackage{sidecap}
 \usepackage[font=small,labelfont=bf]{caption}
 \usepackage{float}
\usepackage{titlesec}
\usepackage{makecell}
\usepackage{xcolor,colortbl}
\hypersetup{hidelinks}
\usetikzlibrary{positioning}

\titleformat{\paragraph}
{\normalfont\normalsize\bfseries}{\theparagraph}{1em}{}
\titlespacing*{\paragraph}
{0pt}{3.25ex plus 1ex minus .2ex}{1.5ex plus .2ex}
 
\def\avg#1{\left\langle #1 \right\rangle}

\def\vet#1{{\bm #1}}
\def\reali{\mathbb{R}}
\def\complessi{\mathbb{C}}

\def\interi{\mathbb{Z}}
\def\toro{\mathbb{T}}

\def\e{\mathrm{e}}
\def\i{\mathrm{i}}

\def\C{\mathscr{C}}

\def\F{\mathscr{F}}

\def\Escr{\mathcal{E}}

\def\Hscr{\mathcal{H}}

\def\Oscr{\mathcal{O}}

\def\b{\mathbf{b}}
\def\c{\mathbf{c}}
\def\d{\mathbf{d}}
\def\Ups{$\upsilon$-Andromed{\ae}~}
\def\ups{$\upsilon$-And~}
\def\rho{\varrho}
\def\csi{\xi}

\def\poisson#1#2{\lbrace #1,#2 \rbrace}

\def\t#1{\tilde{#1}}

\definecolor{lavender}{rgb}{0.9, 0.9, 0.98}

\newtheorem{theorem*}{Theorem}
\newtheorem{lemma}{Lemma}[section]

\newtheorem{definition}{Definition}[section]
\newtheorem{remark}{Remark}[section]

\begin{document}

\pagenumbering{arabic}

\begin{center}
{\Large\bf Secular orbital dynamics of the innermost exoplanet of the $\upsilon$-Andromed{\ae} system}
\vskip 1cm
{\it Rita Mastroianni\footnote{Dipartimento di Matematica ``Tullio Levi-Civita'', Universit\`a degli Studi di Padova, via Trieste 63, 35121 Padova, rita.mastroianni@math.unipd.it} and Ugo Locatelli\footnote{Dipartimento di Matematica, Universit\`a
    degli Studi di Roma ``Tor Vergata'', via della Ricerca Scientifica 1, 00133 Roma, locatell@mat.uniroma2.it}}

\end{center}

\begin{abstract}
{\small
We introduce a quasi-periodic restricted Hamiltonian to
  describe the secular motion of a small-mass planet in a
  multi-planetary system. In particular, we refer to the motion of
  \ups $\b$ which is the innermost planet among those discovered in
  the extrasolar system orbiting around the \Ups{A} star. We
  preassign the orbits of the Super-Jupiter exoplanets \ups $\c$ and
  \ups $\d$ in a stable configuration. The Fourier decompositions of
  their secular motions are reconstructed by using the well known
  technique of the (so called) Frequency Analysis and are injected in
  the equations describing the orbital dynamics of \ups $\b$ under the
  gravitational effects exerted by those two external exoplanets (that
  are expected to be major ones in such an extrasolar
  system). Therefore, we end up with a Hamiltonian model having
  $2+3/2$ degrees of freedom; its validity is confirmed by the
  comparison with several numerical integrations of the complete
  $4$-body problem. Furthermore, the model is enriched by taking into
  account also the effects due to the relativistic corrections on the
  secular motion of the innermost exoplanet. We focus on the problem
  of the stability of \ups $\b$ as a function of the parameters that
  mostly impact on its orbit, that are the initial values of its
  inclination and the longitude of its node (as they are measured with
  respect to the plane of the sky). In particular, we study the
  evolution of its eccentricity, which is crucial to exclude orbital
  configurations with high probability of (quasi)collision with the
  central star in the long-time evolution of the system. Moreover, we
  also introduce a normal form approach, that is based on the complete
  average of our restricted model with respect to the angles
  describing the secular motions of the major exoplanets. Therefore,
  our Hamiltonian model is further reduced to a system with $2$
  degrees of freedom, which is integrable because it admits a constant
  of motion that is related to the total angular momentum. This allows
  us to very quickly preselect the domains of stability for \ups $\b$,
  with respect to the set of the initial orbital configurations that
  are compatible with the observations.}
  \end{abstract}


\section{Introduction}
\label{sec:introduction}

The \Ups{system} was the first ever to be discovered among the ones that
host at least two exoplanets. In fact, a few years after the discovery of the first exoplanet, the
evidence for multiple companions of the $\upsilon$-Andromed{\ae} system was
announced (see~\cite{mayque1995} and~\cite{butetal1999},
respectively). In particular, the observations made with the detection
tecnique of the Radial Velocity (hereafter, RV) revealed the existence
of orbital objects with three different periods, $4.6$, $241$ and
$1267$ days, which revolve around \Ups{A}, that is the brightest star
of a binary hosting also the red dwarf \Ups{B}. Such exoplanets were named \ups$\b$, $\c\,$
and $\d\,$ in increasing order with respect to their distance from
the main star. Since \Ups{B} is very far with respect to these other
bodies (i.e., $\sim 750\,$AU), then it is usual to not consider its
negligible gravitational effects when the dynamical behavior of the
planetary system orbiting around \Ups{A} is studied.

None of the present detection methods allows us to know all the
orbital parameters of an extrasolar planet.  For instance, the RV
technique does not provide any information about both the inclination
and the longitude of node. In the \Ups case, these two orbital elements were measured
(although with rather remarkable error bars) for both \ups$\c$ and
\ups$\d$ thanks to observational data taken from the Hubble Space
Telescope (see~\cite{mcaetal2010}). The information provided by such
an application of astrometry significantly complemented the knowledge about this extrasolar system; in fact, it has led to the evaluation of the masses of \ups$\c$ and \ups$\d$ (ranging in
$13.98^{+2.3}_{-5.3}\>M_J$ and $10.25^{+0.7}_{-3.3}\>M_J$,
respectively) and of their mutual inclination ($29.9^\circ\pm 1^\circ$).
It is well known that only minimum limits for the masses can be
deduced by observations made with the RV method (due to the intrinsic
limitations of such a technique). Moreover, the mutual inclination
between planetary orbits plays a crucial role for what concerns the
stability of extrasolar systems (see, e.g.,~\cite{voletal2019}). Thus,
the orbital configuration of \Ups is probably one  of the most accurately
known among the extrasolar multi-planetary systems which have been
discovered so far.

The question of the orbital stability of \Ups is quite challenging.
Numerical integrations revealed that unstable orbits are frequent. Moreover, these extensive explorations allowed to locate
four different regions of initial values of the orbital parameters
(consistent with all observational constraints) yielding dynamically stable orbital configurations for the three planets of
the system (see~\cite{deietal2015}). All these sets of parameters
correspond to values of the mass of \ups$\c$ that are relatively small,
in the sense that they are much closer to the lower bound of the range
$13.98^{+2.3}_{-5.3}\>M_J$ than to the upper one. On the other hand,
according to a numerical criterion inspired from normal form theory
and introduced in~\cite{locetal2022a}, the most robust orbital
configurations correspond to the largest possible value of the
mass of \ups$\c$ in the above range. In the vicinity of the initial
conditions giving rise to the most robust orbital
configuration, the existence of KAM tori for the dynamics of a
secular three-body problem including \ups$\c$ and \ups$\d$ was proved in a rigorous
computer--assisted way (see~\cite{caretal2022}).

In the present paper we aim to extend the study of the stability to
the orbital dynamics of \ups$\b$, still adopting a hierarchical
approach. In the case of the particular extrasolar system under
consideration, this means that we assume the mass of \ups$\b$ so
small\footnote{Actually, the mass of \ups$\b$ is unknown, but it has
  been determined its minimal value $m\,\sin \i$, which is about $1/3$ of the one
  of \ups$\c$ (that, in turn, is less than $1/2$ of the minimal mass of
  \ups$\d$). Moreover, our modelization is further motivated by the
  fact that the semi-major axis of the innermost exoplanet is (more
  than) one order of magnitude smaller than the other ones; thus, the
  gravitational interactions between \ups$\b$ and \ups$\c$ or \ups$\d$
  are expected to be negligible with respect to the one between the outer
  planets. We recall that the RV measurements also detected the
  presence of a fourth exoplanet in the system, namely \ups{\bf e}
  (see~\cite{curetal2011}). However, since it is expected to be in a
  $3:1$ external resonance with \ups$\d$ and the minimal mass of
  \ups{\bf e} is nearly equal to the Jupiter one, its effects on the
  orbital dynamics of the innermost exoplanet of the system look to be
  negligible.}  with respect to the ones of \ups$\c$ and \ups$\d$, that
the motion of the innermost planet $\b$ can be modeled with a good
approximation via a restricted four-body problem. More precisely, in
order to study the dynamical behavior, we preassign the secular motions of
the Super-Jupiter exoplanets $\c$ and $\d$ in correspondence to the
quasi-periodic orbit that is expected to be the most robust. The
Fourier decompositions of the secular motions of $\c$ and $\d$ are reconstructed by using the well known technique of the (so
called) Frequency Analysis (see e.g.~\cite{las2005}) and are injected
in the equations describing the orbital dynamics of \ups$\b$, under
the gravitational effects exerted by those two external exoplanets.
This way to introduce a quasi-periodic restricted model has been
recently used to study the long-term dynamics of our Solar System
(see~\cite{moglas2022} and~\cite{hoaetal2022}).

The advantage of introducing a \textit{secular quasi-periodic
  restricted Hamiltonian} looks evident. In our present case
(referring to the \Ups{system} with planets $\b\,$, $\c\,$, $\d\,$),
we start with a Hamiltonian model having $9$ degrees of freedom,
ending up with a simpler one with $2+3/2$ degrees of freedom, where
the short periods are dropped. This explains why numerical
explorations of the restricted Hamiltonian model are much
faster. Our main purpose is further promoting this procedure, in such
a way to introduce a new simplified model with just two degrees of
freedom. Indeed, in the present paper, we show that this can be done
by adopting a suitable normal form approach, which is so accurate to
produce an integrable Hamiltonian that can be used to efficiently
characterize the \textit{stability domain} with respect to the unknown
orbital parameters of \ups$\b$, i.e., the inclination and the
longitude of the node.

The present work is organized as follows. In
Section~\ref{subsec:AINF}, Frequency Analysis is used to reconstruct
the Fourier decompositions of the secular motions of the outer
exoplanets \ups$\c$ and \ups$\d$.  In Section~\ref{sec:QPR}, the
secular quasi-periodic restricted Hamiltonian model (with $2+3/2$
degrees of freedom) is introduced and validated through the comparison
with several numerical integrations of the complete four-body problem,
hosting planets $\b\,$, $\c\,$, $\d\,$ of the \Ups{system}. The double
normalization procedure allowing to perform a sort of averaging which
further simplifies the model is described with a rather general
approach in Section~\ref{sub:forme_normali_astro}. In
Section~\ref{sec:risult_astro} this normal form procedure is applied
to the quasi-periodic restricted Hamiltonian, in such a way to derive
an integrable model with $2$ degrees of freedom describing the secular
orbital dynamics of \ups$\b\,$. Such a simplified model is used to
study \ups $\b$ stability domain in the parameters space of the
initial values of the inclination and the longitude of node. All this
computational procedure is repeated in Section~\ref{sec:QPR_REL}
starting from a version of the secular quasi-periodic restricted
Hamiltonian model which includes also relativistic corrections; this
allows us to appreciate the effects on the orbital dynamics due to
General Relativity.

\section{Determination of the outer planets motion}
\label{subsec:AINF}

To prescribe the orbits of the giant planets \ups$\c$ and \ups$\d$, we
start from the Hamiltonian of the three-body problem (hereafter,
$3$BP) in Poincar\'e heliocentric canonical variables, using the
formulation based on the reduced masses $\beta_2\,$, $\beta_3\,$, that
is
\begin{align}
\label{Ham.3BP.reduced.mass}
\Hscr=&\sum_{j=2}^{3}\left( \frac{\vet{p}_j\cdot\vet{p}_j}{2\, \beta_j}
-\frac{\mathcal{G}\, m_0\, m_j}{r_j}\right)
+\frac{\vet{p_2}\cdot\vet{p_3}}{ m_0}
-\frac{\mathcal{G}\, m_2\, m_3}{|\vet{r_2}-\vet{r_3}|}\, ,
\end{align}
where $m_0$ is the mass of the star, $m_{j}\,$, $\vet{r}_j\,$,
$\vet{p}_j\,$, $j=2,3$, are the masses, astrocentric position vectors
and conjugated momenta of the planets, respectively, $\mathcal{G}$ is
the gravitational constant and $\beta_j=m_0 m_j/(m_0+m_j)$, $j=2,3$,
are the reduced masses. Let us remark that, in the following, we
use the indexes $2$ and $3$ respectively, for the inner (\ups$\c$)
and outer (\ups$\d$) planets between the giant ones, while the
index $1$ is used to refer to \ups$\b$

In order to set up a quasi-periodic restricted model for the
description of the motion of \ups$\b$, we need to characterize the
motion of the giant planets; this can be done through the
\textit{Frequency Analysis} method, starting from the numerical
integration of the complete $3$BP Hamiltonian, reported in
equation~\eqref{Ham.3BP.reduced.mass} (i.e., before any expansions and
averaging\footnote{We remark that, in principle, the \textit{Frequency
    Analysis} method can be performed starting from the
  \textit{secular} $3$BP Hamiltonian at order two in masses; more
  precisely, in order to compute the secular Hamiltonian, the
  dependence on the fast angles $\lambda_2, \lambda_3\,$ need to be
  removed. It can be done by ``averaging by scissors'', that is
  equivalent to do a first order (in the mass ratios) averaging
  (simply meaning to remove from the Hamiltonian the terms depending
  upon the mean anomalies of the planets); otherwise, in order to have
  a more accurate representation, this elimination can be done through
  a canonical transformation, corresponding to a second order (in the
  mass ratios) averaging (see~\cite{locgio2000}). However, we have
  observed that the Fourier decomposition given by the second order in
  masses numerical integration is not enough accurate for such a kind
  of model and that a more detailed approximation of the orbits of the
  outer giant planets \ups$\c$, \ups$\d$ is required.}).  Thus, we
numerically integrate the complete
Hamiltonian~\eqref{Ham.3BP.reduced.mass} using a symplectic method of
type $\mathcal{SBAB}_3\,$, which is described in~\cite{lasrob2001}.
As initial orbital parameters for the outer planets, we adopt those
reported in Table~\ref{tab:param.orb}, corresponding to the most
robust planetary orbit compatible with the observed data available for
\ups$\c$ and \ups$\d$ (see~\cite{mcaetal2010}), according to the
criterion of ``minimal area'' explained in~\cite{locetal2022a}.

\begin{table}[h]
\begin{minipage}{0.5\textwidth}
\begin{center}
\begin{tabular}{lll}
\toprule
 & \ups$\c$ & \ups$\d$ \\
\midrule
$m \, [M_J]$ & $15.9792$ & $9.9578$ \\
$a(0) \, [AU]$ & $0.829$ & $2.53$ \\
$\e(0)$ & $0.239$ & $0.31$ \\
$\i(0)\, [^{\circ}]$ & $6.865$ & $25.074$ \\
$M(0)\, [^{\circ}]$ & $355$ & $335$ \\
$\omega(0)\, [^{\circ}]$ & $245.809$ & $254.302$ \\
$\Omega(0)\, [^{\circ}]$ & $229.325$ & $7.374$ \\
\bottomrule
\end{tabular}
\end{center}
\end{minipage}
\begin{minipage}{0.4\textwidth}
\caption{Chosen values of the masses and of the initial orbital
  parameters for \ups$\c$ and \ups$\d$, compatible with the observed
  data available, as reported in~\cite{mcaetal2010}.}
\label{tab:param.orb}
\end{minipage}
\end{table}

Having fixed as initial orbital parameters the ones described in
Table~\ref{tab:param.orb}, it is possible to compute their
correspondent values in the Laplace reference frame (i.e., the
invariant reference frame orthogonal to the total angular momentum
vector $\vet{r}_2\times\vet{p}_2+\vet{r}_3\times\vet{p}_3\,$) and to
perform the numerical integration of the full $3$BP corrisponding to
these initial values. Then, it is possible to express the
\textit{discrete} results produced by the numerical integrations in
the canonical Poincaré variables $(\csi_j, \eta_j)$, $(P_j, Q_j)$
(momenta-coordinates) given by\footnote{The definition
    of the Poincaré variables $(\csi_1, \eta_1)$, $(P_1, Q_1)$ will be
    useful in the following Sections.}
\begin{equation}
\label{Poinc.var.U}
\begin{aligned}
&\csi_j=\sqrt{2\Gamma_j}\cos (\gamma_j)=\sqrt{2\Lambda_j}\,\sqrt{1-\sqrt{1-\e_j^2}}\cos (\varpi_j)\, , \\
&\eta_j=\sqrt{2\Gamma_j}\sin (\gamma_j)=-\sqrt{2\Lambda_j}\,\sqrt{1-\sqrt{1-\e_j^2}}\sin (\varpi_j)\, ,\qquad j=1,\,2,\,3\, ,\quad\\
&P_j=\sqrt{2\Theta_j}\cos (\theta_j)=2\sqrt{\Lambda_j}\,\sqrt[4]{ 1-\e_j^2}\,\sin \left(\frac{\i_j}{2}\right)\cos( \Omega_j)\, ,\\
&Q_j=\sqrt{2\Theta_j}\sin (\theta_j)=-2\sqrt{\Lambda_j}\,\sqrt[4]{ 1-\e_j^2}\,\sin \left(\frac{\i_j}{2}\right)\sin( \Omega_j)\, 
\end{aligned}
\end{equation}
where $\Lambda_j=\beta_j\sqrt{\mu_j a_j}\,$, $\beta_j=m_0
m_j/(m_0+m_j)\,$,
$\mu_j=\mathcal{G}\left(m_0+m_{j}\right)\,$, and
$\e_j\,$, $\i_j\,$, $\omega_j\,$, $\Omega_j$,
$\varpi_j=\omega_j+\Omega_j$ refer, respectively,
to the eccentricity, inclination, argument of the periastron, longitudes of the node and of the periastron of
the $j$-th planet.

However, the numerical integrations do not allow to obtain a complete
knowledge of the motion laws $t\mapsto(\csi_j(t), \eta_j(t))$,
$t\mapsto(P_j(t), Q_j(t))$ ($j=2,3$), producing only discrete time
series made by sets of finite points computed on a regular grid in the
interval $[0,T]\,$. The computational method of \textit{Frequency
  Analysis} (hereafter, FA) allows however to reconstruct with a good
accuracy the motion laws by using suitable continuous in the time
variable $t$ functions. This has been done recently
in~\cite{moglas2022} and~\cite{hoaetal2022}, in order express the
motion of the Jovian planets of our Solar System as a Fourier
decomposition including just a few of the main terms. In the present
Section, we basically follow that approach; therefore, here we limit
ourselves to report some definitions which are essential in order to
make our computational procedure well definite (see,
e.g.,~\cite{las2005} for an introduction and a complete exposition
about FA).  We consider analytic quasi-periodic motion laws $t\mapsto
z(t)$. This means that the function $z:\reali\mapsto\complessi$ admits
the following Fourier series decomposition:
\vskip -.25truecm
\begin{align}
\label{AINF1}
z(t)=\sum_{\vet{k}\in\mathbb{Z}^n}a_{\vet{k}}e^{i(\vet{k}\cdot\vet{\omega}t + \vartheta_\vet{k})}\, ,
\end{align}
\vskip -.15truecm
where $\vet{\omega}\in\mathbb{R}^n$ is the so called fundamental
angular velocity vector, while $a_\vet{k}\in\mathbb{R}_+\cup\lbrace 0
\rbrace$ and $\vartheta_\vet{k}\in\mathbb{T}\,$, $\forall\>{\bm
  k}\in\interi^n$; moreover, the sequence $\{a_k\}_{{\bm
    k}\in\interi^n}$ is assumed to satisfy an exponential decay law, i.e.,
$|a_{\vet{k}}|\leq c\, e^{-|\vet{k}|\sigma}$ $\forall\,\vet{k}\in\mathbb{Z}^n$
with $c$ and $\sigma$ real positive parameters.  The FA allows us
to find an approximation of $z(t)$ of the following form:
\begin{align}
\label{AINF2}
z(t)\simeq \sum_{s=1}^{\mathcal{N}_C}a_{s;T}e^{i(\nu_{T}^{(s)}t+\vartheta_{s;T})}
\end{align}
where ${\mathcal{N}_C}$ is the number of components we want to
consider. The equation~\eqref{AINF2} is an approximation of the motion
$z(t)$ in the sense that if ${\mathcal{N}_C}\to +\infty$ and $T\to
+\infty\, $, the right hand side of~\eqref{AINF2} converges
to~\eqref{AINF1}. Moreover, $a_{s;T}\in\mathbb{R_+}\,$ and
$\vartheta_{s;T}\in[0,2\pi)$ are called respectively the amplitude and
the initial phase of the $s$-th component, while $\nu_{T}^{(s)}$ is
a local maximum point of the function
\begin{equation}
\label{tuning}
\nu \mapsto \mathcal{A}(\nu)=\frac{1}{T}\left|\int_{0}^{T} z(t)\,e^{-i\nu t}\,\mathcal{W}(t)\, dt \right|\, ,
\end{equation}
where $\mathcal{W}$ is a suitable weight function such that
${\int_{0}^{T}\mathcal{W}(t)\,dt=T}\,$. Following~\cite{las2005}, we
use the so called ``Hanning filter'', defined (in $[0, T]$) as
$\mathcal{W}(t)=1-\cos[\pi(2t/T-1)]\,$.

The numerical integration of the $3$BP
(equation~\eqref{Ham.3BP.reduced.mass}) produces a discretization of
the signals\footnote{Actually, the numerical integration of the
  complete problem allows to determine also a discretization of the
  fast variables
  $\sqrt{2\Lambda_2}\cos(\lambda_2)+i\,\sqrt{2\Lambda_2}\sin(\lambda_2)$
  and
  $\sqrt{2\Lambda_3}\cos(\lambda_3)+i\,\sqrt{2\Lambda_3}\sin(\lambda_3)\,$;
  however we are not interested in these variables. Thus, we do not
  report their decompositions as they are provided by the FA.}
$t\mapsto \csi_j(t)+i\eta_j(t)$ and $t\mapsto P_j(t)+i Q_j(t)\,$, that
are $\lbrace(\csi_j(s\Delta),
\eta_j(s\Delta))\rbrace_{s=0}^{\mathcal{N}_P}$ and
$\lbrace(P_j(s\Delta), Q_j(s\Delta))\rbrace_{s=0}^{\mathcal{N}_P}\,$
($j=2,3$), where $s = 0, \ldots,\mathcal{N}_P$ and the sampling time
is $\Delta = T /\mathcal{N}_P\,$. These discretizations allow to
compute (by numerical quadrature) the integral in~\eqref{tuning} and,
consequently, a few of local maximum points of the
function~\eqref{tuning} considering, as $z(t)$,
$\csi_2(t)+i\eta_2(t)\,$, $\csi_3(t)+i\eta_3(t)\,$, $P_2(t)+iQ_2(t)$
and $P_3(t)+iQ_3(t)\,$.

Then, we use the FA to compute a quasi-periodic approximation of the
secular dynamics of the giant planets \ups$\c$ and \ups$\d$, i.e.,
\begin{equation}
\label{segnalicd}
\begin{aligned}
&\csi_j(t)+i\eta_j(t)\simeq\sum_{s=1}^{\mathcal{N}_C}A_{j,s}e^{i(\vet{k}_{j,s}\cdot\vet{\theta}(t)+\vartheta_{j,s})}\, , \\ 
&P_j(t)+iQ_j(t)\simeq\sum_{s=1}^{\widetilde{\mathcal{N}}_C}\t{A}_{j,s}e^{i(\tilde{\vet{k}}_{j,s}\cdot\vet{\theta}(t)+\tilde{\vartheta}_{j,s})}\, , 
\end{aligned}
\end{equation}
$j=2,3\,$, where the angular vector 
\begin{equation}
\label{def_angle_theta_FA}
\vet{\theta}(t)=(\theta_3(t), \theta_4(t), \theta_5(t))=(\omega_3\,t, \omega_4\, t, \omega_5\, t):=\vet{\omega}\,t
\end{equation}
depends \textit{linearly on time} and $\vet{\omega}\in\mathbb{R}^3$ is
the fundamental angular velocity vector whose components are listed in
the following:
\begin{align}
\begin{aligned}
\label{freq.fond.CD}
&\omega_3=-2.4369935819462266\cdot10^{-3}\, , \\
&\omega_4=-1.04278712796661375\cdot10^{-3}\, , \\
&\omega_5=\phantom{-}4.88477275490260560\cdot10^{-3}\, .
\end{aligned}
\end{align}

Hereafter, the secular motion of the outer planets $t\mapsto
(\csi_j(t), \eta_j(t), P_j(t), Q_j(t))$, $j=2,3$, is approximated as
it is written in both the r.h.s. of formula~\eqref{segnalicd}. The
numerical values of the coefficients which appear in the
quasi-periodic decompositions\footnote{Of course, the
    exact quasi-periodic decompositions include infinite terms in the
    Fourier series. In order to reduce the computational effort, we
    limit ourselves to consider just a few components, which are the
    main and most reliable ones, according to the following
    criteria. We take into account those terms corresponding to low
    order Fourier armonics, i.e., $\sum_{j=3}^{5} |{k}_j| \leq 5$ or
    $\sum_{j=3}^{5} |\t{k}_j| \leq 5\,$, and such that there are small
    uncertainties on the determination of the frequencies as linear
    combinations of the fundamental ones, i.e.,
    $|\nu_T^{(s)}-\vet{k}^{(s)}\cdot\vet{\omega}|\leq 2\cdot 10^{-7}$
    or $|\nu_T^{(s)}-\t{\vet{k}}^{(s)}\cdot\vet{\omega}|\leq 2\cdot
    10^{-7}\,$.  } of the motions laws are reported in
Tables~\ref{tab:sign1U},~\ref{tab:sign2U}, \ref{tab:sign3U}
and~\ref{tab:sign4U}.

\begin{table}[h]
\centering
\begin{minipage}[b]{0.5\textwidth}
\resizebox{1.25\textwidth}{!}{
\begin{tabular}{|r|r|r|r|r|c|c|c|}
\toprule
$s$ & $\nu_{T}^{(s)}\qquad\qquad\qquad$ & $k_3^{(s)}$ & $k_4^{(s)}$ & $k_5^{(s)}$  & $|\nu_{T}^{(s)} - \vet{k}^{(s)}\cdot \vet{\omega}|$ & $A_{s} $ & $\vartheta_{s}$\\
\midrule
$0$ & $-2.43699358194622660\cdot 10^{-3} $ &   $1$ &   $0$ &   $0$  &  $0.0000$ & $3.8182\cdot 10^{-1}$ & $4.611$\\
 $1$ & $-1.04274752029517815\cdot 10^{-3}$ &   $0$ &  $1$ &   $0$  &  $3.9608\cdot 10^{-8}$ & $1.4219\cdot 10^{-1}$ & $2.434$\\
    $2$ &  $1.22065297958166112\cdot 10^{-2}$ &   $-1$ &   $0$ &   $2$  & $9.2959\cdot 10^{-9}$ & $9.0935\cdot 10^{-2}$ &  $3.898$\\
 $3$ & $-3.83123872535040154\cdot 10^{-3}$ &   $2$ &  $-1$ &   $0$  &  $3.8689\cdot 10^{-8}$ & $4.0358\cdot 10^{-2}$ &  $3.593$\\
\bottomrule
\end{tabular}
}
\phantom{.}
\end{minipage}
\hfill
\begin{minipage}[b]{0.35\textwidth}
\caption{\small Decomposition of the signal $\csi_2(t)+i\,\eta_2(t)\, $ as it is provided by the FA.}
\label{tab:sign1U}
\end{minipage}
\\[1.2ex]
\begin{minipage}[b]{0.5\textwidth}
\resizebox{1.25\textwidth}{!}{
\begin{tabular}{|r|r|r|r|r|c|c|c|}
\toprule
$s$ & $\nu_{T}^{(s)}\qquad\qquad\qquad$ & $k_3^{(s)}$ & $k_4^{(s)}$ & $k_5^{(s)}$  & $|\nu_{T}^{(s)} - \vet{k}^{(s)}\cdot \vet{\omega}|$ & $A_{s} $ & $\vartheta_{s}$\\
\midrule
 $0$&  $-2.43699698221569363\cdot10^{-3}$ &    $1$ &   $0 $ &  $0$  & $3.4003\cdot10^{-9}$ & $5.6387\cdot10^{-1}$ & $1.469$ \\
 $1$ & $ -1.04278712796661375\cdot10^{-3}$ &    $0$ &   $1$ &  $0$  & $  0.0000$ & $1.1039\cdot10^{-1}$ &  $2.437$  \\
  $2$  & $-3.83100979083359590\cdot10^{-3}$ &    $2$ &   $-1$ &   $0$  &  $1.9025\cdot10^{-7}$ & $2.7811\cdot10^{-2}$ &  $3.566$ \\
  $ 3$  & $1.22065393849870793\cdot10^{-2}$ &   $-1$ &   $0$ &  $2$  & $2.9324\cdot10^{-10}$ & $2.4050\cdot10^{-3}$ & $7.556\cdot10^{-1}$ \\
\bottomrule
\end{tabular}
}
\phantom{.}
\end{minipage}
\hfill
\begin{minipage}[b]{0.35\textwidth}
\caption{\small Decomposition of the signal $\csi_3(t)+i\,\eta_3(t)\, $ as it is provided by the FA.}
\label{tab:sign2U}
\end{minipage}%
\\[1.2ex]
\begin{minipage}[b]{0.5\textwidth}
\resizebox{1.25\textwidth}{!}{
\begin{tabular}{|r|r|r|r|r|c|c|c|}
\toprule
$s$ & $\nu_{T}^{(s)}\qquad\qquad\qquad$ & $\t{k}_3^{(s)}$ & $\t{k}_4^{(s)}$ & $\t{k}_5^{(s)}$ & $|\nu_{T}^{(s)} - \t{\vet{k}}^{(s)}\cdot \vet{\omega}|$ & $\t{A}_{s} $ & $\t{\vartheta}_{s}$\\
\midrule
 $0$&  $4.88477275490260560\cdot10^{-3}$ &    $0$ &   $0 $ &  $1$  & $0.0000$ & $5.5389\cdot10^{-1}$ & $2.670$ \\
 $1$ & $ -9.75856551797929864\cdot10^{-3}$ &    $2$ &   $0$ &  $-1$  & $  1.944\cdot10^{-7}$ & $4.9772\cdot10^{-2}$ &  $1.914\cdot 10^{-1}$ \\
 $2$ & $ -8.36452054946431114\cdot10^{-3}$ &    $1$ &  $1$ &   $-1$  &  $3.2915\cdot10^{-8}$ & $2.2433\cdot10^{-2}$ & $4.351$\\
 $3$ & $6.27899221605471031\cdot10^{-3}$ &   $ -1$ &   $1$ &  $1$  & $1.3007\cdot10^{-8}$ & $1.2854\cdot10^{-2}$ &  $5.208\cdot10^{-1}$ \\
  $4$  & $3.49055804503076465\cdot10^{-3}$ &    $1$ &   $-1$ &   $1$  &  $8.2559\cdot10^{-9}$ & $1.0041\cdot10^{-2}$ &  $1.678$ \\
\bottomrule
\end{tabular}
}
\phantom{.}
\end{minipage}
\hfill
\begin{minipage}[b]{0.35\textwidth}
\caption{\small Decomposition of the signal $P_2(t)+i\,Q_2(t)\, $ as it is provided by the FA.}
\label{tab:sign3U}
\end{minipage}%
\\[1.2ex]
\begin{minipage}[b]{0.5\textwidth}
\resizebox{1.25\textwidth}{!}{
\begin{tabular}{|r|r|r|r|r|c|c|c|}
\toprule
$s$ & $\nu_{T}^{(s)}\qquad\qquad\qquad$ & $\t{k}_3^{(s)}$ & $\t{k}_4^{(s)}$ & $\t{k}_5^{(s)}$ & $|\nu_{T}^{(s)} - \t{\vet{k}}^{(s)}\cdot \vet{\omega}|$ & $\t{A}_{s} $ & $\t{\vartheta}_{s}$\\
\midrule
 $0$&  $4.88477277322339754\cdot10^{-3}$ &    $0$ &   $0 $ &  $1$  & $1.8321\cdot10^{-11}$ & $5.6348\cdot10^{-1}$ & $5.812$ \\
 $1$ & $ -9.75856522554671181\cdot10^{-3}$ &    $2$ &   $0$ &  $-1$  & $  1.9469\cdot10^{-7}$ & $5.1543\cdot10^{-2}$ &  $3.333$ \\
 $2$ & $ -8.36452090216070580\cdot10^{-3}$ &    $1$ &  $1$ &   $-1$  &  $3.2563\cdot10^{-8}$ & $2.3352\cdot10^{-2}$ & $1.209$\\
 $3$ & $3.49054260511432292\cdot10^{-3}$ &   $ 1$ &   $-1$ &  $1$  & $2.3696\cdot10^{-8}$ & $1.3434\cdot10^{-2}$ &  $4.821$ \\
  $4$  & $6.27897429080374707\cdot10^{-3}$ &    $-1$ &   $1$ &   $1$  &  $4.9181\cdot10^{-9}$ & $9.7673\cdot10^{-3}$ &  $3.664$ \\
\bottomrule
\end{tabular}
}
\phantom{.}
\end{minipage}
\hfill
\begin{minipage}[b]{0.35\textwidth}
\caption{\small Decomposition of the signal $P_3(t)+i\,Q_3(t)\, $ as it is provided by the FA.}
\label{tab:sign4U}
\end{minipage}%
\vskip -.3truecm
\end{table}

\begin{figure}[!h]
\begin{minipage}{0.6\textwidth}
\begin{minipage}{0.4\textwidth}
\resizebox{1.5\textwidth}{!}{
{
\includegraphics[scale=0.56]{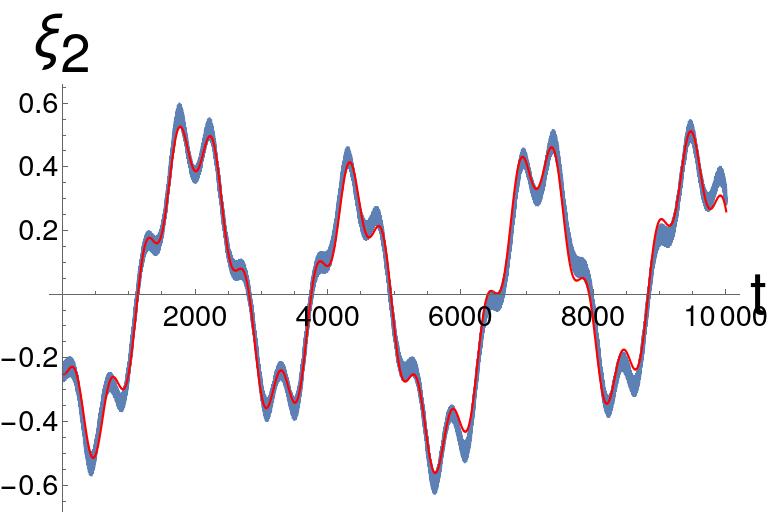}}
}
\end{minipage}
\quad\qquad\qquad
\begin{minipage}{0.4\textwidth}
\resizebox{1.5\textwidth}{!}{
{
\includegraphics[scale=0.56]{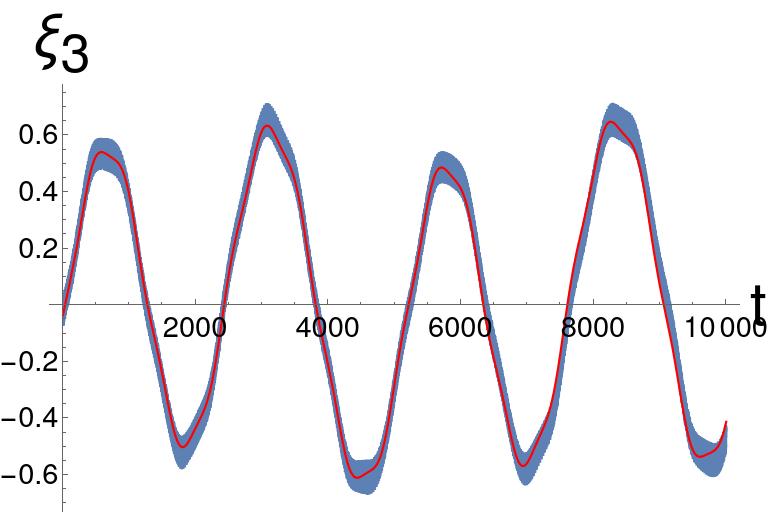}}
}
\end{minipage}
\\
\begin{minipage}{0.4\textwidth}
\resizebox{1.5\textwidth}{!}{
{
\includegraphics[scale=0.56]{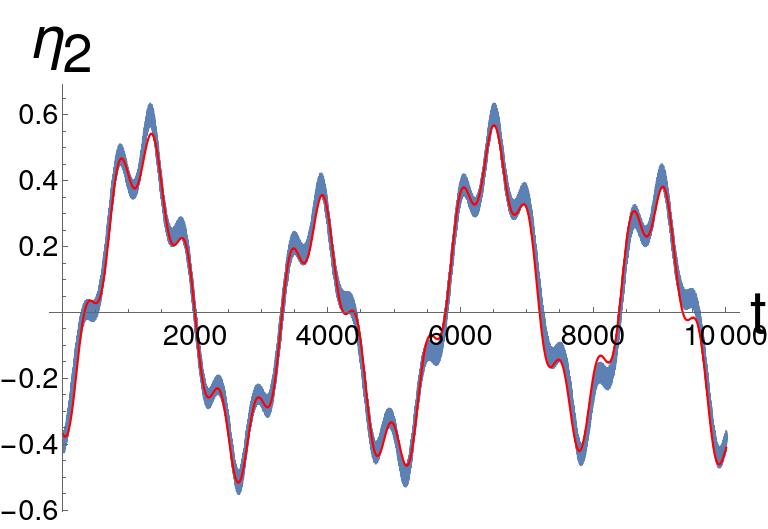}}
}
\end{minipage}
\quad\qquad\qquad
\begin{minipage}{0.4\textwidth}
\resizebox{1.5\textwidth}{!}{
{
\includegraphics[scale=0.56]{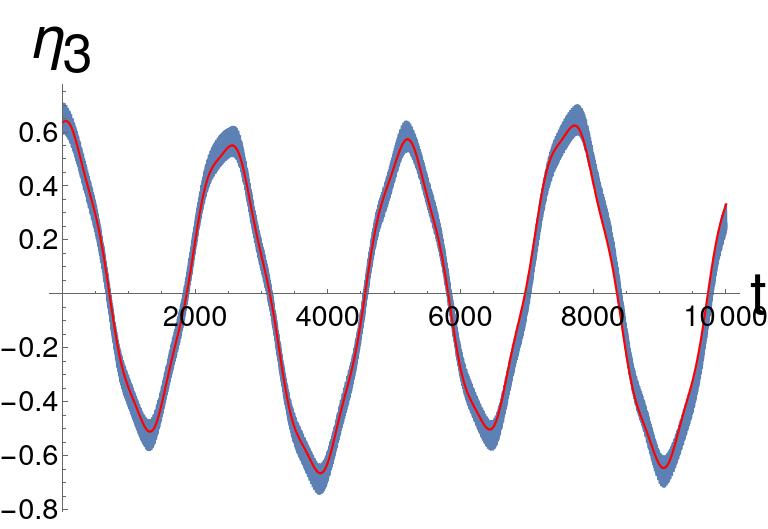}}
}
\end{minipage}\\
\begin{minipage}{0.4\textwidth}
\resizebox{1.5\textwidth}{!}{
{
\includegraphics[scale=0.56]{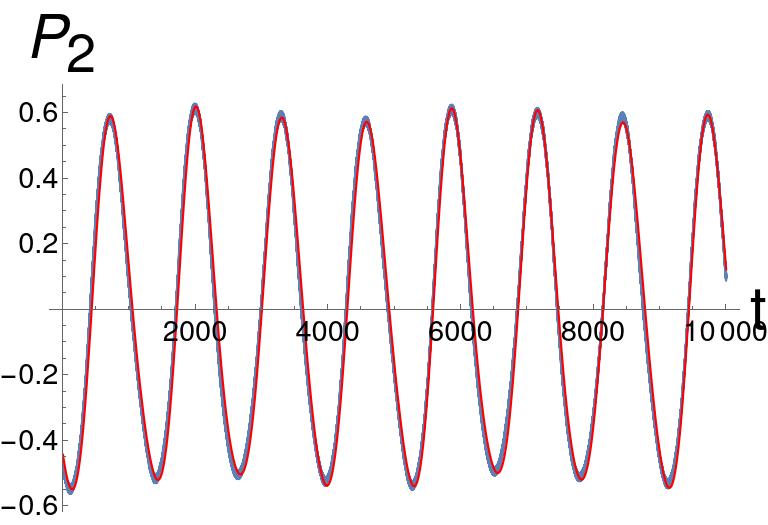}}
}
\end{minipage}
\qquad\qquad\quad
\begin{minipage}{0.4\textwidth}
\resizebox{1.5\textwidth}{!}{
{
\includegraphics[scale=0.56]{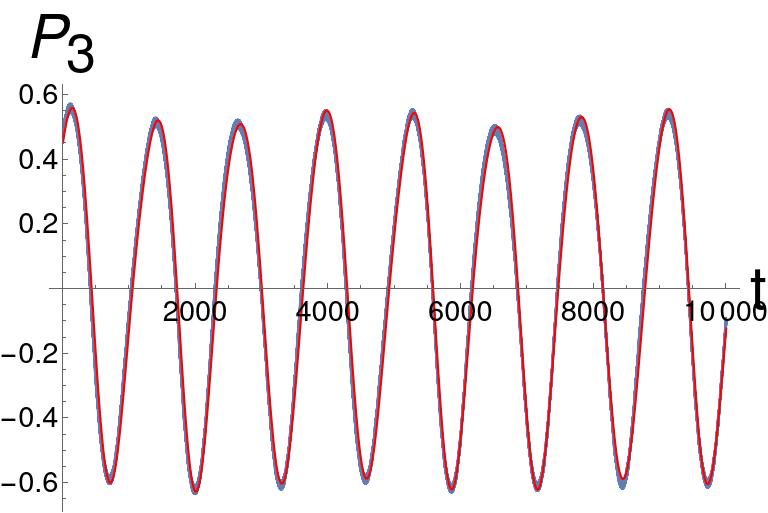}}
}
\end{minipage}
\\
\begin{minipage}{0.4\textwidth}
\resizebox{1.5\textwidth}{!}{
{
\includegraphics[scale=0.56]{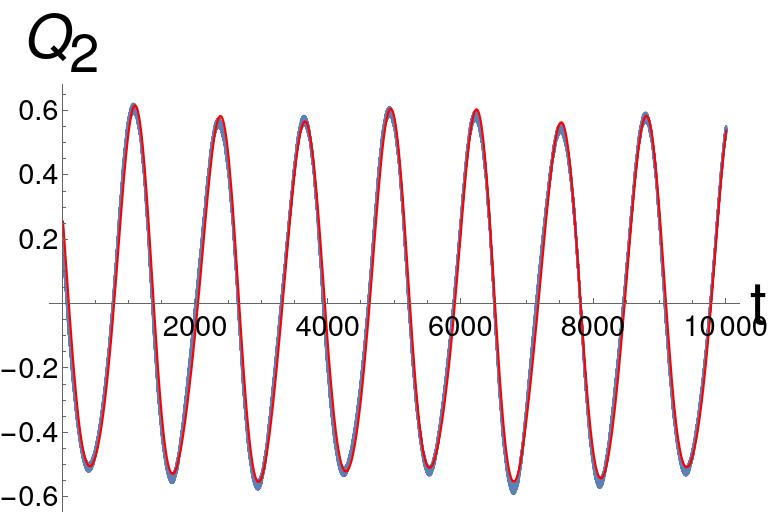}}
}
\end{minipage}
\qquad\qquad\quad
\begin{minipage}{0.4\textwidth}
\resizebox{1.5\textwidth}{!}{
{
\includegraphics[scale=0.56]{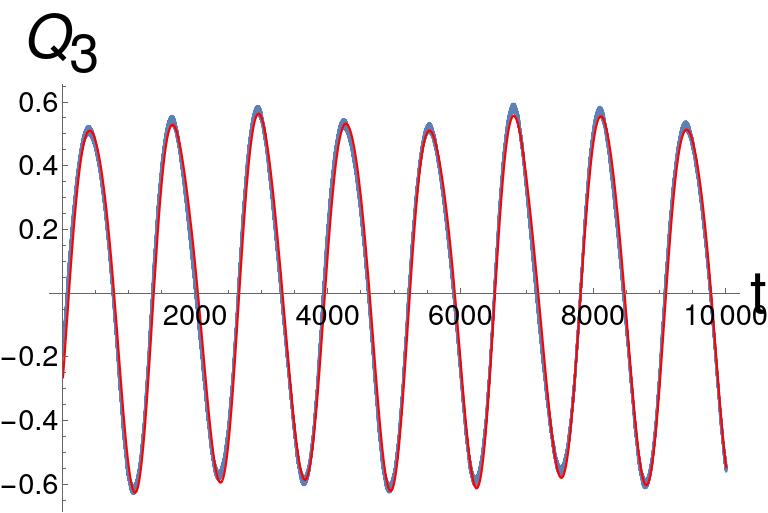}}
}
\end{minipage}
\end{minipage}
\hfill
\begin{minipage}{0.25\textwidth}
\caption{Dynamical behavior of the variables $\csi_2$, $\csi_3$,
  $\eta_2$, $\eta_3$, $P_2$, $P_3$, $Q_2$, $Q_3$ (their definition is
  reported in~\eqref{Poinc.var.U}) as it is computed by numerical
  integration of the complete (non secular) three-body problem and by
  the quasi-periodic approximation provided by the FA; the
  corresponding plots are in blue and red, respectively.}
\label{fig:AINFvstime_behav}
\end{minipage}
\end{figure}

In order to verify that the numerical solutions are well approximated
by the quasi-periodic decompositions computed above, we compare the
time evolution of the variables $\csi_2$, $\csi_3$, $\eta_2$,
$\eta_3$, $P_2$, $P_3$, $Q_2$, $Q_3$ as obtained by the numerical
integration and by the FA. Figure~\ref{fig:AINFvstime_behav} shows
that the quasi-periodic approximations nearly perfectly superpose to
the plots of the numerical solutions.

\section{The secular quasi-periodic restricted (SQPR) Hamiltonian}
\label{sec:QPR}
Having preassigned the motion of the two outer planets \ups$\c$ and \ups$\d\,$, it is now possible to properly define the secular model for a quasi-periodic restricted four-body problem (hereafter, $4$BP). We start from the Hamiltonian of the $4$BP, given by
\begin{equation}
\label{Ham.4BP.reduced.mass}
\Hscr_{4BP}=
\sum_{j=1}^{3}\left( \frac{\vet{p}_j\cdot\vet{p}_j}{2\, \beta_j}
-\frac{\mathcal{G}\, m_0\, m_j}{r_j}\right)
+\sum_{1\leq  i < j \leq 3}\frac{\vet{p_i}\cdot\vet{p_j}}{ m_0}
-\sum_{1\leq i < j \leq 3}\frac{\mathcal{G}\, m_i\, m_j}{|\vet{r_i}-\vet{r_j}|}\, .
\end{equation}
We recall that the so called secular model of order one in the masses
is given by averaging with respect to the mean motion angles, i.e.,
\begin{equation}
\label{average_Ham4BP}
\Hscr_{sec}(\vet{\csi}, \vet{\eta}, \vet{P}, \vet{Q})=
\>\int_{\toro^3}\!\!\!\!\frac{\Hscr_{4BP}(\vet{\Lambda}, \vet{\lambda}, \vet{\csi}, \vet{\eta}, \vet{P}, \vet{Q})}{8\pi^3}\, d \lambda_1 d\lambda_2 d\lambda_3\, .
\end{equation}
In the l.h.s. of the equation above, we disregard the dependence on
the actions $\vet{\Lambda}$, because in the secular approximation of
order one in the masses their values
$\Lambda_j=\beta_j\sqrt{\mu_j\,a_j}\,$, $j=1,2,3$, are constant. Due
to the d'Alembert rules (see, e.g.,~\cite{murder1999}
and~\cite{mor2002}), it is well known that the secular Hamiltonian can
be expanded in the following way:
\begin{equation}
\label{H4BP_expl}
\Hscr_{sec}(\vet{\csi}, \vet{\eta}, \vet{P}, \vet{Q})=
\sum_{s=0}^{\mathcal{N}/2}\sum_{\substack{|\vet{i}|+|\vet{l}|+\quad\,\:\:\\
    |\vet{m}|+|\vet{n}|=2s}}\!\!\!\!\!\!
c_{\vet{i},\vet{l},\vet{m},\vet{n}}
\prod_{j=1}^{3}\csi_j^{i_j}\eta_j^{l_j} P_{j}^{m_j} Q_j^{n_j}\, ,
\end{equation}
where $\mathcal{N}$ is the order of truncation in powers of
eccentricity and inclination. We fix $\mathcal{N}=8$ in all our
computations.

Since we aim at describing the dynamical secular evolution of the
innermost planet \ups$\b\,$, it is sufficient to consider the
interactions between the two pairs \ups$\b$, \ups$\c$ and \ups$\b$,
\ups$\d$. In more details, let $\Hscr_{sec}^{\mathfrak{i}
  -\mathfrak{j}}$ be the secular Hamiltonian derived from the
three-body problem for the planets $\mathfrak{i}$ and $\mathfrak{j}$,
averaging with respect to the mean longitudes
$\lambda_\mathfrak{i}\,$, $\lambda_\mathfrak{j}\,$. Its expansion
writes as
\begin{equation}
\label{Ham_3BP_scissor}
\Hscr^{\mathfrak{i}-\mathfrak{j}}_{sec}(\csi_{\mathfrak{i}}, \eta_{\mathfrak{i}}, P_{\mathfrak{i}}, Q_{\mathfrak{i}},\csi_{\mathfrak{j}}, \eta_{\mathfrak{j}}, P_{\mathfrak{j}}, Q_{\mathfrak{j}})=\sum_{s=0}^{\mathcal{N}/2}\sum_{\substack{|\vet{i}|+|\vet{l}|+\quad\,\:\:\\
|\vet{m}|+|\vet{n}|=2s}}\!\!\!\!\!\! c_{\vet{i},\vet{l},\vet{m},\vet{n}}\prod_{j=\mathfrak{i},\mathfrak{j}}\csi_j^{i_j}\eta_j^{l_j} P_{j}^{m_j} Q_j^{n_j}\, .
\end{equation}
Therefore, a restricted non-autonomous model which approximates the
secular dynamics of \ups$\b$ can be defined by considering the
terms
\begin{equation*}
 \Hscr^{1-2}_{sec}(\csi_1, \eta_1, P_1, Q_1,\csi_2(t), \eta_2(t), P_2(t), Q_2(t))
  +
  \Hscr^{1-3}_{sec}(\csi_1, \eta_1, P_1, Q_1,\csi_3(t), \eta_3(t), P_3(t), Q_3(t))
\end{equation*}
where $\csi_2(t)$, $\eta_2(t)$, \ldots $P_3(t)$, $Q_3(t)$ are replaced
with the corresponding quasi-periodic approximations written in both
the r.h.s. appearing in formula~\eqref{segnalicd}.  Let us stress
that, having prescribed the motion of the two outermost planets
\ups$\c$ and \ups$\d\,$, at this stage the Hamiltonian
$\Hscr_{sec}^{2-3}$ does not need to be reconsidered; indeed, it would
introduce additional terms that disappear in the equations of motion
(see formula~\eqref{campo.Ham.b} which is written below).

We can finally introduce the quasi-periodic restricted Hamiltonian
model for the secular dynamics of \ups$\b\,$;
it is given by the following $2+3/2$ degrees of freedom Hamiltonian:
\begin{equation}
\label{Ham.b.new}
\begin{aligned}
 \Hscr_{sec,\, 2+\frac{3}{2}}(\vet{p},\vet{q},\csi_1, \eta_1, P_1, Q_1 )&=
 \omega_3\,p_3 + \omega_4\,p_4 + \omega_5\,p_5
 \\
 &\phantom{=}+\Hscr^{1-2}_{sec}(\csi_1, \eta_1, P_1, Q_1, \csi_2(\vet{q}),
 \eta_2(\vet{q}), P_2(\vet{q}), Q_2(\vet{q}))
 \\
&\phantom{=}+\Hscr^{1-3}_{sec}(\csi_1, \eta_1, P_1, Q_1,\csi_3(\vet{q}),
 \eta_3(\vet{q}), P_3(\vet{q}), Q_3(\vet{q}))\, ,
\end{aligned}
\end{equation}
where the pairs of canonical coordinates referring to the planets
\ups$\c$ and \ups$\d$ (that are $\csi_2\,$, $\eta_2\,$, \ldots
$P_3\,$, $Q_3$) are replaced by the corresponding finite Fourier
decomposition written in formula~\eqref{segnalicd} as a function of
the angles $\vet{\theta}$, renamed\footnote{This replacement of
  symbols has been done just in order to write three pairs of
  canonical coordinates as $(p_j,q_j)$, $j=3,\,4,\,5$, in agreement
  with the traditional notation that is adopted in many treatises
  about Hamiltonian mechanics.} as $\vet{q}$, i.e.,
\begin{equation}
\label{rename_theta}
\vet{q}=(q_3\,,\,q_4\,,\,q_5):=
(\theta_3\,,\,\theta_4\,,\,\theta_5)=\vet{\theta}\, .
\end{equation}
Let us focus on the summands appearing in the first row
of~\eqref{Ham.b.new}, i.e., the Hamiltonian term
$\vet{\omega}\cdot\vet{p}\,$, where $\vet{\omega}$ is the fundamental
angular velocity vector (defined in formula~\eqref{freq.fond.CD}) and
$\vet{p}=(p_3, p_4, p_5)$ is made by three so called ``dummy
variables'', which are conjugated to the angles $\vet{q}$. The role
they play is made clear by the equations of motion for the innermost
planet, which write in the following way in the framework of the
restricted quasi-periodic secular approximation:
\begin{equation}
\label{campo.Ham.b}
\begin{cases}
  \dot{q_3}=\partial \Hscr_{sec,\,2+\frac{3}{2}}/\partial p_3=\omega_3\\
  \dot{q_4}=\partial \Hscr_{sec,\,2+\frac{3}{2}}/\partial p_4=\omega_4\\
  \dot{q_5}=\partial \Hscr_{sec,\,2+\frac{3}{2}}/\partial p_5=\omega_5\\
  \dot{\csi}_{1}=-\partial \Hscr_{sec,\,2+\frac{3}{2}} /\partial \eta_1=
  -\partial \big( \Hscr^{1-2}_{sec} + \Hscr^{1-3}_{sec} \big) /\partial \eta_1 \\
  \dot{\eta}_{1}=\partial \Hscr_{sec,\,2+\frac{3}{2}} /\partial \csi_1=
  \partial \big( \Hscr^{1-2}_{sec} + \Hscr^{1-3}_{sec} \big) /\partial \csi_1 \\
  \dot{P}_{1}=-\partial \Hscr_{sec,\,2+\frac{3}{2}} /\partial Q_1=
  -\partial \big( \Hscr^{1-2}_{sec} + \Hscr^{1-3}_{sec} \big) /\partial Q_1 \\
  \dot{Q}_{1}=\partial \Hscr_{sec,\,2+\frac{3}{2}}/\partial P_1=
  \partial \big( \Hscr^{1-2}_{sec} + \Hscr^{1-3}_{sec} \big) /\partial P_1
\end{cases}\, .
\end{equation}
Due to the occurrence of the term $\vet{\omega}\cdot\vet{p}$ in the
Hamiltonian $\Hscr_{sec,\,2+\frac{3}{2}}\,$, the first three equations
admit $\vet{q}(t)=\vet{\omega}t$ as a solution, in agreement with
formul{\ae}~\eqref{def_angle_theta_FA} and~\eqref{rename_theta}.  This
allows to reinject the wanted quasi-periodic time-dependence in the
Fourier approximations $\csi_2(\vet{q})$, $\eta_2(\vet{q})$, $\ldots$
$P_3(\vet{q})$, $Q_3(\vet{q})$.  As a matter of fact, we do not need
to compute the evolution of $({p}_3(t),{p}_4(t),{p}_5(t))$ because
they do not exert any influence on the motion of \ups$\b\,$; they are
needed just if one is interested in checking that the energy is
preserved, because it is given by the evaluation of
$\Hscr_{sec,\,2+\frac{3}{2}}\,$.

We also recall that, in order to produce a restricted quasi-periodic
secular model, it is possible to apply the \textit{closed-form
  averaging}, which is compared in~\cite{mas2023} with the
computational method that is adopted here and is based on the
expansions in Laplace coefficients. Finally, we emphasize what is
discussed below.

\begin{remark}
  \label{nota:Noether->momentoangolare}
The Hamiltonian $\Hscr_{sec,\,2+\frac{3}{2}}$ is invariant with
respect to a particular class of rotations. Thus, it admits a
constant of motion that could be reduced, so to decrease\footnote{This
  reduction is performed in Chap.~6 of~\cite{mas2023} in such a way to
  introduce a further simplified model with $2+2/2$ degrees of
  freedom. In the present work, we prefer not to perform such a
  reduction, in order to make the role of the angular (canonical)
  variables more transparent, clarifying their meaning
  for what concerns the positions of the exoplanets.} by one the number of
degrees of freedom of the model.

In order to clarify the statement above, it is convenient
to introduce a complete set of action-angle variables, defining two
new pairs of canonical coordinates
$\csi_1=\sqrt{2\Gamma_1}\cos(\varpi_1)$,
$\eta_1=\sqrt{2\Gamma_1}\sin(-\varpi_1)$,
$P_1=\sqrt{2\Theta_1}\cos(\Omega_1)$,
$Q_1=\sqrt{2\Theta_1}\sin(-\Omega_1)$, referring to a pair of
orbital angles of \ups$\b$, i.e., $\varpi_1$ and $\Omega_1\,$, that
are the longitudes of the pericenter and of the node,
respectively (see definition~\eqref{Poinc.var.U} of the Poincar\'e
canonical variables). Thus, it is possible to verify the following
invariance law:
\begin{equation*}
\vcenter{\openup1\jot\halign{
 \hbox {\hfil $\displaystyle {#}$}
&\hbox {$\displaystyle {#}$\hfil}\cr
  &-\frac{\partial\,\Hscr_{sec,\,2+\frac{3}{2}}}{\partial\xi_1}\,
  \,\frac{\partial\,\xi_1}{\partial\varpi_1}-
  \frac{\partial\,\Hscr_{sec,\,2+\frac{3}{2}}}{\partial\eta_1}\,
  \frac{\partial\,\eta_1}{\partial\varpi_1}-
  \frac{\partial\,\Hscr_{sec,\,2+\frac{3}{2}}}{\partial P_1}\,
  \frac{\partial P_1}{\partial\Omega_1}-
  \frac{\partial\,\Hscr_{sec,\,2+\frac{3}{2}}}{\partial Q_1}\,
  \frac{\partial Q_1}{\partial\Omega_1}
  \cr
  &\qquad\qquad\ \ +
  \frac{\partial\,\Hscr_{sec,\,2+\frac{3}{2}}}{\partial q_3}+
  \frac{\partial\,\Hscr_{sec,\,2+\frac{3}{2}}}{\partial q_4}+
  \frac{\partial\,\Hscr_{sec,\,2+\frac{3}{2}}}{\partial q_5}\,=\,0\,\,.
  \cr
}}
\end{equation*}
Therefore, $p_3+p_4+p_5+\Gamma_1+\Theta_1$ is preserved; such a quantity, apart from an inessential
additional constant, is
equivalent to the total angular momentum. 

The above invariance law is better
understood, recalling that $q_3$ and $q_4$
correspond to the longitudes of the pericenters of \ups$\c$ and
\ups$\d$, respectively, while $q_5$ refers to the longitude of the
nodes of \ups$\c$ and \ups$\d$ (that in the
Laplace frame, determined by taking into account only these two
exoplanets, are opposite one to the other). This identification is due to the way we have determined
$(q_3,q_4,q_5)$ by decomposing some specific signals of the secular
dynamics of the outer exoplanets (this is made by using the Frequency
Analysis, as it is explained in Section~\ref{subsec:AINF}).
Thus, the aforementioned invariance law is due to the fact that the
dynamics of the model we are studying does depend just on the
pericenters arguments of the three exoplanets and on the difference
between the longitude of the nodes of \ups$\b$ and \ups$\c$, i.e.,
$\Omega_1-\Omega_2=\Omega_1-\Omega_3-\pi$. Therefore, the Hamiltonian
is invariant with respect to any rotation of the same angle that is
applied to all the longitudes of the nodes; as it is well known, by Noether theorem, this
is equivalent to the preservation of the total angular momentum.
\end{remark}

\subsection{Numerical validation of the SQPR model}
\label{sub:numerical_comparison}
In order to validate our secular quasi-periodic restricted (hereafter
SQPR) model describing the dynamics of \ups$\b$, we want to compare
the numerical integrations of the complete 4BP with the ones of the
equations of motion~\eqref{campo.Ham.b}. Let us recall that the chosen
values of parameters and initial conditions for the two outer planets
are given in Table~\ref{tab:param.orb}. For what concerns the orbital
elements of the innermost planet \ups$\b$, both the inclination $\i_1$
and the longitude of the ascending node $\Omega_1$ are unknown (see,
e.g.,~\cite{mcaetal2010}). The available data for \ups$\b$ are
reported in Table~\ref{tab:param.orb.poss.b}.  Among the possible
values of the initial orbital parameters of \ups$\b$, we have chosen
$a_1\,$, $\e_1$, $M_1$ and $\omega_1$ as in the stable prograde trial
PRO2 described in~\cite{deietal2015}. They are reported in
Table~\ref{tab:param.orb.b} and are compatible with the available
ranges of values appearing in Table~\ref{tab:param.orb.poss.b}. Let us
recall that the dynamical evolution of the SQPR model does not depend
on the mass of \ups$\b$, therefore, the choice about its value is not
reported in Table~\ref{tab:param.orb.b}. For what concerns the unknown
initial values of the inclination and of the longitude of nodes, we
have decided to vary them so as to cover a $2$D regular grid of values
$(\i_1(0),\,\Omega_1(0))\in I_\i\times I_\Omega=[6.865^\circ,
  34^\circ]\times[0^\circ,360^\circ]$, dividing $I_\i$ and $I_\Omega$,
respectively, in $20$ and $60$ sub-intervals; this means that the
widths of the grid-steps are equal to $1.35675^\circ$ and $6^\circ$ in
inclination and longitude of nodes, respectively.  Let us recall that
the lowest possible value of the interval $I_\i$,
i.e. $\i_2(0)=6.865^\circ$, corresponds to the inclination of
\ups$\c$. Considering values smaller than $\i_2(0)$ could be
incoherent with the assumptions leading to the SPQR model we have just
introduced; indeed, the factor $1/\sin(\i_1(0))$ increases the mass of
the exoplanet by one order of magnitude with respect to the minimal
one. Therefore, for small values of $\i_1(0)$ the mass of \ups$\b$
could become so large that the effects exerted by its gravitational
attraction on the outer exoplanets could not be neglected anymore. On
the other hand, it will be shown in the sequel that the stability
region for the orbital motion of \ups$\b$ nearly completely disappears
for values of $\i_1(0)$ larger than $34^\circ$. These are the reasons
behind our choice about the lower and upper limits of $I_\i$.

\begin{table}[!h]
\begin{minipage}{0.45\textwidth}
\begin{center}
\begin{tabular}{ll}
\toprule
 & $\upsilon$-And $\bf{b}$  \\
\midrule
$m \,\sin(\i)\, [M_J]$ & $0.69 \pm 0.016$ \\
$a(0)\,[AU] $ & $0.0594\pm 0.0003$ \\
$\e(0)$ & $0.012\pm 0.005$ \\
$\omega(0)\, [^{\circ}]$ & $44.106\pm 25.561$ \\
\bottomrule
\end{tabular}
\caption{Available data for the orbital parameters of the exoplanet
  \ups$\b$. The values above are reported from Table~$13$
  of~\cite{mcaetal2010}.}
\label{tab:param.orb.poss.b}
\end{center}
\end{minipage}
\hfill
\begin{minipage}{0.47\textwidth}
\begin{center}
\begin{tabular}{ll}
\toprule
 & $\upsilon$-And $\bf{b}$  \\
\midrule
$a(0) \, [AU]$ & $0.0594$  \\
$\e(0)$ & $0.011769$ \\
$M(0)\, [^{\circ}]$ & $103.53$ \\
$\omega(0)\, [^{\circ}]$ & $51.14$ \\
\bottomrule
\end{tabular}
\caption{Values of the initial orbital parameters for \ups$\b$ as they
  have been selected in the stable prograde trial PRO2
  of~\cite{deietal2015} (Table~$3\,$).}
\label{tab:param.orb.b}
\end{center}
\end{minipage}
\end{table}

We emphasize that the study of the stability domain, as it is deduced
by the numerical integrations, can help us to obtain information about
the possible ranges of the unknown values
$(\i_1(0),\Omega_1(0))$. Moreover, the comparisons between the
numerical integrations of the complete $4$BP and the ones of the SQPR
model aim to demonstrate that the agreement is sufficiently good so
that it becomes possible to directly work with the latter Hamiltonian
model, that has to be considered easier than the former, because the
degrees of freedom are $2+3/2$ instead of $9$.

\subsubsection{Numerical integration of the complete $4$-body problem}
\label{subsub:num_integ_compl}
For each pair of values $(\i_1(0)\,,\Omega_1(0))\in I_\i\times
I_\Omega$ ranging in the $20\times 60$ regular grid we have previously
prescribed, we first compute the corresponding initial orbital
elements of the three exoplanets in the Laplace-reference frame, then
we perform the numerical integration of the complete 4BP
Hamiltonian~\eqref{Ham.4BP.reduced.mass} by using the symplectic
method of type $\mathcal{SBAB}_3\,$. Contrary to the SPQR model, the
numerical integrations of the 4BP are affected by the mass of
\ups$\b$; its value is simply fixed in such a way that
$m_1=0.674/\sin(\i_1(0))\,$.

\begin{figure}[h]
\subfloat[]{
\begin{minipage}{.45\textwidth}
\includegraphics[scale=0.295]
{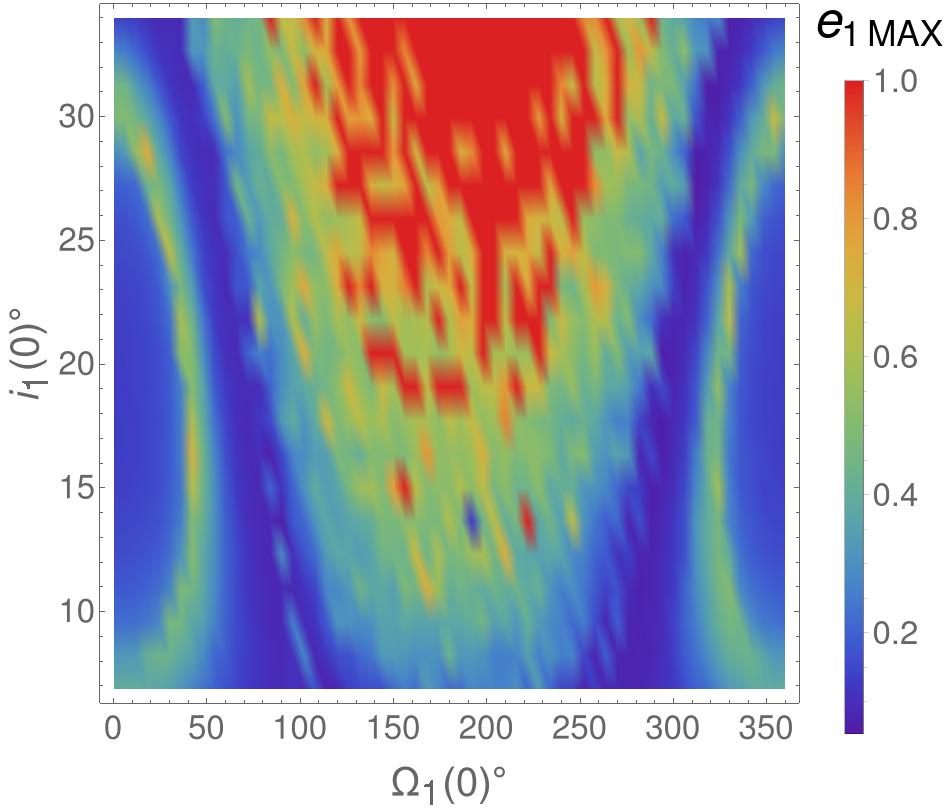}
\vskip -.05truecm
\label{fig.Color_compl_e1}
\end{minipage}}\qquad
\subfloat[]
{
\begin{minipage}{.45\textwidth}
\includegraphics[scale=0.295]
{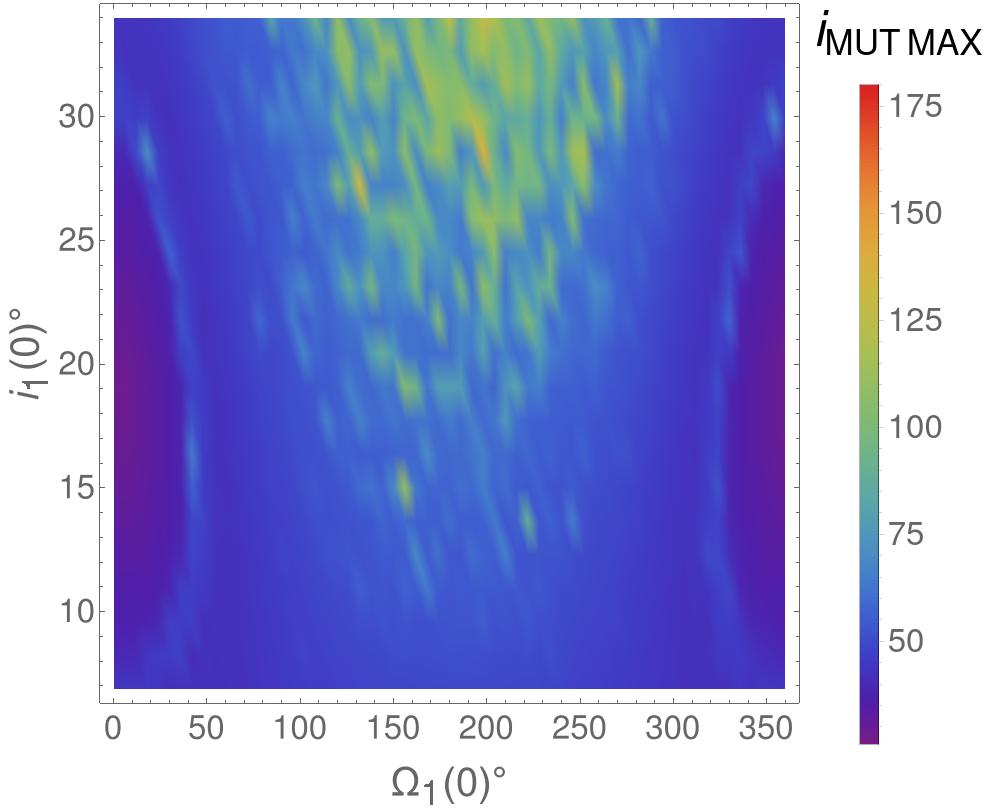}
\vskip -.05truecm
\label{fig.Color_compl_imut}
\end{minipage}
}
\vskip -.2truecm
\caption{Color-grid plots of the maximal value reached by the
  eccentricity of \ups$\b$ (on the left) and by the mutual inclination
  between \ups$\b$ and \ups$\c$ (on the right). The maxima are
  computed during the symplectic numerical integrations of the 4BP
  which cover a timespan of $10^5$~yr.}
\label{fig.Color_compl_e1_imut}
\end{figure}

The largest value reached by the eccentricity can be considered as a
very simple numerical indicator about the stability of the orbital
configurations. The maximum eccentricity obtained along each of the
$21\times 60$ numerical integrations is reported in the left panel of
Figure~\ref{fig.Color_compl_e1_imut}. In particular, since we are
interested in initial conditions leading to regular behavior, i.e.,
avoiding quasi-collisions, every time that the eccentricity $\e_1$
exceeds a threshold value (fixed to be equal to $0.85$), in the
color-grid plots its maximal value is arbitrarily set equal to
one. Moreover, since we expect that \ups$\c$ is the most massive
planet in that extrasolar system and being it the closest one to
\ups$\b$, it is natural to focus the attention also on the mutual
inclination between \ups$\b$ and \ups$\c\,$. Let us recall that it is
defined in such a way that
\begin{equation}
\label{cosimut.bc}
\cos(\i_{{mut}_{\b\c}})=
\cos(\i_1)\cos(\i_2)+\sin(\i_1)\sin(\i_2)\cos(\Omega_1-\Omega_2)\, ;
\end{equation}
therefore, for each numerical integration it is also possible to
compute the maximal value reached by $\i_{{mut}_{\b\c}}\,$. The
results are reported in the right panel of
Figure~\ref{fig.Color_compl_e1_imut}.  In both those panels the
color-grid plots are provided as functions of the initial values of
the longitude of nodes $\Omega_1$ and the inclination $\i_1\,$, which
are reported on the $x$ and $y$ axes, respectively. By comparing the
two plots in the Figure panels~\ref{fig.Color_compl_e1}
and~\ref{fig.Color_compl_imut}, one can easily appreciate that the
regions which have to be considered as dynamically unstable, because
the eccentricity of $\e_1$ can grow to large values, correspond also to
large mutual inclinations of the planetary orbits of \ups$\b$ and
\ups$\c$.

\begin{figure}[h]
\begin{minipage}{.32\textwidth}
\includegraphics[scale=0.23]{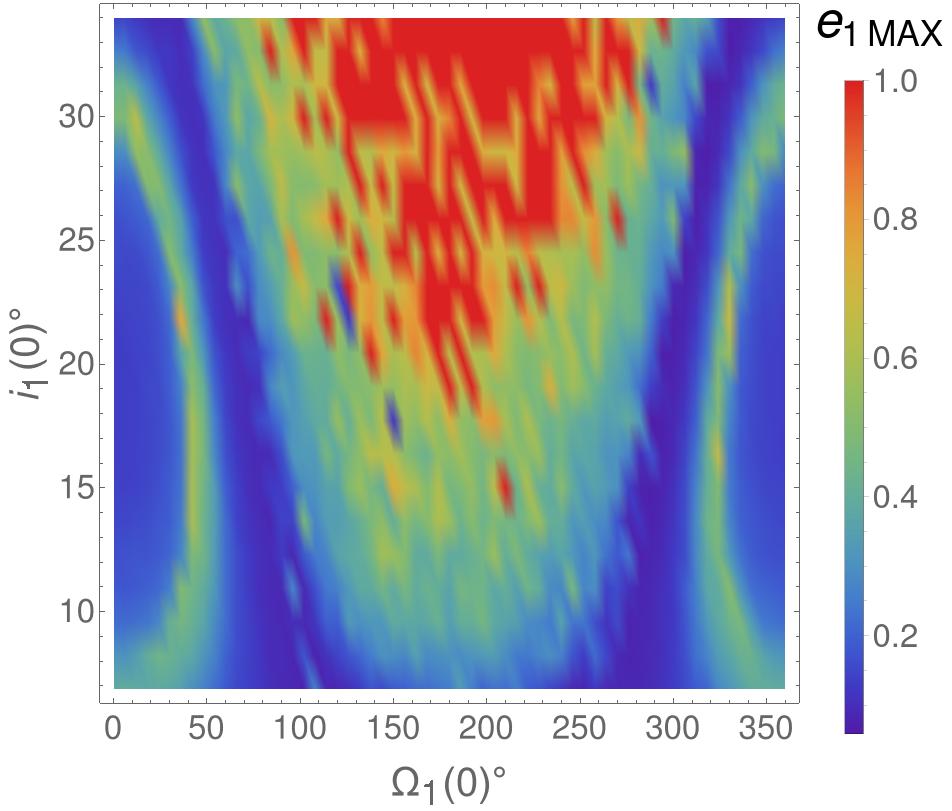}
\end{minipage}
\begin{minipage}{.32\textwidth}
\includegraphics[scale=0.23]{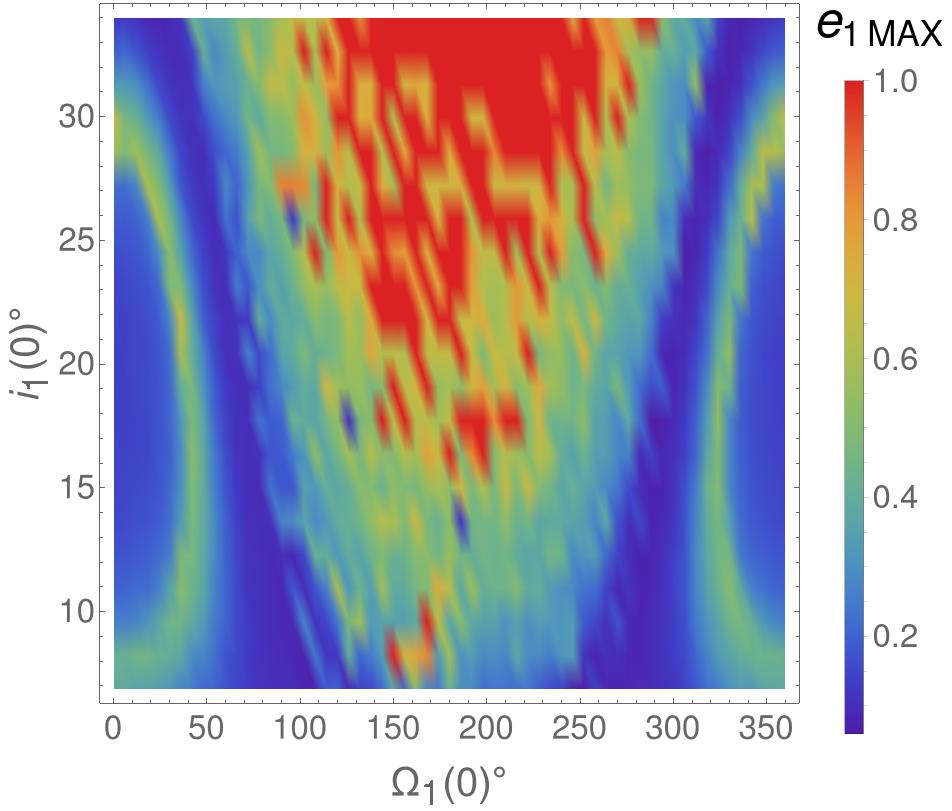}
\end{minipage}
\begin{minipage}{.32\textwidth}
\includegraphics[scale=0.23]{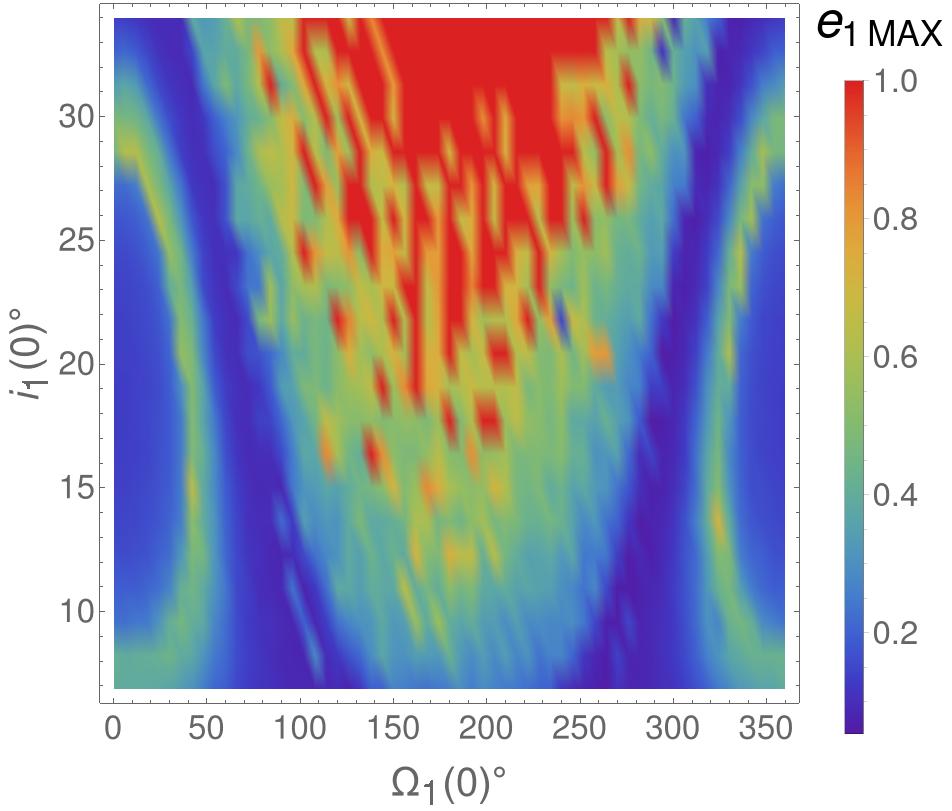}
\end{minipage}
\vskip -.2truecm
\caption{Color-grid plots of the maximal value reached by the
  eccentricity of \ups$\b$. The maxima are computed during the
  symplectic numerical integrations of the 4BP which cover a timespan
  of $10^5$~yr. The results are obtained by numerical integrations
  which refer to sets of initial conditions that differ (passing from
  one panel to another) just because of the choice of the initial
  values of the mean anomaly; from left to right the plots refer to
  $M_1(0)$ equal to $51.4286^\circ$, $205.714^\circ$, and
  $308.571^\circ$, respectively.}
\label{fig.Color_compl_e1_M}
\end{figure}

We remark that the value of the initial mean anomaly $M_1(0)$ is
missing among the available observational data reported in
Table~\ref{tab:param.orb.poss.b}. As a matter of fact, mean anomalies
of exoplanets are in general so poorly known that usually their values
are not reported in the public databases.\footnote{See, e.g., {\tt
    http://exoplanet.eu/}} However, in order to understand if (and up
to what extent) the initial value $M_1(0)$ can affect the dynamics of
\ups$\b\,$, we repeat all the numerical integrations of the $4$BP for
four different initial values of $M_1\,$, chosen so as to have one of
them belonging to each of the quadrants $[0^\circ, 90^\circ]\,$,
$[90^\circ, 180^\circ]\,$, $[180^\circ, 270^\circ]\,$ and $[270^\circ,
  360^\circ]\,$. In Figure~\ref{fig.Color_compl_e1_M} we report three
examples; in particular, they show the color-grid plots of the maximal
value reached by the eccentricity $\e_1\,$, taking $M_1(0)$ as
$360^\circ/7= 51.4286^\circ$, $4\cdot 360^\circ/7= 205.714^\circ$ and
$6\cdot 360^\circ/7=308.571^\circ$, respectively. For what concerns
the region $[90^\circ, 180^\circ]$, let us recall that
Figure~\ref{fig.Color_compl_e1} refers to $M_1(0)=103.53^\circ$. The
comparison between Figures~\ref{fig.Color_compl_e1}
and~\ref{fig.Color_compl_e1_M} shows that the choice of the value of
$M_1(0)$ does not seem to produce any remarkable impact on the global
structure of the dinamical stability of these exoplanetary
orbits. 

\begin{figure}[h]
\subfloat[]{
\begin{minipage}{.45\textwidth}
\includegraphics[scale=0.295]
{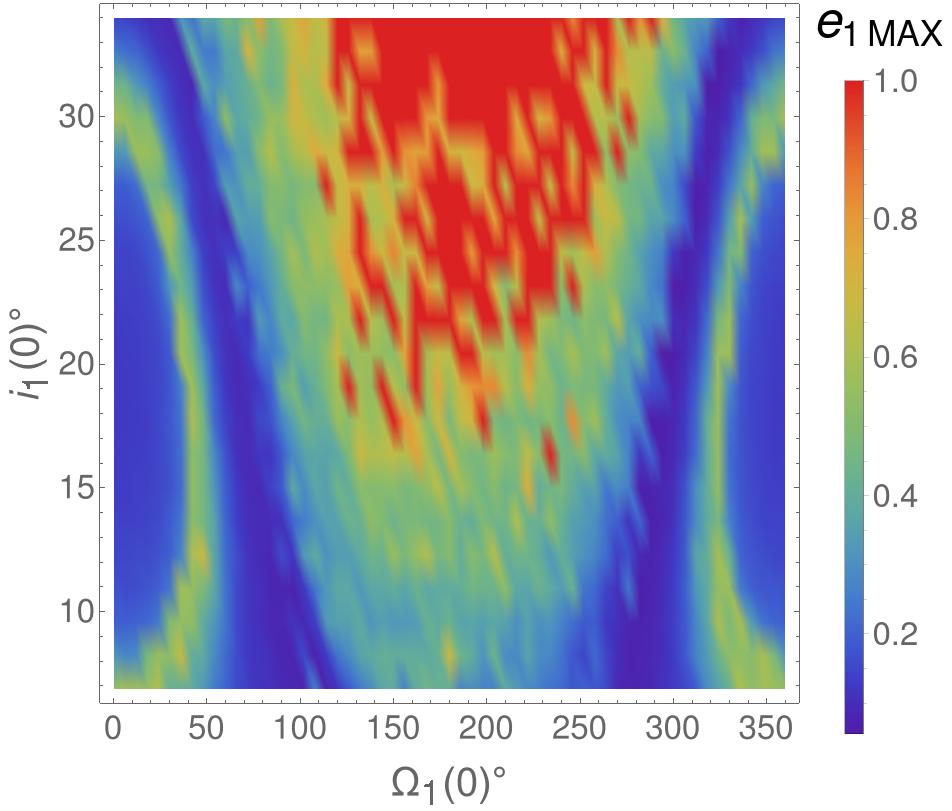}
\vskip -.05truecm
\label{fig.Color_compl_e1_mfix}
\end{minipage}}\qquad
\subfloat[]
{
\begin{minipage}{.45\textwidth}
\includegraphics[scale=0.295]
{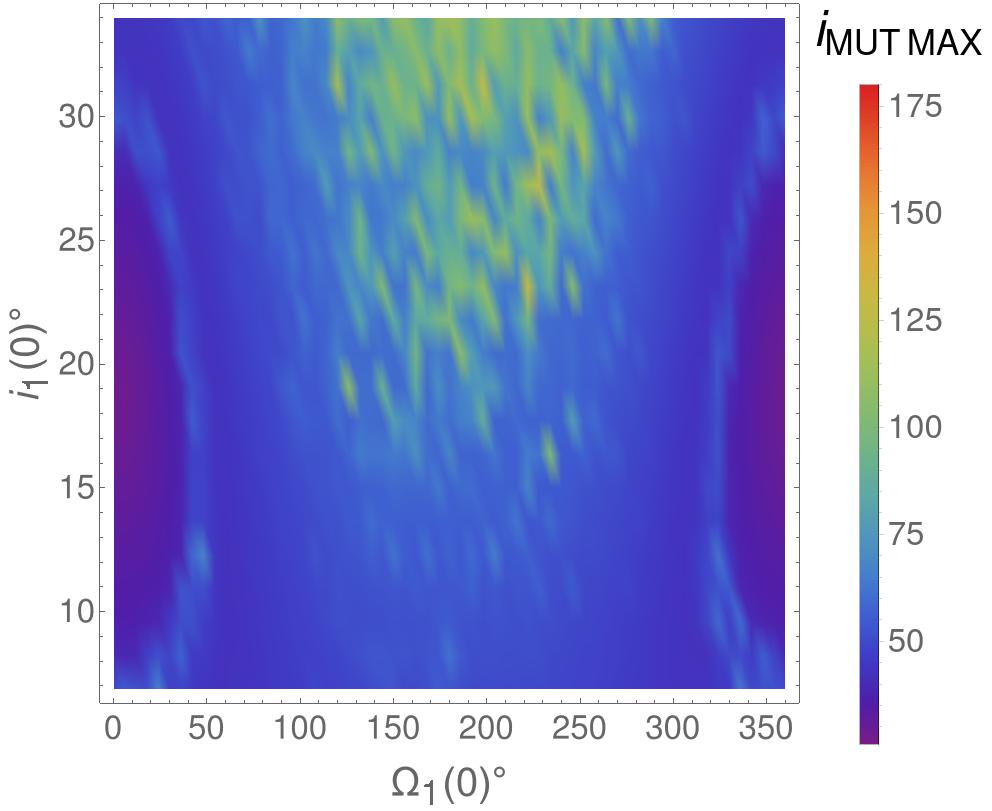}
\vskip -.05truecm
\label{fig.Color_compl_imut_mfix}
\end{minipage}
}
\vskip -.2truecm
\caption{Color-grid plots of the maximal value reached by the
  eccentricity of \ups$\b$ (on the left) and by the mutual inclination
  between \ups$\b$ and \ups$\c$ (on the right). The maxima are
  computed during the symplectic numerical integrations of the 4BP which
  cover a timespan of $10^5$~yr. As the only difference with respect
  to the numerical integration whose results are reported in
  Figure~\ref{fig.Color_compl_e1_imut}, here the mass of \ups$\b$ is
  always kept fixed so as to be equal to its minimal value
  $m_1=0.674\,$.}
\label{fig.Color_compl_e1_imut_mfix}
\end{figure}

Moreover, the same conclusion applies also to the increasing
factor $1/\sin(\i_1(0))$ (with $\i_1(0)\in I_\i$) which multiplies the
minimal mass of \ups$\b\,$ in such a way to determine the value of
$m_1\,$. In fact, substantial differences are not observed between
Figures~\ref{fig.Color_compl_e1_imut}
and~\ref{fig.Color_compl_e1_imut_mfix}.

\subsubsection{Numerical integration of the secular quasi-periodic restricted model}
\label{subsub:num_integ_qper}

We want now to compare the previous results with those found in the
SQPR approximation of the $4$-body problem, performing numerical
integrations of the system of equations~\eqref{campo.Ham.b}. In order
to make these comparisons coherent, also here we consider the data
listed in Table~\ref{tab:param.orb.b} as initial conditions for the
orbital elements of \ups$\b$ which are completed with the values of
$(\i_1(0)\,,\Omega_1(0))$ ranging in the $20\times 60$ regular grid
that covers $I_\i\times I_\Omega=[6.865^\circ,
  34^\circ]\times[0^\circ,360^\circ]$. At the beginning of the
computational procedure, the initial values of the orbital elements
are determined in the Laplace reference frame, which is fixed by
taking into account \textit{only} the two outermost planets (i.e., the
total angular momentum of the system is given \textit{only} by the sum
of the angular momentum of \ups$\c$ and \ups$\d\,$). Of course, this
is made in agreement with our choice to consider a \textit{restricted}
framework, because we are assuming that the mass of \ups$\b$ is so
small that can be neglected.

\begin{figure}[h]
\subfloat[]{
\begin{minipage}{.45\textwidth}
\includegraphics[scale=0.295]
{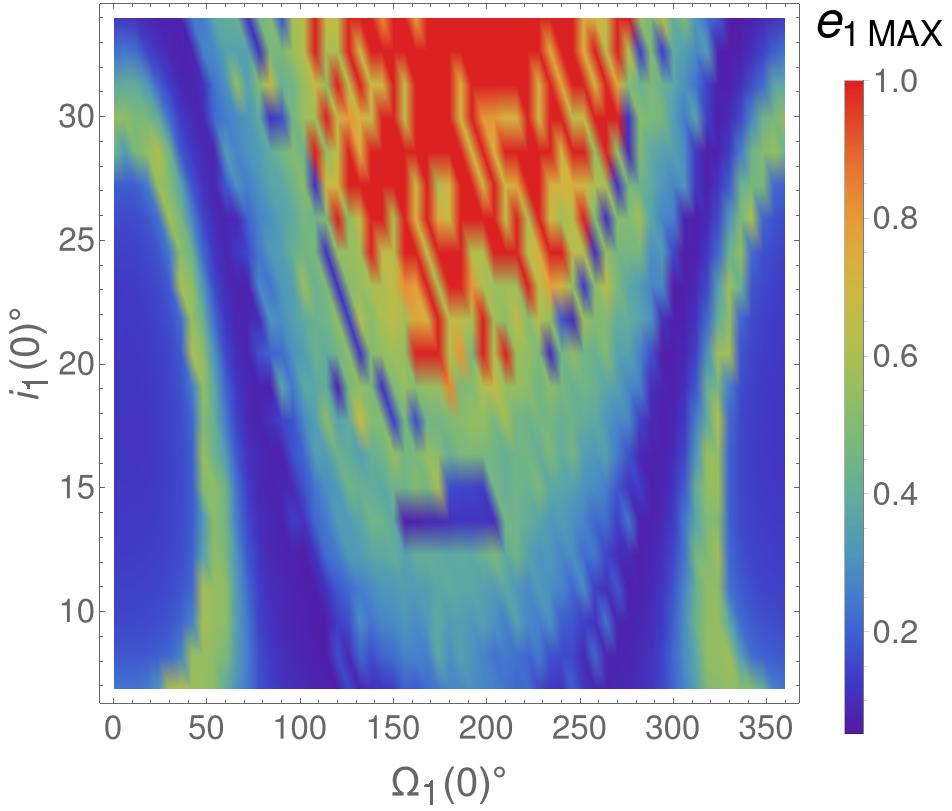}
\vskip -.05truecm
\label{fig.Color_qper_e1}
\end{minipage}}\qquad
\subfloat[]
{
\begin{minipage}{.45\textwidth}
\includegraphics[scale=0.295]
{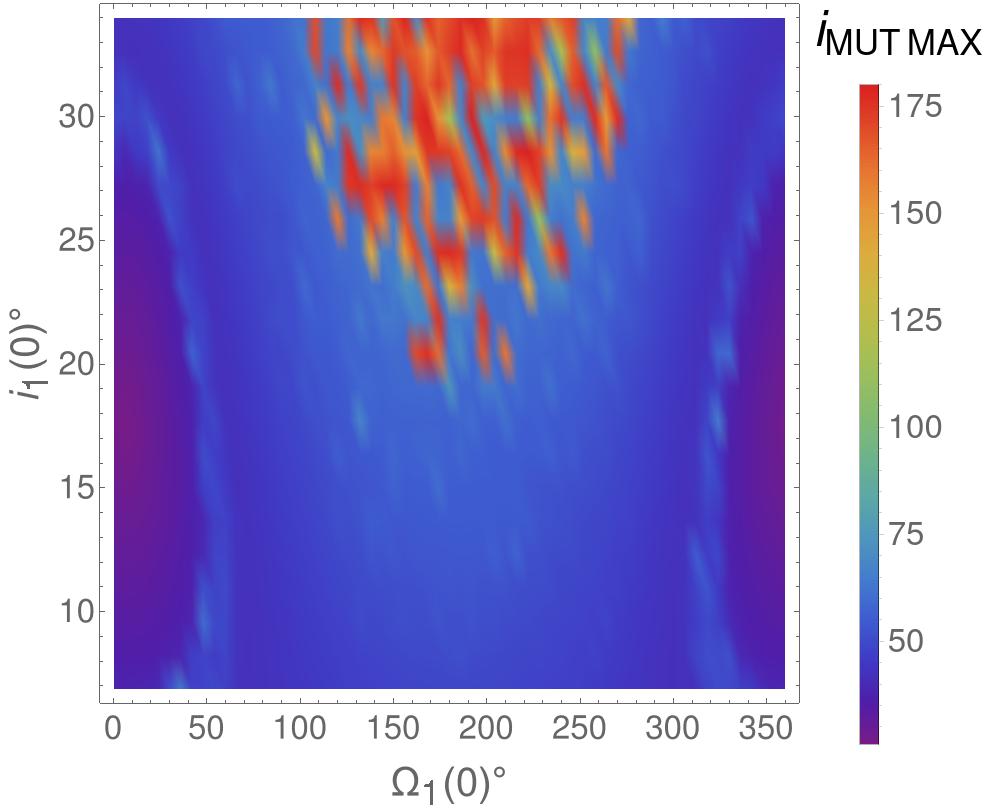}
\vskip -.05truecm
\label{fig.Color_qper_imut}
\end{minipage}
}
\caption{Color-grid plots of the maximal value reached by the
  eccentricity of \ups$\b$ (on the left) and by the mutual inclination
  between \ups$\b$ and \ups$\c$ (on the right). The maxima are
  computed along the RK4 numerical integrations of the equations of
  motion~\eqref{campo.Ham.b} of the SQPR model, covering
  a timespan of $10^5$ yr.}
\label{fig.Color_qper_e1_imut}
\end{figure}

For each numerical integration we compute the maximal value reached by
the eccentricity $\e_1$ and the mutual inclination
$\i_{{mut}_{\b\c}}\,$. The results are reported in the color-grid
plots of the left and right panels of
Figure~\ref{fig.Color_qper_e1_imut}, respectively. Once again, they
are provided as functions of $\Omega_1(0)$ and $\i_1(0)\,$, whose
values appear on the $x$ and $y$ axes, respectively.

Comparing Figures~\ref{fig.Color_compl_e1}
with~\ref{fig.Color_qper_e1} and~\ref{fig.Color_compl_imut}
with~\ref{fig.Color_qper_imut}, respectively, we can immediately
conclude the striking similarity of the color-grid plots, implying the
same dependence of the dynamics on the initial values of the orbital
elements $\i_1(0)$ and $\Omega_1(0)\,$ in either model. In particular,
the regions of stability located at the two lateral sides of the plots, where the orbit of \ups$\b$ does not become very
eccentric, are identical. This occurs also for what concerns the plots of the maximal
mutual inclination. However, some discrepancies are evident in the
central parts of the panels, i.e. for values of $\Omega_1(0)$ ranging
between $90^\circ$ and $270^\circ$. We stress that this lack of
agreement between the results provided by the two models is expected
in these central regions of the panels. Indeed, let us recall that the
SQPR model has been introduced starting from some classical expansions
in powers of eccentricities and inclinations. Therefore, it is
reasonable to expect a deterioration of the accuracy of the SQPR model
in the orbital dynamics depicted in the central regions of the plots
where large values of the eccentricity $\e_1$ and the mutual
inclination are attained. We emphasize that similar remarks about the
very strong impact of the initial value  $\Omega_1(0)$ on the orbital
stability of \ups$\b$ can be found in Section~4.2
of~\cite{pisetal2017}.

\begin{figure}[h]
\subfloat[]{
\begin{minipage}{.45\textwidth}
\includegraphics[scale=0.42]
{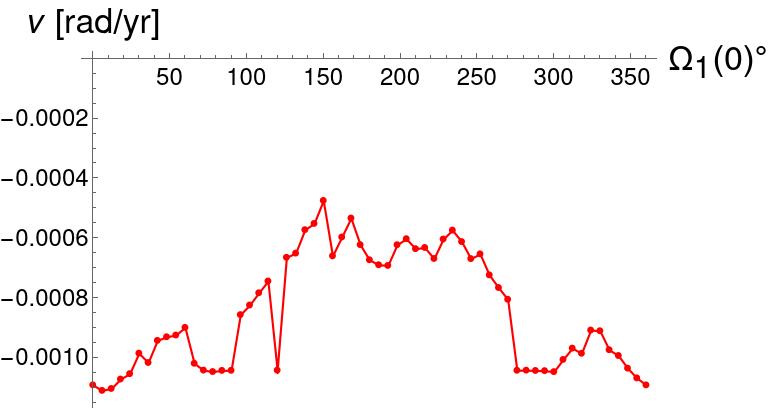}
\vskip -.05truecm
\label{fig.plotfreqNOREL_short}
\end{minipage}}\qquad
\subfloat[]
{
\begin{minipage}{.45\textwidth}
\includegraphics[scale=0.42]
{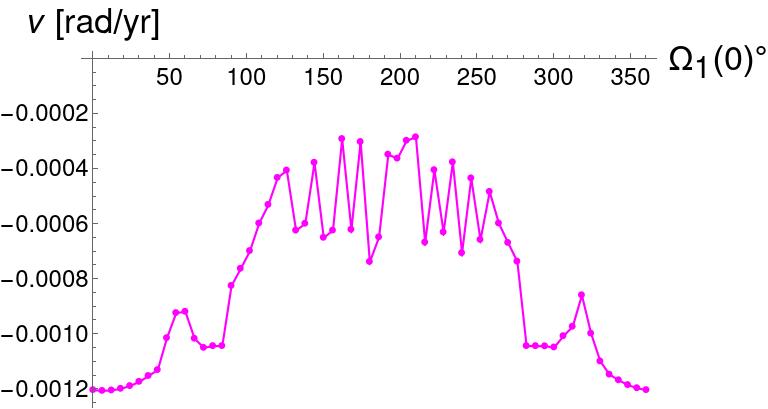}
\vskip -.05truecm
\label{fig.plotfreqNOREL_high}
\end{minipage}
}
\caption{Behaviour of the fundamental angular velocity
    $\nu$ as obtained by applying the Frequency Map Analysis method to
    the signal $\xi_1(t)+i\eta_1(t)\,$, computed through the RK4
    numerical integration of the SQPR model~\eqref{campo.Ham.b},
    covering a timespan of $1.31072\cdot 10^{5}$ yr. We take, as
    initial conditions, $(\i_1(0),\Omega_1(0))\in\lbrace 6.865^\circ
    \rbrace \times I_\Omega$ for the left panel and
    $(\i_1(0),\Omega_1(0))\in\lbrace 10.9353^\circ \rbrace \times
    I_\Omega$ for the right one.}
\label{fig.plotfreqNOREL}
\end{figure}

A further exploration of the stable and chaotic
  regions of Figure~\ref{fig.Color_qper_e1} can be done by applying
  the so called Frequency Map Analysis method (see,
  e.g.,~\cite{las1999}), in order to study the signal
  $\xi_1(t)+i\eta_1(t)$ produced by the numerical integration of the
  system~\eqref{campo.Ham.b}, i.e., in the SQPR approximation. We
  perform the numerical integrations as prescribed at the beginning of
  the present Section, taking into account only a few values in
  $I_\i\,$ for the initial inclinations, i.e., $\i_1(0)=6.865^\circ,\,
  8.22175^\circ,\,9.5785^\circ,\,10.9353^\circ$ and $\Omega_1(0)\in
  I_\Omega\,$. In Figure~\ref{fig.plotfreqNOREL} we report the
  behaviour of the angular velocity corresponding to the first
  component of the approximation of $\xi_1(t)+i\eta_1(t)$, as obtained
  by applying the FA computational algorithm; therefore, this quantity
  is related to the precession rate of $\varpi_1\,$. As initial value
  for the inclination $\i_1(0)$ we fix $6.865^\circ$ for
  Figure~\ref{fig.plotfreqNOREL_short} and $10.9353^\circ$ for
  Figure~\ref{fig.plotfreqNOREL_high}. We do not report the cases
  $(\i_1(0),\Omega_1(0))\in \lbrace 8.22175^\circ,\, 9.5785^\circ
  \rbrace \times I_\Omega\,$, since the behaviour of those plots is
  similar to the ones in Figure~\ref{fig.plotfreqNOREL}. 

\noindent
The situation is well described in
  Figure~\ref{fig.plotfreqNOREL_high} and analogous considerations can
  be done for Figure~\ref{fig.plotfreqNOREL_short}. For what concerns
  the values of $\Omega_1(0)$ in the range $[0, \sim 50^\circ]$ and
  $[\sim 325^\circ, \, 360^\circ]$ we can observe a regular behaviour
  of the angular velocity $\nu$ which is also monotone with the only
  exception of the local minimum. According to the interpretation of
  the Frequency Map Analysis (in the light of KAM theory), such a
  regular regime is due to the presence of many invariant tori which
  fill the stability region located at the two lateral sides of the
  plot~\ref{fig.Color_qper_e1}. Instead for values of $\Omega_1(0)$ in
  $[\sim 50^\circ, \sim 70^\circ ] \cup [\sim 300^\circ, \, \sim
    325^\circ]$ and $\Omega_1(0)$ in $[\sim 120^\circ, \, \sim
    270^\circ]$ we observe a strongly irregular behaviour, which
  corresponds to the lateral green stripes and the internal green region of
  Figure~\ref{fig.Color_qper_e1}. Thus, they represent
  chaotic regions in proximity of a secular resonance. Indeed, in
  Figure~\ref{fig.plotfreqNOREL_high} the angular velocity is constant
  for values of $\Omega_1(0)$ in $[\sim 70^\circ, \, \sim 85^\circ] $
  and $[\sim 280^\circ, \, \sim 300^\circ] $ (corresponding to part of
  the blue central stripes of Figure~\ref{fig.Color_qper_e1}). More
  precisely, the value of $\nu$ is equal to $\simeq -1.04\cdot
  10^{-3}\,$, that is $\omega_4\,$, i.e., one of the fundamental
  angular velocities which characterize the quasi-periodic motion of
  the outer planets (see Eq.~\eqref{freq.fond.CD}). This allows us to
  conclude that they represent the stable central part of a resonant
  region.

\section{Introduction of a secular model by a normal form approach}
\label{sub:forme_normali_astro}
This Section aims at manipulating the Hamiltonian with normal form
algorithms in order to define a new model that is more compact; this
allows us to simulate the secular dynamics of \ups$\b$ with much
faster numerical integrations. In fact, we describe a
reduction of the number of degrees of freedom (DOF) of our Hamiltonian
model. For such a purpose, we apply two normal form methods: first,
we perform the construction of an \textit{elliptic torus}, hence,
we proceed removing the angles ($q_3,q_4,q_5\,$) whose evolution
is linearly depending on time. The latter elimination is made by
applying a normalization method {\it \`a la} Birkhoff in
such a way to introduce a so called resonant normal
form\footnote{Resonant normal forms play a relevant role in the
  proof of the celebrated Nekhoroshev theorem (see,
  e.g.,~\cite{gio2022}).} that includes, at least partially, the
long-term effects due to the outer planets motion.

\subsection{Normal form algorithm constructing elliptic tori}
\label{sub:toro_ellittico}

In~\cite{giolocsan2014} the existence of invariant elliptic tori in 3D
planetary problems with $n$~bodies has been proved by using a normal
form method which is explicitly constructive. However, such an
approach does not look suitable to be directly applied to Hamiltonian
secular models, because in this latter case the separation between
fast and slow dynamics is lost. Therefore, we follow the explanatory
notes~\cite{locetal2022}, where the algorithm constructing the normal
form for elliptic tori is compared with the classical one {\it \`a la
  Kolmogorov}, which is at the basis of the original proof scheme of
the KAM theorem. We first summarize this general procedure leading to
the construction of elliptic tori. We then add some comments
explaining how this general method can be suitably adapted to our
problem.

We start considering a Hamiltonian $\Hscr^{(0)}$ written as follows:
\begin{equation}
\label{Ham.T.E.0}
\begin{aligned}
\Hscr^{(0)}(\vet{p},\vet{q}, \vet{I},\vet{\alpha})&= \Escr^{(0)}+\vet{\omega}^{(0)}\cdot\vet{p}+\vet{\Omega}^{(0)}\cdot\vet{I}+\sum_{s\geq 0}\sum_{l\geq 3} f_{l}^{(0,s)}(\vet{p},\vet{q},\vet{I},\vet{\alpha})\\
&\phantom{=}+\sum_{s\geq 1} f_{0}^{(0,s)}(\vet{q})+\sum_{s\geq 1} f_{1}^{(0,s)}(\vet{q},\vet{I},\vet{\alpha})+\sum_{s\geq 1} f_{2}^{(0,s)}(\vet{p},\vet{q},\vet{I},\vet{\alpha}) \, ,
\end{aligned}
\end{equation}
where $\Escr^{(0)}$ is a constant term, representing the energy,
$(\vet{p},\vet{q})\in\reali^{n_1}\times\toro^{n_1}\,$,
$(\vet{I},\vet{\alpha})\in\reali^{n_2}_{\geq 0}\times \toro^{n_2}$ are
action-angle variables and
$(\vet{\omega}^{(0)},\vet{\Omega}^{(0)})\in\reali^{n_1}\times
\reali^{n_2}$ is the angular velocity vector. The symbol
$f_{l}^{(r,s)}$ is used to denote a function of the variables
$(\vet{p},\vet{q}, \vet{I},\vet{\alpha})\,$, such that $l$ is the
total degree in the square root of the actions $(\vet{p}\,, \vet{I})$,
$s$ is the index such that the maximum trigonometric degree, in the
angles $(\vet{q},\vet{\alpha})\,$, is $sK$ (for a fixed positive
integer $K$) and $r$ refers to the normalization step. In more
details, we can say that $f_{l}^{(0,s)}\!\!\in\mathfrak{P}_{l,sK}\,$,
which is a class of functions that we introduce as follows.
\begin{definition}
\label{def:pol.T.E.}
We say that $g\in\mathfrak{P}_{l,sK}$ if 
$\displaystyle{g\in\!\!\!\!\bigcup_{\substack{\widehat{m}\geq 0, \,\widehat{l}\geq 0 \\ 2\widehat{m}+\widehat{l}=l}}\!\!\!\!\widehat{\mathfrak{P}}_{\widehat{m},\widehat{l},sK}\, ,}$ where
\begin{align*}
\widehat{\mathfrak{P}}_{\widehat{m},\widehat{l},sK}=\Bigg\lbrace  &g:\mathbb{R}^{n_1}\times\mathbb{T}^{n_1}\times\mathbb{R}_{\geq 0}^{n_2}\times\mathbb{T}^{n_2}\to \mathbb{R}\, :\\
 &g(\vet{p},\vet{q}, \vet{I},\vet{\alpha})\!=\!\!\!\!\!\sum_{\substack{\vet{m}\in\mathbb{N}^{n_1}\\ |\vet{m}|=\widehat{m}}}\sum_{\substack{\vet{l}\in\mathbb{N}^{n_2}\\ |\vet{l}|=\widehat{l}}}\sum_{\substack{\vet{k}\in\mathbb{Z}^{n_1}\\ |\vet{k}|+|\widehat{\vet{l}}|\leq sK}}\!\!\!\!\!\!\!\!\!\!\!\!\!\!\!\!\sum_{\substack{\qquad\quad\widehat{l}_j=-l_j,-l_j+2,\ldots,l_j\\ j=1,\ldots,n_2}}\!\!\!\!\!\!\!\!\!\!\!c_{\vet{m},\vet{l},\vet{k},\widehat{\vet{l}}}\,\,\,\vet{p}^{\vet{m}}\left(\sqrt{\vet{I}}\right)^{\vet{l}}e^{i\left(\vet{k}\cdot\vet{q}+\widehat{\vet{l}}\cdot\vet{\alpha}\right)}\Bigg\rbrace\, .
\end{align*}
\end{definition}
\noindent
A few remarks about the above definition are in order. First, since
we deal with \emph{real} functions, the complex coefficients
must be such that
$c_{\vet{m},\vet{l},-\vet{k},-\widehat{\vet{l}}}=\bar{c}_{\vet{m},\vet{l},\vet{k},\widehat{\vet{l}}}\,$. Moreover,
the rules about the integer coefficients vector $\widehat{\vet{l}}$
are such that, $\forall\,j=1,\ldots,n_2\,$, the $j$-th component of
the Fourier harmonic $\widehat{l}_j$ (that refers to the angle
$\alpha_j$) must have the same parity with respect to the
corresponding degree $l_j$ of $\sqrt{I_j}$ and must satisfy the
inequality\footnote{These rules are inherited from the polynomial
  structure of the canonical coordinates describing the small
  oscillations that are transverse to the elliptic torus. For istance,
  it is easy to verify that the restrictions on the indexes appearing
  in definition~\ref{def:pol.T.E.} is satisfied when the change of
  variables~\eqref{coord:I_alpha} is plugged into the
  Hamiltonian~\eqref{Ham.b.new}. } $|\widehat{l}_j|\leq
l_j\,$.

Let us here emphasize that our SQPR model of the secular dynamics of
\ups$\b$ can be reformulated in such a way to be described by a
Hamiltonian of the type~\eqref{Ham.T.E.0}; this will be explained in
detail at the beginning of Section~\ref{sec:risult_astro}.

The following statement plays a substantial
role, since it ensures that the structure of the functions
$f_{l}^{(r,s)}$ is preserved while the normalization algorithm is
iterated.
\begin{lemma}
\label{lemma:pol_T.E}
Let us consider two generic functions $g\in\mathfrak{P}_{l,sK}$ and $h\in\mathfrak{P}_{m,rK}\,$, where $K$ is a fixed positive integer number. Then
$$
\poisson{g}{h}=L_{h}\,g\,\in\,\mathfrak{P}_{l+m-2,\,(r+s)K}\qquad\forall\,l,\,m,\,r,\,s\,\in\,\mathbb{N}\, ,
$$
where we trivially extend the definition~\ref{def:pol.T.E.} in such
a way that $\mathfrak{P}_{-2,\,sK}=\mathfrak{P}_{-1,\,sK}=\{0\}$
$\forall\ s\in\mathbb{N}$.
\end{lemma}

The algorithm constructing the normal form for elliptic tori is
applied to Hamiltonians of the type~\eqref{Ham.T.E.0}, where the terms
appearing in the second row (namely, $\sum_{s\geq 1}\sum_{l=0}^{2}
f_{l}^{(0,s)}(\vet{p},\vet{q},\vet{I},\vet{\alpha})\,$) are considered
as the perturbation to remove. Therefore, one can easily realize that
such a perturbation must be sufficiently small so that the procedure behaves well as regards convergence. There are general situations where this
essential smallness condition is satisfied. For instance, this occurs
for Hamiltonian systems in the neighborhood of a stable equilibrium
point; in fact, it is possible to prove that,
$f_{l}^{(0,s)}=\Oscr(\varepsilon^s)\,$, where $\varepsilon$ is a small
parameter which denotes the first approximation of the distance
(expressed in terms of the actions) between the elliptic torus and the
stable equilibrium point. The elimination of the small perturbing
terms can be done through a sequence of canonical transformations,
leading the Hamiltonian in the following final form:
\begin{equation}
\label{Ham.T.E.goal}
\begin{aligned}
\Hscr^{(\infty)}(\t{\vet{p}}, \t{\vet{q}}, \t{\vet{I}},\t{\vet{\alpha}})&= \Escr^{(\infty)}+\vet{\omega}^{(\infty)}\cdot\t{\vet{p}}+\vet{\Omega}^{(\infty)}\cdot\t{\vet{I}}+\sum_{s\geq 0}\sum_{l\geq 3} f_{l}^{(\infty,s)}(\t{\vet{p}},\t{\vet{q}},\t{\vet{I}},\t{\vet{\alpha}})\, ,
\end{aligned}
\end{equation}
with $f_{l}^{(\infty, s)}\in\mathfrak{P}_{l,sK}\,$. Therefore, for any
initial conditions of type $(\vet{0},\t{\vet{q}}_0, \vet{0},
\t{\vet{\alpha}})$ (where $\t{\vet{q}_0}\in\mathbb{T}^{n_1}$ and the
value of $\t{\vet{\alpha}}\in\mathbb{T}^{n_2}$ does not play any
role\footnote{Indeed, when $\t{\vet{I}}=\vet{0}$
  $\forall\>\t{\vet{\alpha}}\in\mathbb{T}^{n_2}$, the canonical
  coordinates $(\sqrt{2\tilde
    I_j}\cos(\tilde\alpha_j)\,,\,\sqrt{2\tilde
    I_j}\sin(\tilde\alpha_j))$ are mapped into the origin of the
  $j$-th subspace that is transversal to the elliptic torus. This
  fictitious singularity of the action-angle variables
  $(\vet{I},\vet{\alpha})$ is completely harmless just because all the
  normalization algorithm can be performed working on Hamiltonians
  whose expansions are made by terms belonging to sets of functions
  of type $\mathfrak{P}_{l,sK}\,$. We stress that all the algorithm
  could be reformulated using polynomial canonical coordinates to
  describe the dynamics in the subspaces transversal to the elliptic
  torus; in particular, this is done with complex pairs of canonical
  coordinates in~\cite{car2022}. In the sequel, we adopt an exposition entirely based on the use of the action-angle coordinates, which makes the
  algorithm easier to understand.}), the motion law $(\t{\vet{p}}(t),
\t{\vet{q}}(t),
\t{\vet{I}}(t),\t{\vet{\alpha}}(t))=(\vet{0},\t{\vet{q}}_0+\vet{\omega}^{(\infty)}t,
\vet{0}, \t{\vet{\alpha}})$ is a solution of the Hamilton's equations
related to $\Hscr^{(\infty)}\,$. This quasi-periodic solution (having
$\vet{\omega}^{(\infty)}$ as constant angular velocity vector) lies on
the $n_1$-dimensional invariant torus such that the values of the
action coordinates are $\t{\vet{p}} = \vet{0},$
$\t{\vet{I}}=\vet{0}\,$.

The generic $r$-th step of the algorithm is defined as follows. Let us
assume that after $r-1$ normalization steps the expansion of the
Hamiltonian can be written as
\begin{equation}
\label{Ham.T.E.r-1_0}
\begin{aligned}
\Hscr^{(r-1)}(\vet{p},\vet{q}, \vet{I},\vet{\alpha})&= \Escr^{(r-1)}+\vet{\omega}^{(r-1)}\!\!\cdot\vet{p}+\vet{\Omega}^{(r-1)}\!\!\cdot\vet{I}+\!\sum_{s\geq 0}\sum_{l\geq 3} f_{l}^{(r-1,s)}(\vet{p},\vet{q},\vet{I},\vet{\alpha})\\
&\phantom{=}+\sum_{s\geq r} f_{0}^{(r-1,s)}(\vet{q})+\sum_{s\geq r} f_{1}^{(r-1,s)}(\vet{q},\vet{I},\vet{\alpha})+\sum_{s\geq r} f_{2}^{(r-1,s)}(\vet{p},\vet{q},\vet{I},\vet{\alpha}) ,
\end{aligned}
\end{equation}
with $f_{l}^{(r-1, s)}\in\mathfrak{P}_{l,sK}\,$. By comparing
formula~\eqref{Ham.T.E.0} with~\eqref{Ham.T.E.r-1_0}, one immediately realizes that
the assumption above is satisfied in the case with $r=1$ for what
concerns the expansion of the initial Hamiltonian $\Hscr^{(0)}$.

The $r$-th normalization step consists of three substeps, each of them involving a canonical transformation which is expressed in terms of
the Lie series having $\chi_{0}^{(r)}$, $\chi_{1}^{(r)}$,
$\chi_{2}^{(r)}$ as generating function, respectively. Therefore, the
new Hamiltonian that is introduced at the end of the $r$-th
normalization step is defined as follows:
\begin{align}
\label{Ham.T.E.r.goal}
\Hscr^{(r)}=\exp\left(L_{\chi_2^{(r)}}\right)
\exp\left(L_{\chi_1^{(r)}}\right)\exp\left(L_{\chi_0^{(r)}}\right)
\Hscr^{(r-1)}\, ,
\end{align}
where $\exp \left( L_{\chi} \right)\cdot = \sum_{j \geq 0 }
(L_{\chi}^{j} \cdot)/j!$ is the Lie series operator, $ L_{\chi}
\cdot=\poisson{\cdot}{\chi}$ is the Lie derivative with respect to the
dynamical function $\chi\,$, and $\poisson{\cdot}{\cdot}$ represents
the Poisson bracket.

\subsubsection*{First substep (of the r-th normalization step)}

The first substep aims to remove the term depending only on the
angles\footnote{This first substep of the algorithm is basically
  useless when the explicit construction of the normal form related
  to an elliptic torus is started from the Hamiltonian $\Hscr_{sec,\, 2+3/2}$ described in~\eqref{Ham.T.E.0.the.prequel}. Indeed, in the
  case under study just the so called dummy actions are affected by
  this kind of canonical change of variables, which is defined by a
  Lie series with a generating function depending on the angles
  $\vet{q}$ only. Aiming to make a rather general discussion of
  the computational procedure, we keep in the algorithm
  the description of this first normalization substep.}  $\vet{q}$ up
to trigonometric degree $rK\,$, i.e., $f_0^{(r-1,r)}$ (included in the
first sum of the second row of~\eqref{Ham.T.E.r-1_0}), which has to be
considered as $\Oscr(\varepsilon^r)\,$. The first generating function
$\chi_{0}^{(r)}(\vet{q})$ is determined by solving the following
homological equation:
\begin{equation}
\label{homo.T.E.0}
\poisson{\vet{\omega}^{(r-1)}\cdot\vet{p}}{\chi_{0}^{(r)}}+f_{0}^{(r-1,r)}(\vet{q})=\avg{f_{0}^{(r-1,r)}}_{\vet{q}}\, .
\end{equation}
Since $f_{0}^{(r-1,r)}\in\mathfrak{P}_{0,rK}\,$, its Fourier expansion can
be written $f_{0}^{(r-1,r)}(\vet{q})=\sum_{|k|\leq rK}c_{\vet{k}}^{(r-1)}
e^{i\vet{k}\cdot\vet{q}}$. Because of the homological
equation~\eqref{homo.T.E.0}, we find 
\begin{equation}
\label{def.chi0.T.E.}
\chi_{0}^{(r)}(\vet{q})=\sum_{0<|k|\leq rK}\frac{c_{\vet{k}}^{(r-1)}}{i\,\vet{k}\cdot\vet{\omega}^{(r-1)}} e^{i\vet{k}\cdot\vet{q}}\, ;
\end{equation}
the above solution is well defined if the non-resonance condition 
$$
\vet{k}\cdot\vet{\omega}^{(r-1)}\neq 0 \quad\quad\forall\,0<\vet{k}\leq rK
$$ is satisfied.  We can now apply the Lie series operator
$\exp\big(L_{\chi_0^{(r)}}\big)$ to $\Hscr^{(r-1)}\,$. This allows
us to write the expansion of the new intermediate Hamiltonian as
follows:
\begin{equation}
\label{Ham.T.E.r_I}
\begin{aligned}
\Hscr^{(I;\,r)}(\vet{p},\vet{q}, \vet{I},\vet{\alpha})&=\exp\left(L_{\chi_0^{(r)}}\right)\Hscr^{(r-1)}\\
& = \Escr^{(r)}+\vet{\omega}^{(r-1)}\cdot\vet{p}+\vet{\Omega}^{(r-1)}\cdot\vet{I}+\sum_{s\geq 0}\sum_{l\geq 3} f_{l}^{(I;\, r,s)}(\vet{p},\vet{q},\vet{I},\vet{\alpha})\\
&\phantom{=}+\sum_{s\geq r} f_{0}^{(I;\,r,s)}(\vet{q})+\sum_{s\geq r} f_{1}^{(I;\,r,s)}(\vet{q},\vet{I},\vet{\alpha})+\sum_{s\geq r} f_{2}^{(I;\,r,s)}(\vet{p},\vet{q},\vet{I},\vet{\alpha}) \, ,
\end{aligned}
\end{equation}
where (by abuse of notation) for the new canonical coordinates we
adopt the same symbols as the old ones. From a practical point of
view, the new Hamiltonian terms can be conveniently defined in such a
way to mimic what is usually done in any programming language. First,
we introduce the new summands as the old ones, so that
$f_{l}^{(I;\,r,s)}=f_l^{(r-1,s)}$ $\forall \,l\geq 0\,$, $s\geq 0\,$.
Hence, each term generated by Lie derivatives with respect to
$\chi^{(r)}_0$ is added to the corresponding class of functions. By a
further abuse of notation, this is made by the following
sequence\footnote{From a practical point of view, since we have to
  deal with finite series, that are truncated in such a way that the
  index $s$ goes up to a fixed order called $\mathcal{N}_S$, we have
  to require also that $1\leq j\leq \min\big\{\lfloor l/2
  \rfloor,\lfloor (\mathcal{N}_S-s)/r \rfloor\big\}\,$.}  of
redefinitions:
\begin{align}
\label{formule.T.E.0}
f_{l-2j}^{(I;\,r,s+jr)} \hookleftarrow
\frac{1}{j!} L_{\chi_{0}^{(r)}}^j f_{l}^{(r-1, s)}
\qquad\forall\,l\geq 0 ,\,1\leq j\leq \lfloor l/2 \rfloor,\,s\geq 0\, ,
\end{align}
where with the notation $a \hookleftarrow b$ we mean that the quantity
$a$ is redefined so as to be equal $a + b\,$. In fact, since
$\chi_0^{(r)}\in\mathfrak{P}_{0,rK}\,$, Lemma~\ref{lemma:pol_T.E}
ensures that each application of the Lie derivative operator
$L_{\chi_{0}^{(r)}}$ decreases by $1$ the degree in $\vet{p}$ (that is
obviously equivalent to $2$ in the square root of the actions), while
the trigonometrical degree in the angles $\vet{q}$ is increased by
$rK\,$. By using repeatedly such a simple rule, one can easily verify
that $f_{l}^{(I;\,r,s)}\in\mathfrak{P}_{l,sK}$ $\forall\, l\geq
0,\,s\geq0\,$. Moreover, due to the homological
equation~\eqref{homo.T.E.0}, we set $f_{0}^{(I;\,r,r)}=0$ and update
the energy value in such a way that
$\Escr^{(r)}=\Escr^{(r-1)}+\avg{f_{0}^{(r-1,r)}}_{\vet{q}}\,$, where
$\avg{\cdot}_{\vet{q}}$ is used to denote the angular average with
respect to $\vet{q}$.

\subsubsection*{Second substep (of the r-th normalization step)}

The second substep aims to remove the term that is linear in
$\sqrt{\vet{I}}$ and independent on $\vet{p}\,$, i.e. $f_1^{(I;
  r,r)}$, which is included in the second sum appearing in the second
row of~\eqref{Ham.T.E.r_I}. The second generating function
$\chi_{1}^{(r)}(\vet{q}, \vet{I}, \vet{\alpha})$ is determined solving
the following homological\footnote{In the
  r.h.s. of~\eqref{homo.T.E.1} we do not need to put any term produced
  by an angular average (similar to that appearing, for instance, in
  the r.h.s. of the homological equation~\eqref{homo.T.E.0}), because
  $\avg{f_{1}^{(I; \, r,r)}}_{\vet{q},\vet{\alpha}}=0\,$. In fact,
  since $f_{1}^{(I; \, r,r)}$ is linear in $\sqrt{\vet{I}}$ and
  belongs to $\mathfrak{P}_{1,rK}\,$, from
  definition~\eqref{def:pol.T.E.} it easily follows that in the
  expansion of $f_{1}^{(I; \, r,r)}$ all the terms include the
  dependence on $e^{\pm i\alpha_j}$ with $j=1,\ldots,\,n_2\,$, leading
  to a null mean over the angles.}  equation:
\begin{equation}
\label{homo.T.E.1}
\poisson{\vet{\omega}^{(r-1)}\cdot\vet{p}+ \vet{\Omega}^{(r-1)}\cdot\vet{I}}{\chi_{1}^{(r)}}+f_{1}^{(I;\,r,r)}(\vet{q}, \vet{I}, \vet{\alpha})=0\, .
\end{equation}
Since $f_{1}^{(I;\,r,r)}\in\mathfrak{P}_{1,rK}\,$, we can write its expansion as 
\begin{equation*}
  f_{1}^{(I;\,r,r)}(\vet{q}, \vet{I}, \vet{\alpha})=
  \sum_{ 0\leq  |\vet{k}|\leq rK-1}\ \sum_{j=1}^{n_2}
  \sqrt{I_j}\left[c_{\vet{k},j}^{(+)}e^{i\left(\vet{k}\cdot\vet{q}+\alpha_j\right)}
    +c_{\vet{k},j}^{(-)}e^{i\left(\vet{k}\cdot\vet{q}-\alpha_j\right)}\right]\, ; 
\end{equation*}
due to the homological equation~\eqref{homo.T.E.1}, we find 
\begin{equation}
\label{def.chi1.T.E.}
 \chi_{1}^{(r)}(\vet{q},\vet{I},\vet{\alpha})=\sum_{ 0\leq  \vet{k}\leq rK-1}\sum_{j=1}^{n_2}\sqrt{I_j}\Bigg[\frac{c_{\vet{k},j}^{(+)}}{i\,\left(\vet{k}\cdot\vet{\omega}^{(r-1)}+\Omega_j^{(r-1)}\right)}e^{i\left(\vet{k}\cdot\vet{q}+\alpha_j\right)}
 \ +\frac{c_{\vet{k},j}^{(-)}}{i\,\left(\vet{k}\cdot\vet{\omega}^{(r-1)}-\Omega_j^{(r-1)}\right)}e^{i\left(\vet{k}\cdot\vet{q}-\alpha_j\right)}\Bigg]\, .
\end{equation}
The above expression is well defined provided that the first Melnikov
non-resonance condition is satisfied, i.e.,
\begin{equation}
  \label{Melnikov-1}
  \min_{\substack{0<|\vet{k}|\leq rK-1 \\ |\vet{l}|=1}}
  \left|\vet{k}\cdot\vet{\omega}^{(r-1)}+\vet{l}\cdot\vet{\Omega}^{(r-1)}\right|
  \geq \frac{\gamma}{(rK)^\tau}
  \quad\quad \mathrm{and}\quad\quad
  \min_{|\vet{l}|=1}\left|\,\vet{l}\cdot\vet{\Omega}^{(r-1)}\right|\geq \gamma\, ,
\end{equation}
for a pair of fixed values of $\gamma> 0$ and $\tau>n_1-1\,$
(see~\cite{locetal2022} and reference therein).

We can now apply the transformation
$\exp\left(L_{\chi_1^{(r)}}\right)$ to the Hamiltonian
$\Hscr^{(I;\,r)}\,$. By the usual abuse of notation (i.e., the new
canonical coordinates are denoted with the same symbols of the old
ones), the expansion of the new Hamiltonian can be written as
\begin{equation}
\label{Ham.T.E.r_II}
\begin{aligned}
\Hscr^{(II;\,r)}(\vet{p},\vet{q}, \vet{I},\vet{\alpha})&=\exp\left(L_{\chi_1^{(r)}}\right)\Hscr^{(I;\,r)}\\
&= \Escr^{(r)}+\vet{\omega}^{(r-1)}\cdot\vet{p}+\vet{\Omega}^{(r-1)}\cdot\vet{I}+\sum_{s\geq 0}\sum_{l\geq 3} f_{l}^{(II;\, r,s)}(\vet{p},\vet{q},\vet{I},\vet{\alpha})\\
&\phantom{=}+\sum_{s\geq r+1} f_{0}^{(II;\,r,s)}(\vet{q})+\sum_{s\geq r} f_{1}^{(II;\,r,s)}(\vet{q},\vet{I},\vet{\alpha})+\sum_{s\geq r} f_{2}^{(II;\,\,r,s)}(\vet{p},\vet{q},\vet{I},\vet{\alpha}) \, ,
\end{aligned}
\end{equation}
where in the last row of the previous formula, it is possible to start
the first sum from $r+1$ instead of $r\,$, being
$f_{0}^{(II;\,r,r)}=f_{0}^{(I;\,r,r)}=0\,$. In an analogous way as in the first substep, it is convenient to first
define the new Hamiltonian terms as the old ones, i.e.,
$f_{l}^{(II;\,r,s)}=f_l^{(I;\,r,s)}$ $\forall \,l\geq 0\,$, $s\geq
0\,$. Hence, each term generated by the Lie derivatives with respect
to $\chi^{(r)}_1$ is added to the corresponding class of functions.
This is made by the following sequence\footnote{From a practical point of view, since we have to deal again with series truncated in such a way that the index $s$ goes up to a fixed order called $\mathcal{N}_S$, we have to require also
  that $1\leq j\leq \min\left\{l,\lfloor (\mathcal{N}_S-s)/r
  \rfloor\right\}\,$.} of redefinitions:
\begin{equation}
\label{formule.T.E.I}
\begin{aligned}
&f_{l-j}^{(II;\,r,s+jr)} \hookleftarrow \frac{1}{j!} L_{\chi_{1}^{(r)}}^j f_{l}^{(I;\,r, s)} \qquad\forall\,l\geq 0 ,\,1\leq j\leq l,\,s\geq 0\, ,\\
&f_{0}^{(II;\,r,2r)} \hookleftarrow  \frac{1}{2} L_{\chi_{1}^{(r)}}^2 \left(\vet{\omega}^{(r-1)}\cdot\vet{p}+\Omega^{(r-1)}\cdot\vet{I}\right)\, .
\end{aligned}
\end{equation}
In fact, since $\chi_1^{(r)}\in\mathfrak{P}_{1,rK}$ is linear in
$\sqrt{\vet{I}}$, each application of the Lie derivative operator
$L_{\chi_{1}^{(r)}}$ decreases by $1$ the degree in square root of the
actions, while the trigonometrical degree in the angles is increased
by $rK\,$; such a rule holds because of
Lemma~\ref{lemma:pol_T.E}. Moreover, thanks to the homological
equation~\eqref{homo.T.E.1}, one can easily remark that
$f_{1}^{(II;\,r,r)}=0\,$. A repeated application of
Lemma~\ref{lemma:pol_T.E} allows us to verify also that
$f_{l}^{(II;\,r,s)}\in\mathfrak{P}_{l,sK}\,$, $\forall l\geq
0,\,s\geq0\,$.

\subsubsection*{Third substep (of the r-th normalization step)}

The last substep aims to remove the term $f_{2}^{(II;\, r,r)}$ which
is quadratic in the square root of the actions (i.e., either quadratic
in $\sqrt{\vet{I}}$ or linear in $\vet{p}\,$) and included in the
third sum appearing in the second row of~\eqref{Ham.T.E.r_II}. The
third generating function $\chi_{2}^{(r)}(\vet{p},\vet{q}, \vet{I},
\vet{\alpha})$ is determined by solving the following homological
equation:
\begin{equation}
\label{homo.T.E.2}
\poisson{\vet{\omega}^{(r-1)}\cdot\vet{p}+ \vet{\Omega}^{(r-1)}\cdot\vet{I}}{\chi_{2}^{(r)}}+f_{2}^{(II;\,r,r)}(\vet{p},\vet{q}, \vet{I}, \vet{\alpha})=\avg{f_{2}^{(II;\,r,r)}}_{\vet{q},\vet{\alpha}}\, .
\end{equation}
Since $f_{2}^{(II;\,r,r)}\in\mathfrak{P}_{2,rK}\,$, we can write it
(according to definition~\ref{def:pol.T.E.} with
$2\widehat{m}+\widehat{l}=2$ and $s=r\,$) as follows:
\begin{align*}
f_{2}^{(II;\,r,r)}(\vet{p},\vet{q}, \vet{I}, \vet{\alpha})
&=\sum_{\substack{\vet{m}\in\mathbb{N}^{n_1}\\ |\vet{m}|=1}}\sum_{\substack{\vet{k}\in\mathbb{Z}^{n_1}\\ |\vet{k}|\leq rK}}c_{\vet{m},\vet{k}}\,\,\vet{p}^{\vet{m}}e^{i\,\vet{k}\cdot\vet{q}}
+\sum_{\substack{\vet{l}\in\mathbb{N}^{n_2}\\ |\vet{l}|=2}}\sum_{\substack{\vet{k}\in\mathbb{Z}^{n_1}\\ |\vet{k}|+|\widehat{\vet{l}}|\leq rK}}\!\!\!\!\!\!\!\!\!\!\!\!\!\!\!\!\!\sum_{\substack{\quad\qquad\widehat{l}_j=-l_j,-l_j+2,\ldots,l_j\\ \:\:\,\,\,j=1,\ldots,n_2}}\!\!\!\!\!\!\!\tilde{c}_{\vet{l},\vet{k},\widehat{\vet{l}}}\,\,\left(\sqrt{\vet{I}}\right)^{\vet{l}}e^{i\left(\vet{k}\cdot\vet{q}+\widehat{\vet{l}}\cdot\vet{\alpha}\right)}\, .
\end{align*}
Due to the homological equation~\eqref{homo.T.E.2}, we obtain 
\begin{equation}
\label{def.chi2.T.E}
\begin{aligned}
\chi_{2}^{(r)}(\vet{p},\vet{q}, \vet{I}, \vet{\alpha})
&=\sum_{\substack{\vet{m}\in\mathbb{N}^{n_1}\\ |\vet{m}|=1}}\sum_{\substack{\vet{k}\in\mathbb{Z}^{n_1}\\ 0<|\vet{k}|\leq rK}}\frac{c_{\vet{m},\vet{k}}\,\,\vet{p}^{\vet{m}}e^{i\,\vet{k}\cdot\vet{q}}}{i\,\vet{k}\cdot\vet{\omega}^{(r-1)}}\\
& \phantom{=}+\sum_{\substack{\vet{l}\in\mathbb{N}^{n_2}\\ |\vet{l}|=2}}\sum_{\substack{\vet{k}\in\mathbb{Z}^{n_1}\\ 0<|\vet{k}|+|\widehat{\vet{l}}|\leq rK}}\!\!\!\!\!\!\!\!\!\!\!\!\!\!\!\!\!\sum_{\substack{\quad\qquad\widehat{l}_j=-l_j,-l_j+2,\ldots,l_j\\ \:\:\,\,\,j=1,\ldots,n_2}}\frac{\tilde{c}_{\vet{l},\vet{k},\widehat{\vet{l}}}\,\,\left(\sqrt{\vet{I}}\right)^{\vet{l}}e^{i\left(\vet{k}\cdot\vet{q}+\widehat{\vet{l}}\cdot\vet{\alpha}\right)}}{i\left(\vet{k}\cdot\vet{\omega}^{(r-1)}+\widehat{\vet{l}}\cdot\vet{\Omega}^{(r-1)}\right)}\, ,
\end{aligned}
\end{equation}
provided that both the non-resonance condition and the Melnikov one of
second kind are satisfied, i.e.,
\begin{equation}
  \label{Melnikov-2}
\vet{k}\cdot\vet{\omega}^{(r-1)}\neq 0 \quad\forall\,0<\vet{k}\leq rK \, , \ \,
\min_{\substack{0<|\vet{k}|\leq rK-2\\|\vet{l}|=2}}\left|\vet{k}\cdot\vet{\omega}^{(r-1)}+\vet{l}\cdot\vet{\Omega}^{(r-1)} \right|\geq \frac{\gamma}{(rK)^\tau} \, ,
\end{equation}
with the same values of the constant parameters $\gamma> 0$ and
$\tau>n_1-1$ appearing in~\eqref{Melnikov-1}.

We can now apply the transformation
$\exp\left(L_{\chi_2^{(r)}}\right)$ to the Hamiltonian
$\Hscr^{(II;\,r)}\,$. By the usual abuse of notation (i.e., the new
canonical coordinates are denoted with the same symbols as the old
ones), the expansion\footnote{In the third row of~\eqref{Ham.T.E.r},
  it is possible to start the second sum from $r+1$ instead of $r\,$,
  being $f_{1}^{(r,r)}=f_{1}^{(II;\,r,r)}=0\,$.} of the new
Hamiltonian can be written as
\begin{equation}
\label{Ham.T.E.r}
\begin{aligned}
\Hscr^{(r)}(\vet{p},\vet{q}, \vet{I},\vet{\alpha})&=\exp\left(L_{\chi_2^{(r)}}\right)\Hscr^{(II;\,r)}\\
&= \Escr^{(r)}+\vet{\omega}^{(r-1)}\cdot\vet{p}+\vet{\Omega}^{(r-1)}\cdot\vet{I}+\sum_{s\geq 0}\sum_{l\geq 3} f_{l}^{(r,s)}(\vet{p},\vet{q},\vet{I},\vet{\alpha})\\
& \phantom{=}+\sum_{s\geq r+1} f_{0}^{(r,s)}(\vet{q})+\sum_{s\geq r+1} f_{1}^{(r,s)}(\vet{q},\vet{I},\vet{\alpha})+\sum_{s\geq r} f_{2}^{(r,s)}(\vet{p},\vet{q},\vet{I},\vet{\alpha}) \, .
\end{aligned}
\end{equation}
Once again, it is convenient to first define the new Hamiltonian terms
as the old ones, i.e., $f_{l}^{(r,s)}=f_l^{(II;\,r,s)}$ $\forall
\,l\geq 0\,$, $s\geq 0\,$. Hence, each term generated by the Lie
derivatives with respect to $\chi^{(r)}_2$ is added to the
corresponding class of functions.  This is made by the following
sequence\footnote{From a practical point of view, since we have to
  deal with series truncated in such a way that the
  index $s$ goes up to a fixed order called $\mathcal{N}_S$, we
  have to require also that $1\leq j\leq \lfloor
  (\mathcal{N}_S-s)/r \rfloor\,$.} of redefinitions:
\begin{equation}
\label{formule.T.E.I}
\begin{aligned}
&f_{l}^{(\,r,s+jr)} \hookleftarrow \frac{1}{j!} L_{\chi_{1}^{(r)}}^j f_{l}^{(II;\,r, s)} \qquad\forall\,l\geq 0 ,\, j\geq 1 ,\,s\geq 0\, ,\\
&f_{2}^{(r,jr)} \hookleftarrow  \frac{1}{j!} L_{\chi_{2}^{(r)}}^j \left(\vet{\omega}^{(r-1)}\cdot\vet{p}+\vet{\Omega}^{(r-1)}\cdot\vet{I}\right)\qquad\forall\,j\geq1\, .
\end{aligned}
\end{equation}
In fact, since $\chi_2^{(r)}\in\mathfrak{P}_{2,rK}$ is either
quadratic in $\sqrt{\vet{I}}$ or linear in $\vet{p}$, each application
of the Lie derivative operator $L_{\chi_{2}^{(r)}}$ does not modify
the degree in the square root of the actions, while the trigonometric
degree in the angles is increased by $rK\,$. By applying Lemma~\ref{lemma:pol_T.E} one can verify also that $f_{l}^{(r,s)}\in\mathfrak{P}_{l,sK}\,$, $\forall l\geq 0,\,s\geq0\,$.

Because of the homological equation~\eqref{homo.T.E.2}, it immediately
follows that the term that cannot be removed, that is
$f_{2}^{(r,r)}=\avg{f_{2}^{(II;\,r,r)}}_{\vet{q},\vet{\alpha}}\in\mathfrak{P}_{2,0}\,$,
is exactly of the same type with respect to the main term that is
linear in the actions, i.e.,
$\vet{\omega}^{(r-1)}\cdot\vet{p}+\vet{\Omega}^{(r-1)}\cdot\vet{I}$. It
looks then natural to update the angular velocity vectors so that
\begin{equation}
\label{update_freq}
\vet{\omega}^{(r)}=\vet{\omega}^{(r-1)}+\nabla_{\vet{p}}\left(\avg{f_{2}^{(II;\,r,r)}}_{\vet{q},\vet{\alpha}}\right) , \quad\, \vet{\Omega}^{(r)}=\vet{\Omega}^{(r-1)}+\nabla_{\vet{I}}\left(\avg{f_{2}^{(II;\,r,r)}}_{\vet{q},\vet{\alpha}}\right) ,
\end{equation}
where, as usual, the symbols $\nabla_{\vet{p}}$ and $\nabla_{\vet{I}}$
denote the gradient with respect to the action variables $\vet{p}$ and
$\vet{I}$, respectively, and to set $f_{2}^{(r,r)}=0\,$. Therefore,
the expansion of the Hamiltonian $\Hscr^{(r)}$ can be rewritten as
\begin{equation}
\label{Ham.T.E.r.final}
\begin{aligned}
  \Hscr^{(r)}(\vet{p},\vet{q}, \vet{I},\vet{\alpha})&= \Escr^{(r)}+\vet{\omega}^{(r)}\cdot\vet{p}+\vet{\Omega}^{(r)}\cdot\vet{I}+\sum_{s\geq 0}\sum_{l\geq 3} f_{l}^{(r,s)}(\vet{p},\vet{q},\vet{I},\vet{\alpha})+\sum_{s\geq r+1}\sum_{l=0}^2 f_{l}^{(r,s)}(\vet{p},\vet{q},\vet{I},\vet{\alpha}) \, ,
\end{aligned}
\end{equation}
where $f_{l}^{(r, s)}\in\mathfrak{P}_{l,sK}\,$ and
$\Escr^{(r)}\in\mathfrak{P}_{0,0}\,$ is a constant.

It is now possible to iterate the algorithm, by performing the (next)
$(r+1)$-th normalization step. The convergence of
this normal form algorithm is proved in~\cite{car2022} under suitable
conditions.

In order to implement such a kind of normalization algorithm with the aid of a computer, we have to deal with Hamiltonians
including a finite number of summands in their expansions in
Taylor-Fourier series. To fix the ideas, let us suppose that we set a
truncation rule in such a way as to neglect all the terms with a
trigonometric degree greater than $\mathcal{N}_SK$, for a fixed
positive integer value of the parameter $\mathcal{N}_S\,$. After
iteratively performing $\mathcal{N}_S$ steps of the constructive algorithm, we end up with an approximation of the Hamiltonian
which is in the normal form corresponding to an elliptic torus, i.e.,
\begin{equation}
\label{Ham.T.E.r.final.S}
\begin{aligned}
\Hscr^{(\mathcal{N}_S)}(\vet{p},\vet{q}, \vet{I},\vet{\alpha})&= \Escr^{(\mathcal{N}_S)}+\vet{\omega}^{(\mathcal{N}_S)}\cdot\vet{p}+\vet{\Omega}^{(\mathcal{N}_S)}\cdot\vet{I}+\sum_{s = 0}^{\mathcal{N}_S}\sum_{l\geq 3} f_{l}^{(\mathcal{N}_S,s)}(\vet{p},\vet{q},\vet{I},\vet{\alpha})\, .
\end{aligned}
\end{equation}
The Hamiltonian $\Hscr^{(\mathcal{N}_S)}$ represents the natural
starting point for the application of a second (Birkhoff-like)
algorithm, which aims to produce a new normal form in such a way
to remove the dependence on the angles $\vet{q}$, as explained in the
next Subsection.

\subsection{Construction of the resonant normal form in such a way to average with respect to the angles $\mathit{\bf{q}}$}
\label{sub:media_q3q4q5}
Consider a Hamiltonian\footnote{We use the symbol
  $\Hscr_B^{(0)}$ instead of $\Hscr^{(0)}$ to distinguish this
  starting Hamiltonian from the one of the previous normalization
  algorithm, which is written in equation~\eqref{Ham.T.E.0}.}
$\Hscr_B^{(0)}$ of the form:
\begin{equation}
\label{Ham.Birkh.0}
\begin{aligned}
\Hscr_B^{(0)}(\vet{p},\vet{q}, \vet{I},\vet{\alpha})&= \Escr_B+\vet{\omega}_B\cdot\vet{p}+\vet{\Omega}_B\cdot\vet{I}+\sum_{s= 0}^{\mathcal{N}_S}\sum_{l\geq 3} g_{l}^{(0,s)}(\vet{p},\vet{q},\vet{I},\vet{\alpha})\, ,
\end{aligned}
\end{equation}
where $\Escr_B$ is a constant term, representing the energy,
$(\vet{p},\vet{q})\in\reali^{n_1}\times\toro^{n_1}\,$,
$(\vet{I},\vet{\alpha})\in\reali^{n_2}_{\geq 0}\times \toro^{n_2}$ are
action-angle variables,
$(\vet{\omega}_B,\vet{\Omega}_B)\in\reali^{n_1}\times \reali^{n_2}$
are the frequencies, $\mathcal{N}_S$ is a fixed positive integer
(ruling the truncations in the Fourier series) and $g_{l}^{(0,
  s)}\in\mathfrak{P}_{l,sK}\,$, $\forall \,l\geq 0,\,0\leq s\leq
\mathcal{N}_S\,$.  In practice, we are starting from the normalized
Hamiltonian of the previous Subsection $\Hscr^{(\mathcal{N}_S)}\,$,
given by equation~\eqref{Ham.T.E.r.final.S}, where we have defined
$\Hscr^{(0)}_B:=\Hscr^{(\mathcal{N}_S)}\,$,
$\Escr_B:=\Escr^{(\mathcal{N}_S)}\,$,
$(\vet{\omega}_B,\vet{\Omega}_B):=(\vet{\omega}^{(\mathcal{N}_S)},
\vet{\Omega}^{(\mathcal{N}_S)})$ and $g_{l}^{(0,
  s)}:=f_{l}^{(\mathcal{N}_S, s)}\in\mathfrak{P}_{l,sK}\,$ $\forall
\,l\geq 0,\,0\leq s\leq \mathcal{N}_S\,$; this is done also in order
to simplify the notation. By comparison with
equation~\eqref{Ham.T.E.r.final.S}, it is easy to remark that
$g_{l}^{(0, s)}:=f_{l}^{(\mathcal{N}_S, s)}=0\,$, $\forall \, 0\leq l
\leq 2\,$, $1\leq s \leq \mathcal{N}_S\,$.

The aim of the present algorithm is to delete the dependence of
$\Hscr^{(0)}_B$ on the angles $\vet{q}\,$, reducing by $n_1$ the
number of degrees of freedom. In order to do this, we have to act on
the terms $g_{l}^{(0,s)}(\vet{p},\vet{q},\vet{I},\vet{\alpha})$ such
that $s\geq 1$ and $l\ge 3$, removing their dependence on $\vet{q}\,$;
indeed, for $s=0\,$, the sum $\sum_{l\geq 3}
g_{l}^{(0,0)}(\vet{p},\vet{I})\,$ does not depend on the angles, thus
it is already in normal form. This elimination can be done by a
sequence of canonical transformations. If convergent, this would lead
the Hamiltonian to the following final normal form:
\begin{equation}
\label{Ham.Birkh.inf}
\begin{aligned}
  \Hscr_B^{(\infty)}(\tilde{\vet{p}}, \t{\vet{I}},\t{\vet{\alpha}})&=
  \Escr_B+\vet{\omega}_B\cdot\t{\vet{p}}+\vet{\Omega}_B\cdot\t{\vet{I}}+
  \sum_{s= 0}^{\infty}\sum_{l\geq 3}
  g_{l}^{(\infty,s)}(\t{\vet{p}},\t{\vet{I}},\t{\vet{\alpha}})\, ,
\end{aligned}
\end{equation}
where $ (\t{\vet{p}},\t{\vet{I}},\t{\vet{\alpha}})$ denote the
new variables; it is evident that, having removed the dependence on
$\t{\vet{q}}\,$, the conjugate momenta vector $\t{\vet{p}}$ is
constant. However, as typical of the computational procedures {\it \`a la} Birkhoff, the constructive algorithm produces divergent series if the normalization is iterated infinitely many times. For this reason, it is convenient to look for an optimal order of
normalization to which the algorithm is stopped. In our approach, we
have not to consider such a problem, because we are dealing with
truncated series; this is done in order to keep our discussion as
close as possible to the practical implementations where the maximal
degree in actions of the expansions is usually rather low.

The generic $r$-th step of this new normalization algorithm is defined
as follows. After $r-1$ steps, the Hamiltonian~\eqref{Ham.Birkh.0}
takes the form
\begin{equation}
\label{Ham.Birkh.n-1}
\begin{aligned}
\Hscr_B^{(r-1)}(\vet{p},\vet{q}, \vet{I},\vet{\alpha})&= \Escr_B+\vet{\omega}_B\cdot\vet{p}+\vet{\Omega}_B\cdot\vet{I}+\sum_{l\geq 3} g_{l}^{(r-1,0)}(\vet{p},\vet{I})\\
&\phantom{=}+\sum_{s=1}^{\mathcal{N}_S}\sum_{3\leq l\leq r+1} g_{l}^{(r-1,s)}(\vet{p},\vet{I},\vet{\alpha})+\sum_{s= 1}^{\mathcal{N}_S}\sum_{l\geq r+2} g_{l}^{(r-1,s)}(\vet{p},\vet{q},\vet{I},\vet{\alpha}) \, ,
\end{aligned}
\end{equation}
with $g_{l}^{(r-1,s)}\in\mathfrak{P}_{l,sK}\,$.

The $r$-th normalization step consists of a sequence of
$\mathcal{N}_S$ substeps, each of them involving a canonical
transformation which is expressed in terms of the Lie series having
$\chi_B^{(j;\,r)}$ as generating function, with
$j=1,\,\ldots\,,\mathcal{N}_S\,$. Therefore, the new Hamiltonian introduced at the end of the $r$-th normalization step of this
algorithm is defined as follows:
\begin{equation}
\label{Ham.Birkh.n.S.formula}
{\Hscr}_{B}^{(r)}=
\exp\left(L_{\chi_B^{(\mathcal{N}_S;\,r)}}\right)\ldots\exp\left(L_{\chi_B^{(3;\,r)}}\right)\exp\left(L_{\chi_B^{(2;\,r)}}\right)\exp\left(L_{\chi_B^{(1;\,r)}}\right)
\Hscr_B^{(r-1)}\, .
\end{equation}
The generating functions $\chi_{B}^{(j;\,r)}$ are defined so as to remove the dependence on $\vet{q}$ from the perturbing term
of lowest order in the square root of the actions, i.e., $\sum_{s=
  1}^{\mathcal{N}_S}
g_{r+2}^{(r-1,s)}(\vet{p},\vet{q},\vet{I},\vet{\alpha})\,$.
\vspace{-.15truecm}

\subsubsection*{j-th substep of the r-th step of the algorithm constructing the resonant normal form}
After $j-1$ substeps, the Hamiltonian can be written as follows:
\begin{equation}
\label{Ham.Birkh.n-1.i}
\begin{aligned}
  \Hscr_B^{(j-1; \,r)}(\vet{p},\vet{q}, \vet{I},\vet{\alpha})&=
  \Escr_B+\vet{\omega}_B\cdot\vet{p}+\vet{\Omega}_B\cdot\vet{I}
  +\sum_{l\geq 3} g_{l}^{(j-1;\,r,0)}(\vet{p},\vet{I})\\
  & \phantom{=}+\sum_{s= 1}^{\mathcal{N}_S}\sum_{3\leq l\leq r+1}
  g_{l}^{(j-1;\,r,s)}(\vet{p},\vet{I},\vet{\alpha})
  +\sum_{s=1}^{j-1} g_{r+2}^{(j-1;\,r,s)}(\vet{p},\vet{I},\vet{\alpha})\\
  & \phantom{=}+\sum_{s= j}^{\mathcal{N}_S}
  g_{r+2}^{(j-1;\,r,s)}(\vet{p},\vet{q},\vet{I},\vet{\alpha})+
  \sum_{s= 1}^{\mathcal{N}_S}\sum_{l\geq r+3}
  g_{l}^{(j-1;\,r,s)}(\vet{p},\vet{q},\vet{I},\vet{\alpha}) ,
\end{aligned}
\end{equation}
where, for $j=1\,$, we set $\Hscr_{B}^{(0;\,r)}:=\Hscr_{B}^{(r-1)}\,$ and
$g_{l}^{(0;\,r,s)}=g_{l}^{(r-1,s)}\,$, $\forall l \geq 0$, $\forall\,
0\leq s\leq \mathcal{N}_S\,$.

The $j$-th substep generating function $\chi_{B}^{(j;\,r)}$ is determined by the following homological equation:
\begin{equation}
\label{homo_Birkh}
\begin{aligned}
  \poisson{\vet{\omega}_{B}\cdot\vet{p}+\vet{\Omega}_{B}\cdot\vet{I}\,}
  {\chi_{B}^{(j;\,r)}}+g_{r+2}^{(j-1;\,r,j)}(\vet{p},\vet{q},\vet{I},\vet{\alpha})=
  \avg{g_{r+2}^{(j-1;\,r,j)}}_{\vet{q}}
\end{aligned}\, .
\end{equation}
Proceeding in a similar way as in the description of
the third substep of the previous Subsection~\ref{sub:toro_ellittico},
first we write the expansion of the perturbing function in the form
\begin{equation}
 g_{r+2}^{(j-1; \,r,j)}(\vet{p},\vet{q}, \vet{I}, \vet{\alpha}) =
 \sum_{2|\vet{m}|+|\vet{l}|=r+2}\,\,
 \sum_{\vet{m}\in\mathbb{N}^{n_1}}\,
 \sum_{\vet{l}\in\mathbb{N}^{n_2}}\,
 \sum_{\substack{\vet{k}\in\mathbb{Z}^{n_1}\\ |\vet{k}|+|\widehat{\vet{l}}|\leq jK}}\!\!\!\!\!\!\!\!\!\!\!\!\!\!\!\!
 \sum_{\substack{\quad\qquad\widehat{l}_{j_2}=-l_{j_2},-l_{j_2}+2,\ldots,l_{j_2}\\ \:\:\,\,\,j_2=1,\ldots,n_2}}
 \!\!\!\!\!\!\!c_{\vet{m},\vet{l},\vet{k},\widehat{\vet{l}}}\,\,
 \vet{p}^{\vet{m}}\left(\sqrt{\vet{I}}\right)^{\vet{l}}
 e^{i\left(\vet{k}\cdot\vet{q}+\widehat{\vet{l}}\cdot\vet{\alpha}\right)}\, .
\end{equation}
The solution of the homological equation~\eqref{homo_Birkh} is then
\begin{equation}
\label{def.chiB}
\begin{aligned}
& \chi_{B}^{(j;\, r)}(\vet{p},\vet{q}, \vet{I}, \vet{\alpha})
=\sum_{2|\vet{m}|+|\vet{l}|=r+2}\,\,
 \sum_{\vet{m}\in\mathbb{N}^{n_1}}\,
 \sum_{\vet{l}\in\mathbb{N}^{n_2}}\,
 \sum_{\substack{\vet{k}\in\mathbb{Z}^{n_1}\,,\,|\vet{k}|>0\\ |\vet{k}|+|\widehat{\vet{l}}|\leq jK}}
 \\
&\ \sum_{\substack{\quad\qquad\widehat{l}_{j_2}=-l_{j_2},-l_{j_2}+2,\ldots,l_{j_2}\\ \:\:\,j_2=1,\ldots,n_2}}
 \!\!\!\!\!\!\!\,\,\,\,\,
 \frac{c_{\vet{m},\vet{l},\vet{k},\widehat{\vet{l}}}}
      {i\left(\vet{k}\cdot\vet{\omega}_B+\widehat{\vet{l}}\cdot\vet{\Omega}_B
        \right)}
      \,\,\,\,
 \vet{p}^{\vet{m}}\left(\sqrt{\vet{I}}\right)^{\vet{l}}
 e^{i\left(\vet{k}\cdot\vet{q}+\widehat{\vet{l}}\cdot\vet{\alpha}\right)}\, .
\end{aligned}
\end{equation}


We can now apply the transformation
$\exp\big(L_{\chi_B^{(j;\,r)}}\big)$ to the Hamiltonian
$\Hscr_B^{(j-1;\,r)}\,$. By the usual abuse of notation (i.e., the new
canonical coordinates are denoted with the same symbols of the old
ones), the expansion of the new Hamiltonian can be written as
\begin{equation}
\label{Ham.Birkh.i.n}
\begin{aligned}
  \Hscr_B^{(j;\,r)}(\vet{p},\vet{q}, \vet{I},\vet{\alpha})&=
  \exp\left(L_{\chi_B^{(j;\,r)}}\right)\Hscr_B^{(j-1;\,r)}\\
  &=\Escr_B+\vet{\omega}_B\cdot\vet{p}+\vet{\Omega}_B\cdot\vet{I}
  +\sum_{l\geq 3} g_{l}^{(j;\,r,0)}(\vet{p},\vet{I})\\
  &\phantom{=}+\sum_{s=1}^{\mathcal{N}_S}\sum_{3\leq l\leq r+1}
  g_{l}^{(j;\,r,s)}(\vet{p},\vet{I},\vet{\alpha})
  +\sum_{s=1}^{j} g_{r+2}^{(j;\,r,s)}(\vet{p},\vet{I},\vet{\alpha})\\
  &\phantom{=}+\sum_{s= j+1}^{\mathcal{N}_S}
  g_{r+2}^{(j;\,r,s)}(\vet{p},\vet{q},\vet{I},\vet{\alpha})
  +\sum_{s= 1}^{\mathcal{N}_S}\sum_{l\geq r+3}
  g_{l}^{(j;\,r,s)}(\vet{p},\vet{q},\vet{I},\vet{\alpha}) \, .
\end{aligned}
\end{equation}
In a similar way to what has been done previously, it is convenient to
first define the new Hamiltonian terms as the old ones, i.e.,
$g_{l}^{(j;\,r,s)}=g_l^{(j-1;\,r,s)}$ $\forall \,l\geq 0\,$, $0\leq
s\leq \mathcal{N}_S\,$; hence, each term generated by the Lie
derivatives with respect to $\chi_B^{(j;\,r)}$ is added to the
corresponding class of functions.  This is made by the following
sequence\footnote{From a practical point of view, since we have to
  deal with series truncated in such a way that the
  indexes $s$ and $l$ do not exceed the threshold values
  $\mathcal{N}_S$ and $\mathcal{N}_L\,$, respectively, then we have to
  require that $1\leq m\leq \min\left((\mathcal{N}_L-l)/r,\lfloor
  (\mathcal{N}_S-s)/j \rfloor\right)\,$, which is more restrictive
  with respect to the corresponding rule appearing
  in~\eqref{formule.Birkh.q3q4q5}. }  of redefinitions
\begin{equation}
\label{formule.Birkh.q3q4q5}
\begin{aligned}
&g_{l+mr}^{(j;\,r,s+mj)} \hookleftarrow
\frac{1}{m!} L_{\chi_{B}^{(j;\,r)}}^m g_{l}^{(j;\,r, s)}
\qquad\forall\,l\geq 0 ,\, 1\leq m\leq \lfloor (\mathcal{N}_S-s)/j \rfloor ,
\,0\leq s\leq \mathcal{N}_S\, ,\\
&g_{2+mr}^{(j;\,r,mj)} \hookleftarrow
\frac{1}{m!} L_{\chi_{B}^{(j;\,r)}}^m
\left(\vet{\omega}_B\cdot\vet{p}+\Omega_B\cdot\vet{I}\right)
\qquad\forall\, 1\leq m\leq \lfloor \mathcal{N}_S/j\rfloor\, .
\end{aligned}
\end{equation}
In fact, since $\chi_B^{(j;\,r)}\in\mathfrak{P}_{r+2,jK}\,$, each
application of the Lie derivative operator $L_{\chi_B^{(j;\,r)}}$
increases the degree in square root of the actions and the
trigonometrical degree in the angles by $r$ and $jK\,$,
respectively. Moreover, thanks to the homological
equation~\eqref{homo_Birkh} and the second rule included in
formula~\eqref{formule.Birkh.q3q4q5} (in the case with $m=1$), one can
easily remark that
$g_{r+2}^{(j;\,r,j)}=\avg{g_{r+2}^{(j-1;\,r,j)}}_{\vet{q}}\,$. By
applying Lemma~\ref{lemma:pol_T.E} one can verify also that
$g_{l}^{(j;\,r,s)}\in\mathfrak{P}_{l,sK}\,$, $\forall l\geq
0,\,s\geq0\,$.

The $r$-th step of the algorithm constructing the resonant normal form
is completed at the end of the iterative repetition of the $j$-th
substep for $j=1,\,\ldots\,,\,\mathcal{N}_S\,$. Therefore, the
expansion of the Hamiltonian can be written in the following form:
\begin{equation}
\label{Ham.Birkh.n}
\begin{aligned}
  {\Hscr}_B^{(r)}(\vet{p},\vet{q}, \vet{I},\vet{\alpha})&=
  \exp\left(L_{\chi_B^{(\mathcal{N}_S;\,r)}}\right)\cdots
  \exp\left(L_{\chi_B^{(1;\,r)}}\right)\Hscr_B^{(r-1)}\\
  &= \Escr_B+\vet{\omega}_B\cdot\vet{p}+\vet{\Omega}_B\cdot\vet{I}
  +\sum_{l\geq 3} g_{l}^{(r,0)}(\vet{p},\vet{I})\\
  &\phantom{=}+\sum_{s= 1}^{\mathcal{N}_S}\sum_{3\leq l\leq r+2}
  g_{l}^{(r,s)}(\vet{p},\vet{I},\vet{\alpha})
  +\sum_{s= 1}^{\mathcal{N}_S}\sum_{l\geq r+3}
  g_{l}^{(r,s)}(\vet{p},\vet{q},\vet{I},\vet{\alpha}) \, ,
\end{aligned}
\end{equation}
where $g_{l}^{(r,s)}:=g_{l}^{(\mathcal{N}_S;\,r,s)}\,$,
$\forall\,l\geq 0\,$, $0\leq s\leq \mathcal{N}_S\,$. Then, the
normalization algorithm can be iteratively repeated. Since we are
interested in the computer implementation, we consider finite sequences of Hamiltonians whose expansion is truncated
up to a finite degree, say, $\mathcal{N}_L$ in the square root of the
actions. Therefore, the iteration of $\mathcal{N}_L-2$ normalization
steps of the algorithm constructing the resonant normal form are
sufficient to obtain
\begin{equation}
  \label{Ham.Birkh.final}
  \Hscr_{B}^{(\mathcal{N}_L-2)}(\vet{p}, \vet{I},\vet{\alpha})
  = \Escr_B+\vet{\omega}_B\cdot\vet{p}+\vet{\Omega}_B\cdot\vet{I}+
  \sum_{s=0}^{\mathcal{N}_S}\sum_{l=3}^{\mathcal{N}_L}
  g_{l}^{(\mathcal{N}_L-2,s)}(\vet{p},\vet{I},\vet{\alpha})\, .
\end{equation}
The Hamiltonian~\eqref{Ham.Birkh.final} does not
depend on the angles $\vet{q}$. Therefore, the corresponding actions
$\vet{p}$ are constant and they can be considered as parameters whose
values are fixed by the initial conditions; this allows us to decrease
the number of degrees of freedom by $n_1\,$, passing from $n_1+n_2$ to
$n_2$.

\section{Application of the normalization algorithms to the secular quasi-periodic restricted model of the dynamics of \ups$\b$}
\label{sec:risult_astro}

The SQPR model can be reformulated in such a way as to resume the form of a Hamiltonian of the type~\eqref{Ham.T.E.0}, to which we can sequentially apply both normalization procedures described in the two previous Subsections. In fact, the
canonical change of variables
\begin{equation}
\label{coord:I_alpha}
\begin{aligned}
&\csi_1=\sqrt{2 I_1}\cos(\alpha_1) \, , \qquad & &\eta_1=\sqrt{2 I_1}\sin(\alpha_1)\, ,\\
&P_1=\sqrt{2 I_2}\cos(\alpha_2)\, , \qquad & &Q_1=\sqrt{2 I_2}\sin(\alpha_2)\, ,
\end{aligned}
\end{equation} 
allows to rewrite the expansion of the SQPR
Hamiltonian~\eqref{Ham.b.new} as follows:
\begin{equation}
\label{Ham.T.E.0.the.prequel}
\begin{aligned}
  \Hscr_{sec,\, 2+3/2}(\vet{p},\vet{q},\vet{I},\vet{\alpha})
  &= \omega_3\,p_3 + \omega_4\,p_4 + \omega_5\,p_5
  \\
  & \phantom{=}+ \sum_{\substack{l_1+l_2=0\\(l_1\,,\,l_2)\in\mathbb{N}^2}}^{\mathcal{N}_L} \sum_{\substack{(k_3\,,\,k_4\,,\,k_5)\in\mathbb{Z}^3\\ |\vet{k}|\leq \mathcal{N}_SK}}
  \!\!\!\! \!\!\!\!\!\!\!\!\!\!\!\!\!\!\!\!\!\!\!\sum_{\substack{\qquad\qquad k_j=-l_j,-l_j+2,\ldots,l_j\\ j=1,\,2}}
  \!\!\!\!\!\!\!\!\!\!\!\!\!\!\!\!\!\!\!\!\! c_{\vet{l},\vet{k}} (\sqrt{I_1})^{l_1}(\sqrt{I_2})^{l_2}
  e^{i(k_1\alpha_1+k_2\alpha_2+k_3 q_3 + k_4 q_4 + k_5 q_5)}
 ,
\end{aligned}
\end{equation}
where $\vet{k}=(k_1,\ldots,k_5)\in\interi^5\,$.  The r.h.s. of the
above equation can be expressed in the general and more compact form
described in equation~\eqref{Ham.T.E.0}, by setting $n_1=3\,$,
$n_2=2\,$,
$\vet{\omega}^{(0)}=\vet{\omega}=(\omega_3,\omega_4,\omega_5)\in\reali^3$,
that are the fundamental frequencies of the two outer planets
(described in equation~\eqref{freq.fond.CD}), while
$\vet{\Omega}^{(0)}\in\reali^2$ can be easily determined by performing
the so called diagonalization of the Hamiltonian part that is
quadratic in the square root of the actions $\vet{I}$ and not
depending on the angles $\vet{q}$ (see, e.g.,~\cite{GioDFGS-1989}). In
the equation above, the parameters $\mathcal{N}_L$ and $\mathcal{N}_S$
define the truncation order of the expansions in Taylor and Fourier
series, respectively, in such a way to represent on the computer just
a finite number of terms that are not too many to handle with; in our
computations we fix $\mathcal{N}_L=6$ as maximal power degree in
square root of the actions and we include Fourier terms up to a
maximal trigonometric degree of $8$, putting $\mathcal{N}_S=4\,$,
$K=2\,$. We recall that setting $K=2$ is quite natural for Hamiltonian
systems close to stable equilibria as it is for models describing the
secular planetary dynamics, see, e.g.,~\cite{giolocsan2017}.  Let us
also remark that a simple reordering of the summands according to the
total trigonometric degree $|\vet{k}|$ in the angles
$(\vet{q},\vet{\alpha})$ allows us to represent the second row of
formula~\eqref{Ham.T.E.0.the.prequel} as a sum of Hamiltonian terms
each of them is belonging to a functions class of type
$\mathfrak{P}_{l,sK}\,$, which is unique for any positive integer $K$
if we ask for the minimality of the index $s$. These comments can be
used all together in order to formally verify that the new expansion
of $\Hscr_{sec,\, 2+3/2}$ in~\eqref{Ham.T.E.0.the.prequel} can be
finally reexpressed in the same form as $\Hscr^{(0)}$
in~\eqref{Ham.T.E.0}.

Furthermore, in the case of our SQPR model of the secular dynamics of
\ups$\b$, the only term depending on the action variables $\vet{p}$
(that are the so called dummy variables) is
$\vet{\omega}^{(0)}\cdot\vet{p}\,$; thus, none of the Hamiltonian term
$f_{l}^{(0,s)}$ depends on $\vet{p}\,$. This fact would allow to
introduce some simplification in the computational algorithm.  For
instance, the value of the angular velocity vector
$\vet{\omega}^{(0)}$ is not modified during the first normalization
procedure (i.e. the algorithmic construction of the elliptic tori) and
it remains equal to its initial value, given by the fundamental
frequencies described in~\eqref{freq.fond.CD}. Therefore, the expansion
of the starting Hamiltonian in the special case of our SQPR model
can be rewritten as
\begin{equation}
\label{Ham_init_our}
\begin{aligned}
\Hscr^{(0)}(\vet{p},\vet{q}, \vet{I},\vet{\alpha})&= \Escr^{(0)}+\vet{\omega}^{(0)}\cdot\vet{p}+\vet{\Omega}^{(0)}\cdot\vet{I}+\sum_{s= 0}^{\mathcal{N}_S}\sum_{l= 3}^{\mathcal{N}_L} f_{l}^{(0,s)}(\vet{q},\vet{I},\vet{\alpha})+\sum_{s= 1}^{\mathcal{N}_S}\sum_{l=0}^{2} f_{l}^{(0,s)}(\vet{q},\vet{I},\vet{\alpha}) \, ;
\end{aligned}
\end{equation}
however, in our opinion, for what concerns the general description of
the previous Subsections it has been worth to consider also an
eventual dependence of $f_{l}^{(0,s)}$ on $\vet{p}\,$ in order to keep
the discussion of the constructive procedure as general as possible.

The first algorithm to be applied aims to construct the normal form
corresponding to an invariant elliptic torus. It starts from the
Hamiltonian $\Hscr_{sec,\, 2+3/2}$ rewritten in the same form as
$\Hscr^{(0)}$ in~\eqref{Ham.T.E.0} (more precisely as
in~\eqref{Ham_init_our}) and its computational procedure is fully
detailed in Subsection~\ref{sub:toro_ellittico}. Therefore, we perform
$\mathcal{N}_S$ normalization steps of this first normalization
algorithm. This allows us to bring the Hamiltonian in the following
(intermediate) normal form:
\begin{align*}
\begin{aligned}
\Hscr^{(\mathcal{N}_S)}(\vet{p},\vet{q}, \vet{I},\vet{\alpha})&= \Escr^{(\mathcal{N}_S)}+\vet{\omega}^{(\mathcal{N}_S)}\cdot\vet{p}+\vet{\Omega}^{(\mathcal{N}_S)}\cdot\vet{I}+\sum_{s= 0}^{\mathcal{N}_S}\sum_{l= 3}^{\mathcal{N}_L} f_{l}^{(\mathcal{N}_S,s)}(\vet{q},\vet{I},\vet{\alpha})\, ,
\end{aligned}
\end{align*}
where $f_{l}^{(\mathcal{N}_S,s)}\in\mathfrak{P}_{l,sK}$
$\forall\ l=3,\,\ldots\,,\,\mathcal{N}_L,\>s=0\,,\,\ldots\,,\,\mathcal{N}_S$
and the angular velocity vector related to the angles $\vet{q}$ is such that
$\vet{\omega}^{(\mathcal{N}_S)}=\vet{\omega}^{(0)}=(\omega_3,
\omega_4, \omega_5)$, whose components are given
in~\eqref{freq.fond.CD}.

It is now possible to apply the second algorithm aiming to construct a
resonant normal form where the dependence on the angles $\vet{q}=(q_3,  q_4, q_5)$ is completely removed. Such a normalization starts from the Hamiltonian $\Hscr^{(\mathcal{N}_S)}$ obtained after the first normalization procedure. Therefore, we perform $\mathcal{N}_L-2$
normalization steps of the above algorithm, each of them involving
$\mathcal{N}_S$ substeps as described in
Subsection~\ref{sub:media_q3q4q5}; this allows us to bring the
Hamiltonian in the following (final) normal form:
\begin{equation}
\label{Ham.astro}
\begin{aligned}
\Hscr_{2DOF}(\vet{p}, \vet{I},\vet{\alpha})
&= \Escr_B+\vet{\omega}_B\cdot\vet{p}+\vet{\Omega}_B\cdot\vet{I}+\sum_{l= 3}^{\mathcal{N}_L} g_{l}^{(\mathcal{N}_L-2,0)}(\vet{I})+\sum_{s= 1}^{\mathcal{N}_S}\sum_{l=3}^{\mathcal{N}_L} g_{l}^{(\mathcal{N}_L-2,s)}(\vet{I},\vet{\alpha})\, ,
\end{aligned}
\end{equation}
where $g_{l}^{(\mathcal{N}_L-2,s)}\in\mathfrak{P}_{l,sK}$
$\forall\ l=3,\,\ldots\,,\,\mathcal{N}_L,\>s=0\,,\,\ldots\,,\,\mathcal{N}_S$
and it still holds true that $\vet{\omega}_B=\vet{\omega}^{(0)}\,$.

All the algebraic manipulations that are prescribed by the normal form
algorithms have been performed by using the symbolic manipulator {\tt
  Mathematica} as a programming framework.

We emphasize that $\Hscr_{2DOF}$ is an \emph{integrable} Hamiltonian.
In fact, due to the preservation of the total angular momentum, discussed in Remark~\ref{nota:Noether->momentoangolare}, the
following invariance
law\footnote{Equation~\eqref{eq:simmetriaxrotazione-H2DOF} can be
  easily checked, by explicitly performing the derivatives on the
  expansions~\eqref{Ham.astro} which are computed using {\tt
    Mathematica}. However, it is worth to sketch also a more
  conceptual justification. Indeed, it would not be difficult to
  verify that all the Lie series introduced in
  Subsections~\ref{sub:toro_ellittico}--\ref{sub:media_q3q4q5}
  preserve the invariance law described in
  Remark~\ref{nota:Noether->momentoangolare}. By comparing the
  definitions of the canonical coordinates in~\eqref{coord:I_alpha}
  and~\eqref{Poinc.var.U}, one can immediately realize that the angles
  $-\alpha_1$ and $-\alpha_2$ are nothing but the longitudes of the
  pericenter and of the node,
  respectively, of \ups$\b$. Therefore, taking into account that $\Hscr_{2DOF}$
  does not depend on the angles $\vet{q}$, the invariance law
  discussed in Remark~\ref{nota:Noether->momentoangolare} can be
  rewritten in the form~\eqref{eq:simmetriaxrotazione-H2DOF}.} is
satisfied:
\begin{equation}
  \label{eq:simmetriaxrotazione-H2DOF}
  \frac{\partial\,\Hscr_{2DOF}}{\partial\alpha_1}+
  \frac{\partial\,\Hscr_{2DOF}}{\partial\alpha_2}=0\,;
\end{equation}
thus, from the Hamilton's equations for $\Hscr_{2DOF}\,$, we can
immediately deduce that $I_1+I_2$ is a constant of motion. Therefore,
$\Hscr_{2DOF}$ is integrable because of the Liouville theorem (see,
e.g.,~\cite{gio2022}), since it admits a complete system of constants
of motion in involution, that are the dummy variables $\vet{p}$
(which could be disregarded in~\eqref{Ham.astro}, reducing the model to 2~DOF), $I_1+I_2$
and the Hamiltonian itself.
  
In view of the numerical explorations of the dynamical evolution of
our new model described by the integrable Hamiltonian
$\Hscr_{2DOF}(\vet{p},\vet{I},\vet{\alpha})$, it is convenient to
introduce the canonical transformations related to the so called
semianalytic method of integration for the equations of motion (see,
e.g.,~\cite{giolocsan2017}). In order to fix the ideas, let us focus
on the second algorithm, designed to construct a resonant normal
form. This normalization procedure can be summarized by the
transformation that is obtained by iteratively applying all the Lie
series to the canonical variables. This is done as follows:
\begin{equation}
  \label{cambio_coord_Birkh-like}
\begin{aligned}
  I_i &=
  \exp\left(L_{\chi_B^{(\mathcal{N}_S;\,\mathcal{N}_L-2)}}\right)\dots
  \exp\left(L_{\chi_B^{(2;\,\mathcal{N}_L-2)}}\right)
  \exp\left(L_{\chi_B^{(1;\,\mathcal{N}_L-2)}}\right)\ldots\dots\\
  &\qquad\qquad\qquad\qquad
  \ldots\exp\left(L_{\chi_B^{(\mathcal{N}_S;\,1)}}\right)\dots
  \exp\left(L_{\chi_B^{(2;\,1)}}\right)\exp\left(L_{\chi_B^{(1;\,1)}}\right)
  I_i\bigg|_{{\vet{I}=\t{\vet{I}}} \atop
      {\vet{\alpha}=\t{\vet{\alpha}}}}\, ,\\
  \alpha_i &=
  \exp\left(L_{\chi_B^{(\mathcal{N}_S;\,\mathcal{N}_L-2)}}\right)\dots
  \exp\left(L_{\chi_B^{(2;\,\mathcal{N}_L-2)}}\right)
  \exp\left(L_{\chi_B^{(1;\,\mathcal{N}_L-2)}}\right)\ldots\dots\\
  &\qquad\qquad\qquad\qquad
  \ldots\exp\left(L_{\chi_B^{(\mathcal{N}_S;\,1)}}\right)\dots
  \exp\left(L_{\chi_B^{(2;\,1)}}\right)\exp\left(L_{\chi_B^{(1;\,1)}}\right)
  \alpha_i\bigg|_{{\vet{I}=\t{\vet{I}}} \atop
      {\vet{\alpha}=\t{\vet{\alpha}}}}\, ,
\end{aligned}
\end{equation} 
for $i=1,2\,$. We introduce the symbol $\C_B$ to denote the change of coordinates\footnote{Since none of the generating functions $\chi_B^{(j;\,r)}$ depends on $\vet{p}\,$, the way that
  these dummy variables are modified by the application of the Lie
 series does not really matter, because they do not enter in Hamilton's equations of motion~\eqref{campo.Ham.b}, under the Hamiltonian $\Hscr_{sec, \,2+3/2}\,$. Since, however, the generating functions do depend on $\vet{q}$ (but not on their conjugate actions $\vet{p}$, as we have remarked just above) in the arguments of
  $\C_B$ we have included also the angles $\vet{q}$ that are not
  affected by any modification due to the application of the Lie series.} defined by the above expressions, i.e.,
$(\vet{I},\vet{\alpha})=\C_B(\vet{q},\t{\vet{I}},\t{\vet{\alpha}})\,$.
We can proceed in the same way for what concerns the algorithm
constructing the normal form corresponding to an invariant elliptic
torus. In fact, we first introduce the application of all the Lie
series to the canonical variables in such a way to write,
$\forall\ i=1,2\,$,
\begin{equation}
\label{cambio_coord_TE}
\begin{aligned}
  I_i=
  &\exp\left(L_{\chi_2^{(\mathcal{N}_S)}}\right)
  \exp\left(L_{\chi_1^{(\mathcal{N}_S)}}\right)
  \exp\left(L_{\chi_0^{(\mathcal{N}_S)}}\right)
  \ldots\\
  &\qquad\qquad\qquad\qquad\qquad\ldots\exp\left(L_{\chi_2^{(1)}}\right)
  \exp\left(L_{\chi_1^{(1)}}\right)\exp\left(L_{\chi_0^{(1)}}\right)
  I_i\bigg|_{{\vet{I}=\hat{\vet{I}}} \atop
      {\vet{\alpha}=\hat{\vet{\alpha}}}}\, ,\\
  \alpha_i=
  &\exp\left(L_{\chi_2^{(\mathcal{N}_S)}}\right)
  \exp\left(L_{\chi_1^{(\mathcal{N}_S)}}\right)
  \exp\left(L_{\chi_0^{(\mathcal{N}_S)}}\right)
  \ldots\\
  &\qquad\qquad\qquad\qquad\qquad\ldots\exp\left(L_{\chi_2^{(1)}}\right)
  \exp\left(L_{\chi_1^{(1)}}\right)\exp\left(L_{\chi_0^{(1)}}\right)
  \alpha_i\bigg|_{{\vet{I}=\hat{\vet{I}}} \atop
      {\vet{\alpha}=\hat{\vet{\alpha}}}}\, ;
\end{aligned}
\end{equation}
finally, we use the symbol $\C$ to summarize the whole change of
coordinates that is defined by the whole expression above, i.e.,
$(\vet{I},\vet{\alpha})=\C(\vet{q},\hat{\vet{I}},\hat{\vet{\alpha}})\,$.
Let us now introduce the function
$\F:\mathbb{T}^3\times\mathbb{R}_{\geq 0}^{2}\times\mathbb{T}^{2}
\to\mathbb{R}_{\geq 0}^{2}\times\mathbb{T}^{2}$,
which is defined so that
\begin{equation}
  \label{def:cal_F}
  \F(\vet{q},\t{\vet{I}},\t{\vet{\alpha}})=
  \C\big(\vet{q},\C_B(\vet{q},\t{\vet{I}},\t{\vet{\alpha}})\big)\,,
\end{equation}
where we have omitted to put the $\t{\phantom{q}}$ symbol on top of
$\vet{q}$ in order to stress that the angles $\vet{q}$ are not
affected by the change of coordinates. Moreover, let also introduce
the symbol ${\cal A}$ to denote the usual canonical transformation
defining the action-angle variables for the harmonic oscillator, i.e.,
by formula~\eqref{coord:I_alpha}, in our case this means that
\begin{equation}
  \label{def:passaggio-in-azione-angolo}
  {\cal A}(\vet{I}\,,\vet{\alpha})=
  \big(\sqrt{2 I_1}\cos(\alpha_1),\,\sqrt{2 I_1}\sin(\alpha_1),\,
  \sqrt{2 I_2}\cos(\alpha_2),\,
  \sqrt{2 I_2}\sin(\alpha_2)\big)\,.
\end{equation}
By applying the Exchange Theorem (see~\cite{gro1967}
and~\cite{gio2003}), the solutions of the equations of motions related
to $\Hscr_{2DOF}$ can be mapped to those for $\Hscr_{sec, \,
  2+3/2}\,$. Indeed, assume that
$t\mapsto\big(\t{\vet{p}}(t),\,\t{\vet{q}}(t),
\,\t{\vet{I}}(t),\,\t{\vet{\alpha}}(t)\big)$ is an orbit corresponding
to the integrable flow induced by $\Hscr_{2DOF}\,$; in particular, in
our model we have that $\t{\vet{q}}(t)=\vet{\omega}_Bt=\vet{\omega}t$,
where the components of the angular velocity vector $\vet{\omega}$ are
given in equation~\eqref{freq.fond.CD}. Therefore, the orbit
\begin{equation}
  \label{frm:motion-law-by-semianalytical-scheme}
  t\mapsto\big(\vet{\omega}t\, , \,
  {\cal A}\big(\F(\vet{\omega}t,\t{\vet{I}}(t),\t{\vet{\alpha}}(t))\big)\big)
\end{equation}
is an approximate\footnote{There are at least two substantial reasons
  for which this motion law, produced by a (so called)
  semi-analytic integration scheme, is not an \emph{exact} solution of
  the equations~\eqref{campo.Ham.b}. Let us recall that Lie series
  define near-to-the-identity canonical transformations that are well
  defined on suitable restrictions of the phase space. However, we are
  always working with finite truncated series; therefore, the
  corresponding changes of variables cannot preserve exactly the
  solutions because infinite tails of summands are
  neglected. Moreover, in the resonant normal form $\Hscr_{2DOF}$
  described in~\eqref{Ham.astro} we do not include the remainder
  terms; let us recall that they become dominant if the Birkhoff
  algorithm is iterated infinitely many times, making the series
  expansion of the normal form to be divergent. Therefore, the
  semi-analytic solutions are prevented to be exact also because of
  this second source of truncations acting on the series expansion of
  the Hamiltonians (instead of the Lie series defining the canonical
  transformations). As a final remark, let us also recall that, in
  order to be canonical, the change of coordinates should include also
  the dummy actions $\vet{p}$, in which we are not intested at all
  because they do not exert any role in the equations of
  motion~\eqref{campo.Ham.b}.} solution of the Hamilton's
equations~\eqref{campo.Ham.b}.

For our purposes, it is also useful to construct the inverse of the
function $\F$, which maps from the original canonical coordinates to
the ones referring to the resonant normal form. Therefore, it is
convenient to replace all the compositions of Lie series appearing in
the r.h.s. of~\eqref{cambio_coord_TE} with the following expressions,
$\forall\ i=1,2\,$:
\begin{equation}
\label{inv_cambio_coord_TE}
\begin{aligned}
  &\hat{I}_i=\exp\left(L_{-\chi_0^{(1)}}\right)
  \exp\left(L_{-\chi_1^{(1)}}\right)\exp\left(L_{-\chi_2^{(1)}}\right)
  \ldots\\
  &\qquad\qquad\qquad\qquad\qquad\ldots\exp\left(L_{-\chi_0^{(\mathcal{N}_S)}}\right)
  \exp\left(L_{-\chi_1^{(\mathcal{N}_S)}}\right)
  \exp\left(L_{-\chi_2^{(\mathcal{N}_S)}}\right)I_i\, ,\\
  &\hat{\alpha}_i=\exp\left(L_{-\chi_0^{(1)}}\right)
  \exp\left(L_{-\chi_1^{(1)}}\right)\exp\left(L_{-\chi_2^{(1)}}\right)
  \ldots\\
  &\qquad\qquad\qquad\qquad\qquad\ldots\exp\left(L_{-\chi_0^{(\mathcal{N}_S)}}\right)
  \exp\left(L_{-\chi_1^{(\mathcal{N}_S)}}\right)
  \exp\left(L_{-\chi_2^{(\mathcal{N}_S)}}\right)
  \alpha_i ;
\end{aligned}
\end{equation}
gathering all the corresponding changes of coordinates allows us to
define\footnote{Of course, since also
  $\C^{-1}:\mathbb{T}^3\times\mathbb{R}_{\geq
    0}^{2}\times\mathbb{T}^{2} \to\mathbb{R}_{\geq
    0}^{2}\times\mathbb{T}^{2}$ (i.e., $\C$ and $\C^{-1}$ share the
  same domains and codomains, which are different between them), then
  $\C^{-1}$ cannot be considered as the inverse function in a strict
  sense. However, if we extend trivially both these functions, in such
  a way to introduce
  $\hat\C(\vet{q},\vet{I},\vet{\alpha})=\big(\vet{q},\C(\vet{q},\vet{I},\vet{\alpha})\big)$
  and
  $\hat\C^{-1}(\vet{q},\vet{I},\vet{\alpha})=\big(\vet{q},\C^{-1}(\vet{q},\vet{I},\vet{\alpha})\big)$,
  then $\hat\C^{-1}$ would really be the inverse function of $\hat\C$
  (where elementary properties of the Lie series described in Chap.~4
  of~\cite{gio2003} are also used and the small effects due to the
  truncations are neglected). Therefore, it is by a harmless abuse of
  notation that we are adopting the symbol $\C^{-1}$. The same abuse
  will be made for what concerns the symbols $\C_B^{-1}$ and
  $\F^{-1}$.}  $\C^{-1}(\vet{q},\vet{I},\vet{\alpha})$. Proceeding
in an analogous way, we can introduce the inverse function of
$\C_B\,$; in more detail, we can start from
formula~\eqref{cambio_coord_Birkh-like}, by reversing the order of all
the Lie series and by changing the sign to all the generating
functions, then we can define
$\C_B^{-1}(\vet{q},\vet{I},\vet{\alpha})$.  Therefore, we can
introduce also
\begin{equation}
  \label{def:inv_cal_F}
  \F^{-1}(\vet{q},\vet{I},\vet{\alpha})=
  \C_B^{-1}\big(\vet{q},\C^{-1}(\vet{q},\vet{I},\vet{\alpha})\big)\,.
\end{equation}

We are now ready to exploit the (cheap) numerical solutions of the
$2$~DOF integrable Hamiltonian, which is described
in~\eqref{Ham.astro}, in order to retrieve information about the
secular dynamics of \ups$\b$ through our SQPR model. This can be done
thanks to the knowledge of the approximate
solution~\eqref{frm:motion-law-by-semianalytical-scheme}. The initial
conditions are selected in the same way as in
Subsection~\ref{subsub:num_integ_qper}: we consider the initial
orbital elements reported in Table~\ref{tab:param.orb.b} and the
minimal possible value of the mass of \ups$\b\,$, i.e.,
$m_1=0.674$~$M_{J}\,$. These data are completed with the values of
$(\i_1(0)\,,\Omega_1(0))$ ranging in the $20\times 60$ regular grid
that covers $I_\i\times I_\Omega=[6.865^\circ,
  34^\circ]\times[0^\circ,360^\circ]$; moreover, all these initial
values of the orbital elements are translated in the Laplace reference
frame, which refers only to the two outermost exoplanets. Hence, we
can compute a set of $21\times 60$ initial conditions of type
$\big(\vet{I}(0),\vet{\alpha}(0)\big)={\cal
  A}^{-1}\big(\csi_1(0),\eta_1(0),P_1(0),Q_1(0)\big)$, by using
formula~\eqref{Poinc.var.U} with $j=1\,$, \eqref{coord:I_alpha} and
the definition~\eqref{def:passaggio-in-azione-angolo}. Finally, we can
translate the initial conditions to initial values of the canonical
coordinates found after the resonant normal form, by computing
$\big(\t{\vet{I}}(0),\t{\vet{\alpha}}(0)\big)=
\F^{-1}\big(\vet{0},\vet{I}(0),\vet{\alpha}(0)\big)$.

As shown below, an important information is obtained by a criterion allowing to identify those domains of initial conditions in which the series are either divergent or slowly converging. We introduce such a criterion to preselect initial conditions that are admissible. From a mathematical point of view, the identity $(\vet{I}(0),\vet{\alpha}(0))=(\vet{I}^{O}(0),\vet{\alpha}^{O}(0)):=
\F\big(\vet{0},\F^{-1}\big(\vet{0},\vet{I}(0),\vet{\alpha}(0)\big)\big)\,$ holds in a domain where the normalization procedure is convergent, provided that no truncations are applied to the series $\F$ and $\F^{-1}$ and that the computation of the series is not affected by any round-off errors. Due to the errors and truncations introduced in the computation, however, in general we obtain that
$(\vet{I}(0),\vet{\alpha}(0))\neq
(\vet{I}^{O}(0),\vet{\alpha}^{O}(0))\,$. In the domain where the series expansions are rapidly converging the difference $(\vet{I}(0),\vet{\alpha}(0))-(\vet{I}^{O}(0),\vet{\alpha}^{O}(0))$ is small. When, instead, we obtain a large difference, this is an indicator that we are outside the domain of convergence of the series. The situation is represented graphically below.

\begin{figure}[h]
\qquad\qquad\qquad
\begin{minipage}{.5\textwidth}
\frame{
\begin{tikzpicture}[node distance = 2cm, thick]%
        \node (1) {$(\vet{I}(0),\vet{\alpha}(0))$};
        \node (3) [below=of 1] {$(\vet{I}^{O}(0),\vet{\alpha}^{O}(0))$};
        \node (2) [right=of 1] {$(\t{\vet{I}}(0),\t{\vet{\alpha}}(0))$};
    \draw[->] (1) -- node [midway,above] {$\F^{-1}(\vet{0},\cdot,\cdot)$} (2);  
   \draw[->] (2) to [bend left] node [midway,below] {$\qquad\quad\F(\vet{0},\cdot,\cdot)$} (3);
\end{tikzpicture}
}
\end{minipage}\!\!\!\!\!\!\!\!\!\!\!
\begin{minipage}{.35\textwidth}
\caption{Graphical representation of the definitions about the initial
  conditions.}
\label{fig:crit}
\end{minipage}
\end{figure}
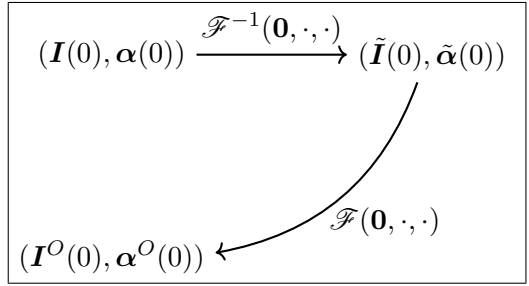

In view of the above, we define the following preselection criterion of \textit{admissible initial conditions}.
For any initial condition $(\vet{I}(0),\vet{\alpha}(0))$ we compute the quantities
\begin{equation}
\label{ratios.1}
\mathfrak{r}_1=\frac{\sqrt{I_1(0)}-\sqrt{I_1^{O}(0)}}{\sqrt{I_1(0)}}\, ,
\qquad\quad
\mathfrak{r}_2=\frac{\sqrt{I_2(0)}-\sqrt{I_2^{O}(0)}}{\sqrt{I_2(0)}}\, .
\end{equation}
The use of the quantities $\sqrt{I_1}$ and $\sqrt{I_2}$ is motivated by the fact that they are of the same order of magnitude as the eccentricity and the inclination of \ups$\b\,$, respectively.
Moreover, it is useful to define also the following ratios
\begin{equation}
\label{astro_ratios}
\mathfrak{R}_1(t)=
\frac{\sqrt{\big(\t{\csi}_1(t)\big)^2+\big(\t{\eta}_1(t)\big)^2}}
     {\sqrt{\big(\t{\csi}_1(0)\big)^2+\big(\t{\eta}_1(0)\big)^2}}
\, ,\qquad \qquad 
\mathfrak{R}_2(t)=
\frac{\sqrt{\big(\t{P}_1(t)\big)^2+\big(\t{Q}_1(t)\big)^2}}
     {\sqrt{\big(\t{P}_1(0)\big)^2+\big(\t{Q}_1(0)\big)^2}}\, ,
\end{equation}
where
$$ t\mapsto
\big(\vet{\omega}t,\,\t{\csi}_1(t),\t{\eta}_1(t),\t{P}_1(t),\t{Q}_1(t)\big)
:= \big(\vet{\omega}t\,,\, {\cal
  A}\big(\F(\vet{\omega}t,\t{\vet{I}}(t),
\t{\vet{\alpha}}(t))\big)\big)
$$ is the approximate solution of Hamilton's
equations~\eqref{campo.Ham.b}, as produced by the semi-analytic
integration scheme summarized in
formula~\eqref{frm:motion-law-by-semianalytical-scheme}. Comparing formula~\eqref{astro_ratios} with the definition of the
Poincar\'e canonical variables in~\eqref{Poinc.var.U}, it is easy to
realize that $\mathfrak{R}_1$ and $\mathfrak{R}_2$ are functions of
the time that describe the behavior of the orbital excursions with
respect to the eccentricity and the inclination of \ups$\b\,$,
respectively. We then investigate the behavior of the
following function:
\begin{equation}
\label{astro_e1new}
\t{\e}_1(t)=\sqrt{\frac{2 \t{I}_1(t)}{\Lambda_1}
  - \frac{\t{I}_1^2(t)}{\Lambda_1^2}}\, .
\end{equation}
Note that $\t{\e}_1$ would be equal to the eccentricity of \ups$\b$ if
$\t{I}_1$ was replaced by $\big(\csi_1^2+\eta_1^2\big)/2$, with
$(\csi_1\,,\,\eta_1)$ defined in~\eqref{Poinc.var.U}. However, the new
action $\t{I}_1$ is conjugated to
$\big(\t{\csi}_1^2+\t{\eta}_1^2\big)/2$ wich is only nearly equal to
$\big({\csi}_1^2+{\eta}_1^2\big)/2\,$, since the composition of the
transformations $\C$ and $\C_B$ is near-to-identity. Therefore, we can
consider $\t{\e}_1$ as an approximate evaluation of the eccentricity
under the resonant normal form model.

For each pair $\big(\i_1(0),\Omega_1(0)\big)\,$ of the $21\times 60$
points definining the grid which covers $I_\i\times
I_\Omega=[6.865^\circ, 34^\circ]\times[0^\circ,360^\circ]$ we
determine the corresponding initial conditions of type
$(\vet{I}(0),\vet{\alpha}(0))$, as explained above, and we proceed as follows:
\begin{itemize}
\item if $\max\{\mathfrak{r}_1\,,\,\mathfrak{r}_2\}> 1$, then the
  corresponding initial condition is considered as ``non-admissible'',
  i.e. outside the domain of applicability of the series. Then, we
  skip the step below and pass directly to consider the next initial
  conditions of the grid;
\item If the initial conditions is admissible, we
  numerically\footnote{In principle, the Liouville theorem ensures us
    that the Hamilton equations for $\Hscr_{2DOF}$ can be solved
    analytically by the quadratures method (see,
    e.g.,~\cite{gio2022}), but, from a practical point of view,
    numerical integrations are much easier to implement.}  solve the
  equations of motion for the integrable Hamiltonian model with
  $2$~DOF described in~\eqref{Ham.astro}, using a RK4 method and
  starting from
  $\big(\vet{0},\,\t{\vet{I}}(0),\t{\vet{\alpha}}(0)\big)$; during
  such a numerical integration, we compute the maximal values attained
  by the three previously defined quantitative indicators, that are
  $$
  \mathfrak{R}_{1\,{\rm MAX}} = \max_t\{\mathfrak{R}_1(t)\} ,
   \ \>
  \mathfrak{R}_{2\,{\rm MAX}} = \max_t\{\mathfrak{R}_2(t)\} ,
   \ \>
  \t{\e}_{1\,{\rm MAX}} = \max_t\{\t{\e}_1(t)\} .
  $$
\end{itemize}

\begin{figure}[h]
\begin{minipage}{.45\textwidth}
\includegraphics[scale=0.28]{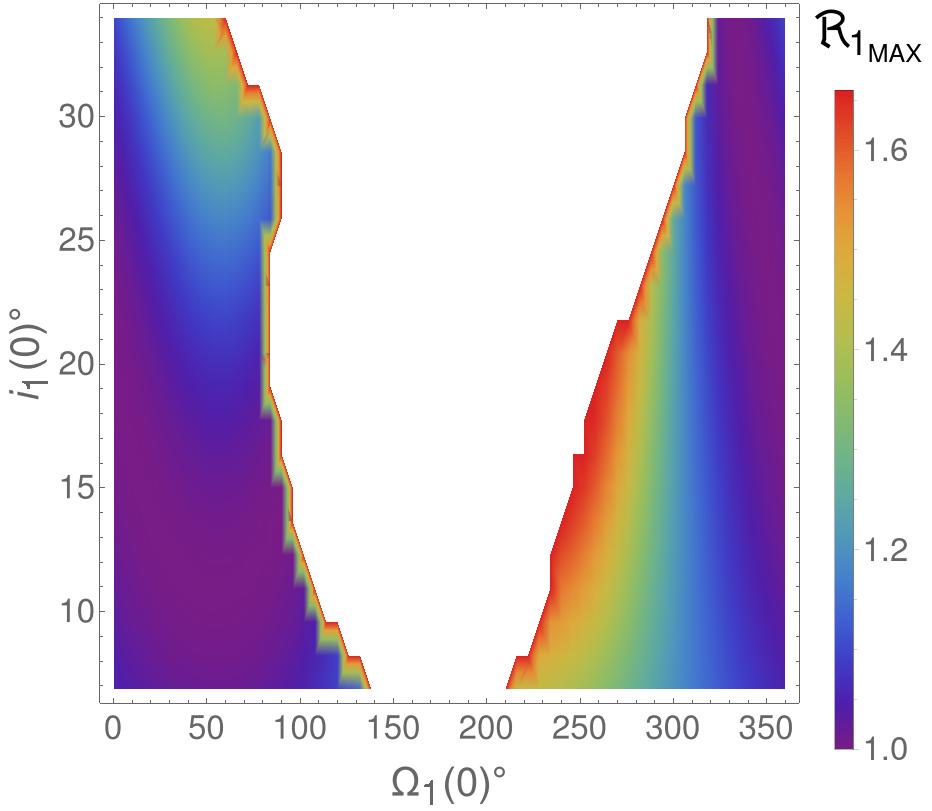}
\end{minipage}
\quad
\begin{minipage}{.45\textwidth}
\includegraphics[scale=0.28]{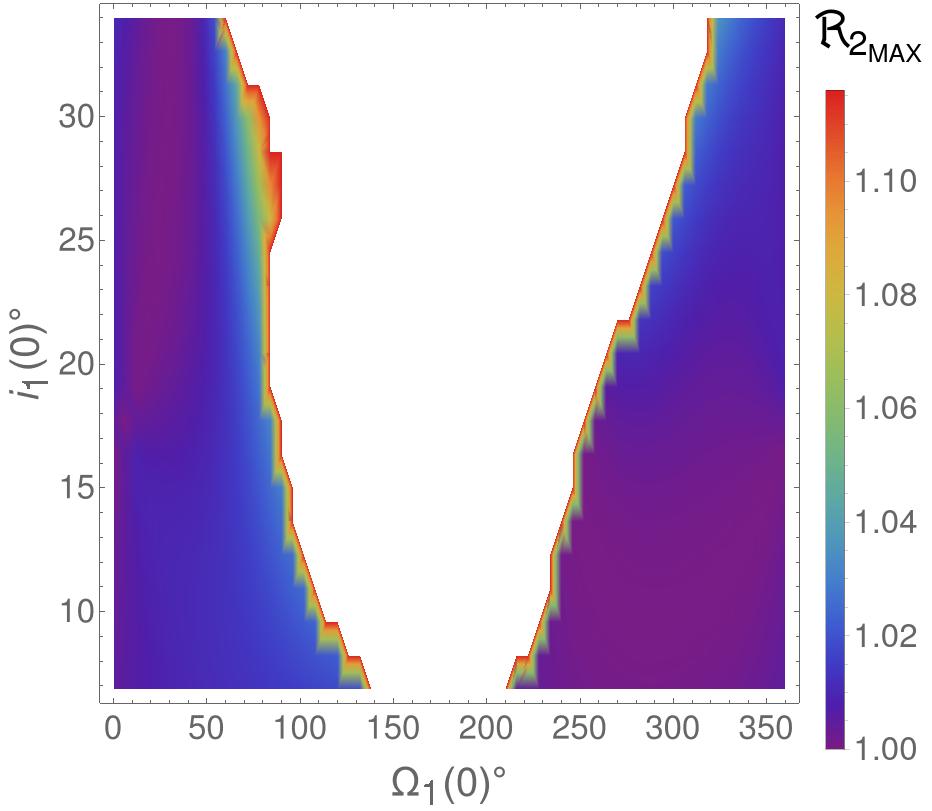}
\end{minipage}
\caption{Color-grid plots of the maximal values reached by the ratio
  $\mathfrak{R}_1$ (on the left) and $\mathfrak{R}_2\,$ (on the
  right); see the text for more details.}
\label{fig.graph_astro}
\end{figure}

\begin{figure}[h]
\begin{minipage}{.5\textwidth}
\includegraphics[scale=0.33]{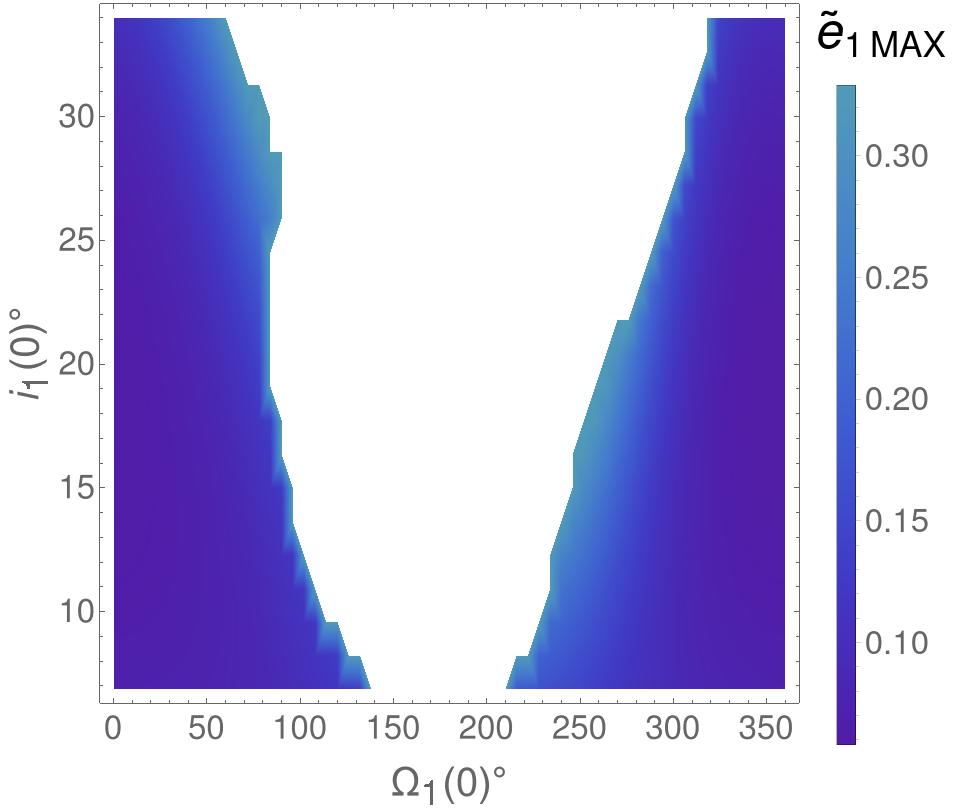}
\end{minipage}
\qquad\quad\quad
\begin{minipage}{.4\textwidth}
\caption{Color-grid plots of the maximum of the function $\t{\e}(t)$,
  which is defined in~\eqref{astro_e1new}.}
\label{fig.graph_astro_ecc1}
\end{minipage}
\end{figure}

\noindent
The results about the maxima of the functions defined
in~\eqref{astro_ratios}--\eqref{astro_e1new} are reported in
Figures~\ref{fig.graph_astro}--\ref{fig.graph_astro_ecc1}. The white
central regions of those pictures correspond to those pairs
$\big(\i_1(0),\Omega_1(0)\big)\,$ for which we obtain failure of the
preliminary test,
i.e. $\max\{\mathfrak{r}_1\,,\,\mathfrak{r}_2\}>1$. We immediately
recognize that the missing part of the plots (where the determination
of the initial conditions is considered so unreliable that the
corresponding numerical integrations are not performed at all) nearly
coincides with the central region of Figure~\ref{fig.Color_compl_e1},
where the orbital eccentricity of \ups$\b$ reaches critical values. We
conclude that the stability domain in the space of the initial values
of $\i_1(0)$ and $\Omega_1(0)$ (which are unknown observational data)
can be reconstructed in a reliable way through the application of the
above criterion, which only involves the series transformations, as
well as through the numerical solutions of our integrable secular
model with $2$~DOF. We emphasize that this allows to reduce
significantly the computational cost with respect to the long-term
symplectic integrations of the complete $4$-body problem, which is a
$9$~DOF Hamiltonian system.

Comparing the regions of the stability domain at the border near the (white) central ones, we see that all three
numerical indicators plotted in
Figures~\ref{fig.graph_astro}--\ref{fig.graph_astro_ecc1} increase
their values when the unstable zone is approached. This is in
agreement with the expectations and the comparison with
Figure~\ref{fig.Color_compl_e1}. On the other hand, the $2$~DOF secular model is unable to capture the region of instability internal to the stable one, highlighted by two green stripes
starting from the bottom of Figure~\ref{fig.Color_compl_e1} in
correspondence with $\Omega_1(0)=0^\circ=360^\circ$. The two curved stripes look rather symmetric and they join each other around the point $\big(\i_1(0)\,,\,\Omega_1(0)\big)\simeq\big(30^\circ,\,0^\circ\big)=\big(30^\circ,\,360^\circ\big)$. Since the dependence of the Hamiltonian on the angles $\vet{q}$ (which describes the dynamics of the outer exoplanets) is removed from the $2$DOF model, it seems reasonable that some of the resonances are not present in the normal form
generated by the algorithm {\it \`a la} Birkhoff, although they play a remarkable role in the dynamics of more complex models.

\section{Secular orbital evolution of \ups$\b$ taking also into account relativistic effects}
\label{sec:QPR_REL}

In this Section we study the dynamics of \ups$\b$ in the
framework of a secular quasi-periodic restricted Hamiltonian model
where also corrections due to general relativity are taken into
account. Since we focus
on the orbital dynamics of the innermost planet of the $\upsilon$-Andromed{\ae} system and it is very close
to a star that is about 30\% more massive than the Sun (let us recall
that the value of the semi-major axis of \ups$\b$ is reported in
Table~\ref{tab:param.orb.b}, i.e., $a_1=0.0594$~AU), it is natural to
expect that the corrections due to general relativity can play a
relevant role for the system under consideration. Similarly as in the previous Sections, we study these effects in the framework of a $2$~DOF secular model. We start by considering the following Hamiltonian:
$$
\Hscr=\Hscr_{4BP}+\Hscr_{GR}\, ,
$$ where $\Hscr_{4BP}$ defines the four body problem (see~\eqref{Ham.4BP.reduced.mass}) and $\Hscr_{GR}$ describes the
general (post-Newtonian) relativistic corrections to the Newtonian
mechanics. Following~\cite{miggoz2009}, the
secular quasi-periodic restricted Hamiltonian which includes
corrections due to the General Relativity (hereafter, GR) is obtained
by removing the dependence of the Hamiltonian on the fast angles.
Therefore, we introduce
\begin{equation}
\label{mean_REL}
\Hscr^{(GR)}_{sec}=
\int_{\toro^3}\frac{\Hscr_{4BP}}{8\pi^3}\,  d\lambda_1 d\lambda_2 d\lambda_3 +
\int_{\toro} \frac{\Hscr_{GR}}{2\pi}\,  d M_1\,
:= \Hscr^{(NG)}_{sec}+\avg{\Hscr_{{GR}}}_{M_1} \, ,
\end{equation}
where the expansion of the mean of the $4BP$ Hamiltonian $\Hscr^{(NG)}_{sec}$ (recall
definition~\eqref{average_Ham4BP}) is explicitely written in
equation~\eqref{H4BP_expl}, while the average of the GR contribution
with respect to the mean anomaly of \ups$\b$ is given by
\begin{equation}
\label{Ham.sec.rel.qpr_1}
\avg{\Hscr_{{GR}}}_{M_1}=
-\frac{3\,\mathcal{G}^2\, m_0^2\, m_1}{ a_1^2 c^2\sqrt{1-\e_1^2}}
+\frac{15\,\mathcal{G}^2\, m_0^2\, m_1}{8 a_1^2 c^2}
-\frac{\mathcal{G}^2 \,m_0\, m_1^2}{8 a_1^2 c^2}\, ,
\end{equation} 
$c$ being the velocity of light in vacuum. In the above expression of
$\avg{\Hscr_{{GR}}}_{M_1}\,$, the summand where the eccentricity of
\ups$\b$ (i.e., $\e_1$) occurs in the denominator is the only to be
untrivial, in the sense that the other two give additional
\emph{constant} contribution to the secular Hamiltonian and, then,
they can be disregarded. By proceeding in a similar way to what has
been already done for the classical expansions of the initial
Hamiltonian~\eqref{Ham.3BP.reduced.mass}, it is possible to express
$\avg{\Hscr_{{GR}}}_{M_1}$ in the Poincar\'e variables $(\csi_1,\eta_1)$,
described in equation~\eqref{Poinc.var.U}.

Thus, keeping in mind the procedure explained in
Section~\ref{sec:QPR}, one easily realizes that the secular
quasi-periodic restricted model of the dynamics of \ups$\b$ which
includes relativistic corrections (hereafter, SQPR-GR) can be
described by the following $2+3/2$~DOF Hamiltonian:
\begin{equation}
\begin{aligned}
\label{Ham.b.new_REL}
\Hscr^{(GR)}_{sec,\, 2+\frac{3}{2}}
(\vet{p},\vet{q}, \csi_1, \eta_1, P_1, Q_1 )&=
\omega_3\,p_3 + \omega_4\,p_4 + \omega_5\,p_5\\
&\phantom{=}+\Hscr^{(NG)}_{sec}(q_3,  q_4, q_5, \csi_1, \eta_1, P_1, Q_1 )
+\avg{\Hscr_{GR}}_{M_1}(\csi_1,\eta_1) \, ,
\end{aligned}
\end{equation}
where the angular velocity vector
$\vet{\omega}=(\omega_3,\omega_4,\omega_5\,)$ is given
in~\eqref{freq.fond.CD} and $\Hscr^{(NG)}_{sec}$ can be replaced by
$\Hscr^{1-2}_{sec}+\Hscr^{1-3}_{sec}$ appearing in
formula~\eqref{Ham.b.new}.  Finally, in the framework of this SQPR-GR
model, the equations for the orbital motion of the innermost planet
can be written as
\begin{equation}
\label{campo.Ham.b_REL}
\begin{cases}
  \dot{q_3}=\partial \Hscr^{(GR)}_{sec,2+\frac{3}{2}}/\partial p_3=\omega_3\\
  \dot{q_4}=\partial \Hscr^{(GR)}_{sec,2+\frac{3}{2}}/\partial p_4=\omega_4\\
  \dot{q_5}=\partial \Hscr^{(GR)}_{sec,2+\frac{3}{2}}/\partial p_5=\omega_5\\
  \dot{\csi}_{1}=
  -\partial \left(\Hscr_{sec}^{(NG)}(q_3, q_4, q_5, \csi_1, \eta_1, P_1, Q_1 )
  +\avg{\Hscr_{GR}}_{M_1}(\csi_1, \eta_1 )\right) /\partial \eta_1 \\
  \dot{\eta}_{1}=
  \partial \left(\Hscr_{sec}^{(NG)}(q_3, q_4, q_5, \csi_1, \eta_1, P_1, Q_1 )
  +\avg{\Hscr_{GR}}_{M_1}(\csi_1, \eta_1)\right) /\partial \csi_1 \\
  \dot{P}_{1}=
  -\partial \Hscr_{sec}^{(NG)}(q_3,  q_4, q_5, \csi_1, \eta_1, P_1, Q_1)
  /\partial Q_1 \\
  \dot{Q}_{1}=
  \partial \Hscr_{sec}^{(NG)}(q_3,  q_4, q_5, \csi_1, \eta_1, P_1, Q_1 )
  /\partial P_1
\end{cases}\, .
\end{equation}

\subsection{Numerical integration of the SQPR-GR model}
\label{sub:num_integ_qper_REL}

\begin{figure}[h]
\subfloat[]{
\begin{minipage}{.45\textwidth}
\includegraphics[scale=0.295]
{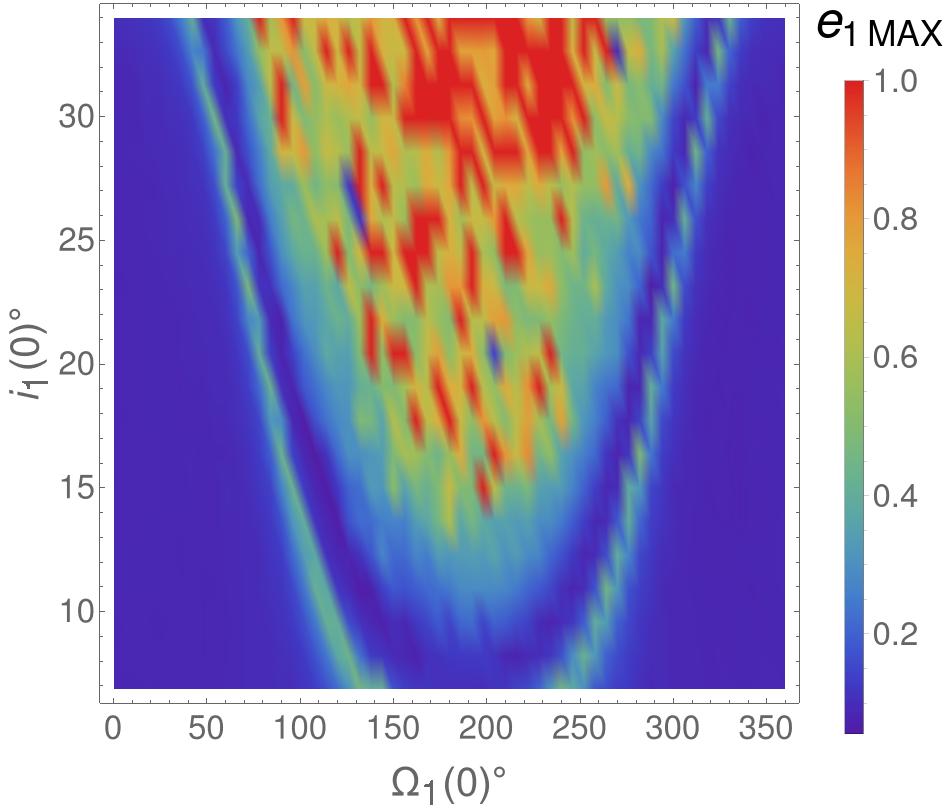}
\vskip -.05truecm
\label{fig.Color_qper_e1_REL}
\end{minipage}}\qquad
\subfloat[]
{
\begin{minipage}{.45\textwidth}
\includegraphics[scale=0.295]
{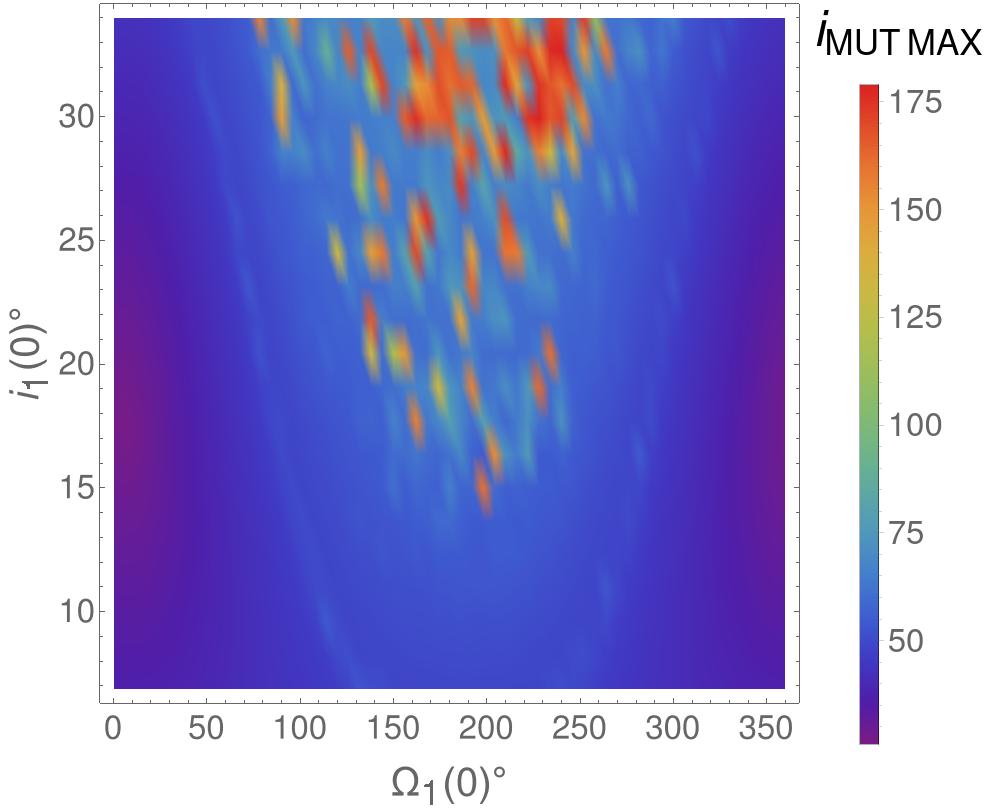}
\vskip -.05truecm
\label{fig.Color_qper_imut_REL}
\end{minipage}
}
\caption{Color-grid plots of the maximal value reached by the
  eccentricity of \ups$\b$ (on the left) and by the mutual inclination
  between \ups$\b$ and \ups$\c$ (on the right). The maxima are
  computed during the RK4 numerical integrations (each of them
  covering a timespan of $10^5$~yr) of the SQPR-GR equations of
  motion~\eqref{campo.Ham.b_REL}, which cover a timespan of $10^5$
  yr.}
\label{fig.Color_qper_e1_imut_REL}
\end{figure}

Similarly as in Subsection~\ref{subsub:num_integ_qper}, we numerically integrate
the equations of motion for the secular quasi-periodic restricted
Hamiltonian with general relativistic corrections, defined in
formula~\eqref{campo.Ham.b_REL}. As initial values of the orbital
parameters $a_1(0)$, $\e_1(0)$, $M_1(0)$ and $\omega_1(0)$ we take
those reported in Table~\ref{tab:param.orb.b}; moreover, we set
$m_1=0.674$ as value for the mass of \ups$\b$ and
$(\i_1(0)\,,\Omega_1(0))$ ranging in the $20\times 60$ regular grid
that covers $I_\i\times I_\Omega=[6.865^\circ,
  34^\circ]\times[0^\circ,360^\circ]$. Hence, it is possible to
compute the corresponding initial values of the orbital elements in
the Laplace reference frame (which is determined taking into account
\ups$\c$ and \ups$\d$ only) and to perform $21\times 60$ numerical
integrations starting from all these initial data.  Once again, for
each numerical integration, we are interested in determining the
maximal values reached by the eccentricity of \ups$\b$ and by the
maximal mutual inclination between \ups$\b$ and \ups$\c\,$. The
results are reported in the color-grid plots of
Figure~\ref{fig.Color_qper_e1_imut_REL}.

By comparing Figure~\ref{fig.Color_qper_e1_REL} with
Figure~\ref{fig.Color_qper_e1}, one can immediately realize that the
regions colored in blue are much wider in the former than in the
latter one. Indeed, the darker regions refer to initial conditions
which generate motions with maximal values of the eccentricity of
\ups$\b$ that are relatively low, while the zones colored in red or
yellow correspond to such large values of the eccentricity implying that
those orbits have to be considered unstable. Therefore, our numerical
explorations highlight that the effects due to general relativity play
a stabilizing role on the orbital dynamics of the innermost planet.
This conclusion is in agreement with was already remarked about the
past evolution of our Solar System, in particular for what concerns
the orbital eccentricity of Mercury (see~\cite{lasgas2009}).

\begin{figure}[h]
\subfloat[]{
\begin{minipage}{.45\textwidth}
\includegraphics[scale=0.42]
{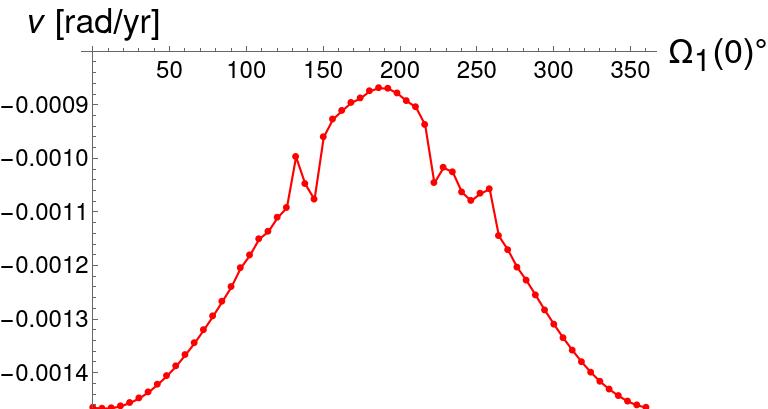}
\vskip -.05truecm
\label{fig.plotfreqREL_short}
\end{minipage}}\qquad
\subfloat[]
{
\begin{minipage}{.45\textwidth}
\includegraphics[scale=0.42]
{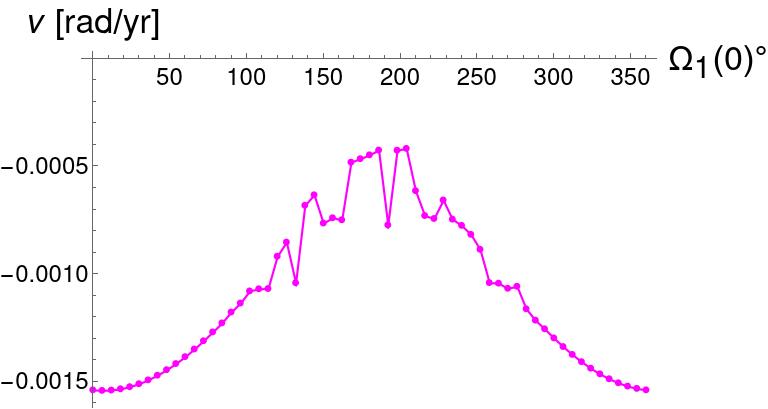}
\vskip -.05truecm
\label{fig.plotfreqREL_high}
\end{minipage}
}
\caption{Behaviour of the fundamental angular velocity
    $\nu$ as obtained by applying the Frequency Map Analysis method to
    the signal $\xi_1(t)+i\eta_1(t)\,$, computed through the RK4
    numerical integration of the SQPR-GR
    model~\eqref{campo.Ham.b_REL}, covering a timespan of
    $1.31072\cdot 10^{5}$ yr. We take, as initial conditions,
    $(\i_1(0),\Omega_1(0))\in\lbrace 6.865^\circ \rbrace \times I_\Omega$
    for the left panel and $(\i_1(0),\Omega_1(0))\in\lbrace 10.9353^\circ
    \rbrace \times I_\Omega$ for the right one.}
\label{fig.plotfreqREL}
\end{figure}

Moreover, as already done in
  Section~\ref{subsub:num_integ_qper}, in order to further explore the
  stable and chaotic regions of Figure~\ref{fig.Color_qper_e1_REL}, we
  apply the Frequency Map Analysis method to the signal
  $\xi_1(t)+i\eta_1(t)$ as produced by the numerical integration of
  the system~\eqref{campo.Ham.b_REL}, i.e., in the SQPR-GR
  approximation. We perform the numerical integrations as described at
  the beginning of the present Section, taking into account only a few
  values in $I_\i\,$ for the initial inclinations,
  i.e. $\i_1(0)=6.865^\circ,\,
  8.22175^\circ,\,9.5785^\circ,\,10.9353^\circ$ and $\Omega_1(0)\in
  I_\Omega\,$. In Figure~\ref{fig.plotfreqREL} we report the behaviour
  of the angular velocity $\nu$ corresponding to the first component
  of the approximation of $\xi_1(t)+i\eta_1(t)$, as obtained by
  applying the FA computational algorithm; we recall that this
  quantity is related to the precession rate of $\varpi_1\,$. As
  initial value for the inclination $\i_1(0)$ we fix $6.865^\circ$ for
  Figure~\ref{fig.plotfreqREL_short} and $10.9353^\circ$ for
  Figure~\ref{fig.plotfreqREL_high}. Also here, we do not report the
  cases $(\i_1(0),\Omega_1(0))\in \lbrace 8.22175^\circ,\,
  9.5785^\circ \rbrace \times I_\Omega\,$, since the behaviour of
  these plots is similar to the ones in Figure~\ref{fig.plotfreqREL}.

 The situation is well described by
  Figure~\ref{fig.plotfreqREL_short} and analogous considerations hold
  for Figure~\ref{fig.plotfreqREL_high} apart a few main
  differences which will be highlighted in the following
  discussion. When the values of $\Omega_1(0)$ are ranging in $[0,
    \sim 120^\circ]$ and $[\sim 260^\circ, \, 360^\circ]$ we can
  observe a regular behaviour of the angular velocity $\nu$ which is
  also nearly monotone, with the only exception around a local
  minimum.  According to the Frequency Map Analysis method, such a
  regular regime is due to the presence of many invariant tori which
  fill the stability region located at the two lateral sides of the
  plot~\ref{fig.Color_qper_e1_REL}. In the case of
  Figure~\ref{fig.plotfreqREL_short}, this also applies when
  $\Omega_1(0)$ is ranging in $[\sim 150^\circ, \sim 220^\circ ]$,
  which corresponds to the stable blue internal area of
  Figure~\ref{fig.Color_qper_e1_REL}. On the other hand, in the case
  of Figure~\ref{fig.plotfreqREL_high}, for the same range of initial
  values of the node longitude of \ups$\b$, the behaviour is not so
  regular; this is in agreement with the fact that in correspondence
  with $\i_1(0)\sim 11^\circ$ the plot of the maximal values of $\e_1$
  in the central region highlights the occurrence of chaotical
  phenomena.  Moreover, for what concerns values of $\Omega_1(0)$ in
  $[\sim 120^\circ, \, \sim 150^\circ] $ and $[\sim 220^\circ, \, \sim
    260^\circ] $ (corresponding to the green stripes of
  Figure~\ref{fig.Color_qper_e1_REL}),
  Figure~\ref{fig.plotfreqREL_short} shows a behaviour typical of the
  crossing of a resonance in the chaotic region surrounding a
  separatrix. The value of the angular velocity for which this
  phenomenon takes place is, again, related to $\omega_4\simeq
  -1.04\cdot 10^{-3}$ (as it can be easily appreciated looking to the
  small plateau appearing in Figure~\ref{fig.plotfreqREL_high}). 

 Comparing Figure~\ref{fig.plotfreqREL} with
  Figure~\ref{fig.plotfreqNOREL} the enlargement of the stable region
  is evident. Moreover, we can also see how much this phenomenon is
  influenced by the modification of the pericenter precession rate of
  the inner planet due to relativistic effects. Indeed, looking at the
  values reported on the $y$-axis of Figures~\ref{fig.plotfreqREL}
  and~\ref{fig.plotfreqNOREL}, one can appreciate that the fundamental
  angular velocity, in the case of the SQPR model, takes values
  remarkably closer to zero with respect to those assumed in the case
  of the SQPR-GR model.

\subsection{Application of the normalization algorithms to the secular quasi-periodic restricted model of the dynamics of \ups$\b$ with relativistic corrections}
\label{sub:risult_astro_REL}

Starting from Hamiltonian~\eqref{Ham.b.new_REL}, we can reapply the
normalization algorithms described in
Subsections~\ref{sub:toro_ellittico} and~\ref{sub:media_q3q4q5}. All
this computational procedure ends up with the introduction of a new
$2$~DOF Hamiltonian\footnote{In the expansion~\eqref{Ham.2DOF_REL},
  the term that is linear in the dummy actions (i.e.,
  $\vet{\omega}_B\cdot\vet{p}$) is removed, because it is irrelevant
  for the present discussion.} model which can be written in the
following form (analogous to the one reported in
formula~\eqref{Ham.astro}):
\begin{align}
\label{Ham.2DOF_REL}
\begin{aligned}
\Hscr_{2DOF}^{(GR)}(\vet{I},\vet{\alpha})
&= \Escr_{B;GR}+\vet{\Omega}_{B;GR}\cdot\vet{I}+
\sum_{s= 0}^{\mathcal{N}_S}\sum_{l=3}^{\mathcal{N}_L}
h_{l}^{(\mathcal{N}_L-2,s)}(\vet{I},\vet{\alpha})\, ,
\end{aligned}
\end{align}
where $\Escr_{B;GR}\in\reali$, $\vet{\Omega}_{B;GR}\in\reali^2$ and
$h_{l}^{(\mathcal{N}_L-2,s)}\in\mathfrak{P}_{l,2s}$
$\forall\ l=3,\,\ldots\,,\,\mathcal{N}_L\,,\>
s=0\,,\,\ldots\,,\,\mathcal{N}_S\,$. We emphasize that also
$\Hscr_{2DOF}^{(GR)}$ is integrable because of the same reasons
already discussed in Section~\ref{sec:risult_astro}; indeed, after
having checked that
${\partial\Hscr_{2DOF}^{(GR)}}/{\partial\alpha_1}\,+\,
{\partial\Hscr_{2DOF}^{(GR)}}/{\partial\alpha_2}\,=\,0\,$, we can apply the
Liouville theorem, because there are two independent constants of
motion, i.e., $I_1+I_2$ and $\Hscr_{2DOF}^{(GR)}$ itself.

\begin{figure}[h]
\begin{minipage}{.45\textwidth}
\includegraphics[scale=0.295]
{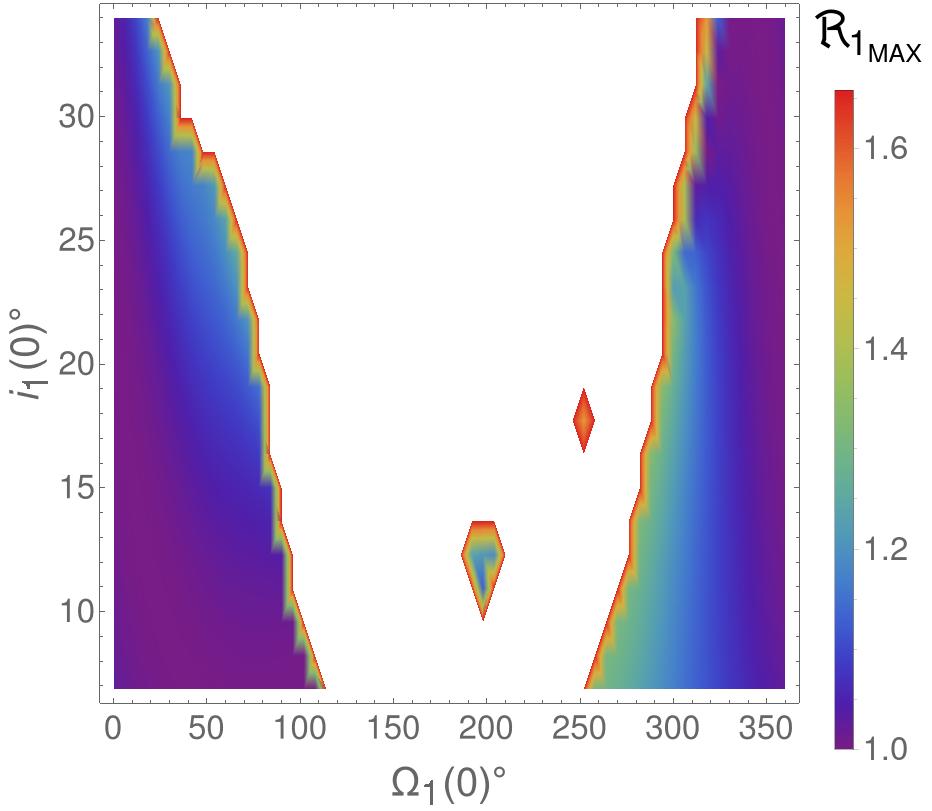}
\vskip -.05truecm
\end{minipage}\quad\,
\begin{minipage}{.45\textwidth}
\includegraphics[scale=0.295]
{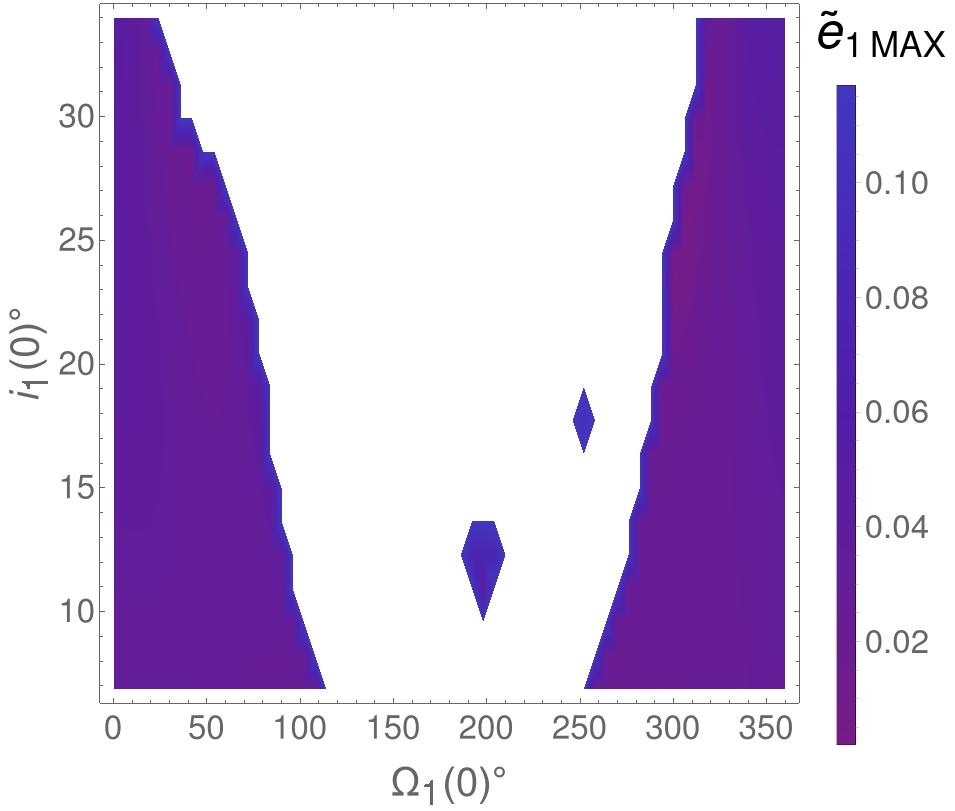}
\vskip -.05truecm
\end{minipage}
\vskip -.05truecm
\caption{Color-grid plots of the maximal values reached by the ratio
  $\mathfrak{R}_1(t)$ (on the left) and the function $\t{\e}(t)$ (on
  the right), which are defined
  in~\eqref{astro_ratios}--\eqref{astro_e1new}. These laws of motion
  are computed along the flow induced by the $2$~DOF Hamiltonian
  $\Hscr_{2DOF}^{(GR)}$ which takes into account also GR corrections
  and is defined in~\eqref{Ham.2DOF_REL}, in the particular case with
  $\mathcal{N}_S=5$ and $\mathcal{N}_L=6$.}
\label{fig.graph_astro_REL_S5_L6}
\end{figure}

\begin{figure}[h]
\begin{minipage}{.45\textwidth}
\includegraphics[scale=0.295]
 {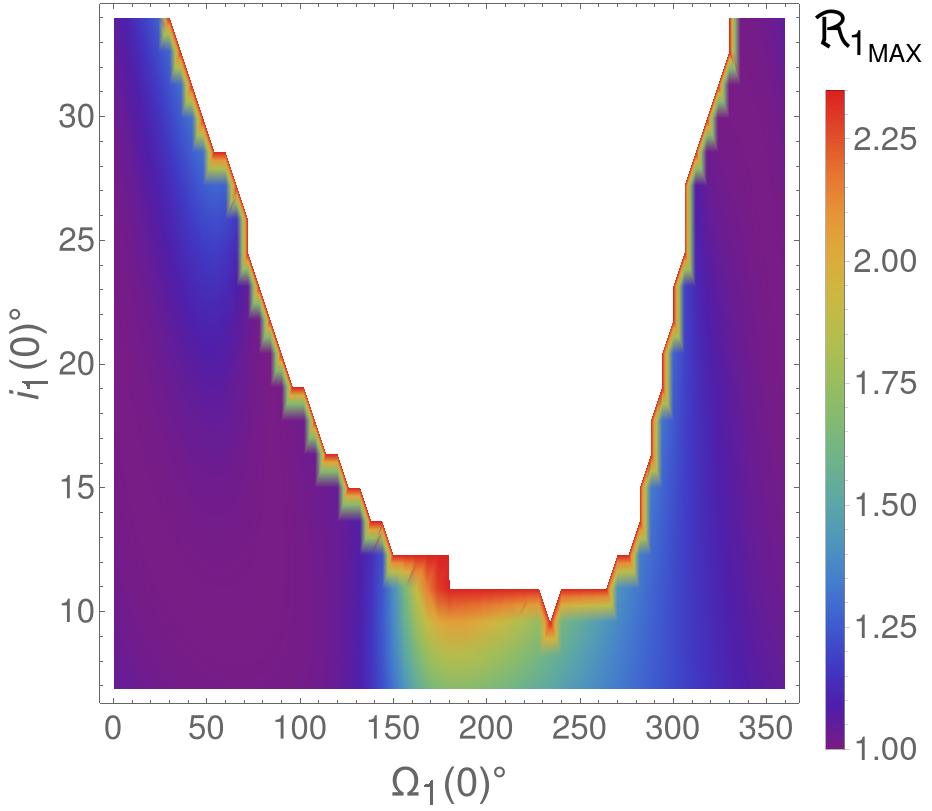}
\vskip -.05truecm
\end{minipage}\quad\,
\begin{minipage}{.45\textwidth}
\includegraphics[scale=0.295]
{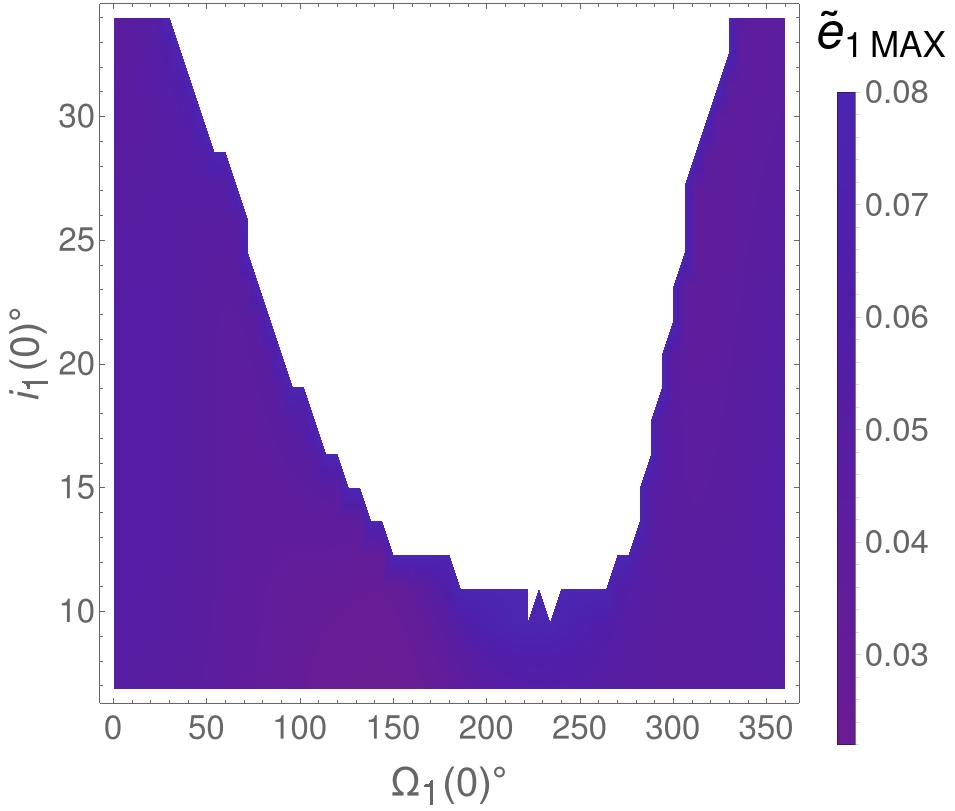}
\vskip -.05truecm
\end{minipage}
\vskip -.05truecm
\caption{Same as in Figure~\ref{fig.graph_astro_REL_S5_L6}, in the
  case with $\mathcal{N}_S=6$ and $\mathcal{N}_L=5$.}
\label{fig.graph_astro_REL_S6_L5}
\end{figure}

\begin{figure}[h]
\begin{minipage}{.45\textwidth}
\includegraphics[scale=0.295]
  {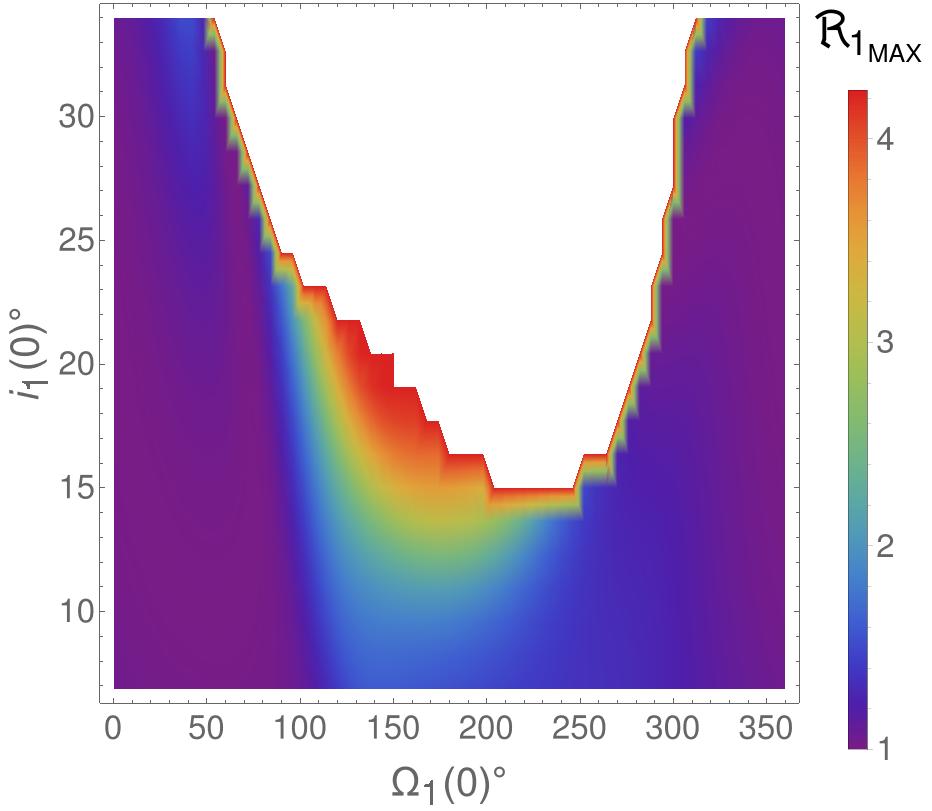}
\vskip -.05truecm
\end{minipage}\quad\,
\begin{minipage}{.45\textwidth}
\includegraphics[scale=0.295]
 {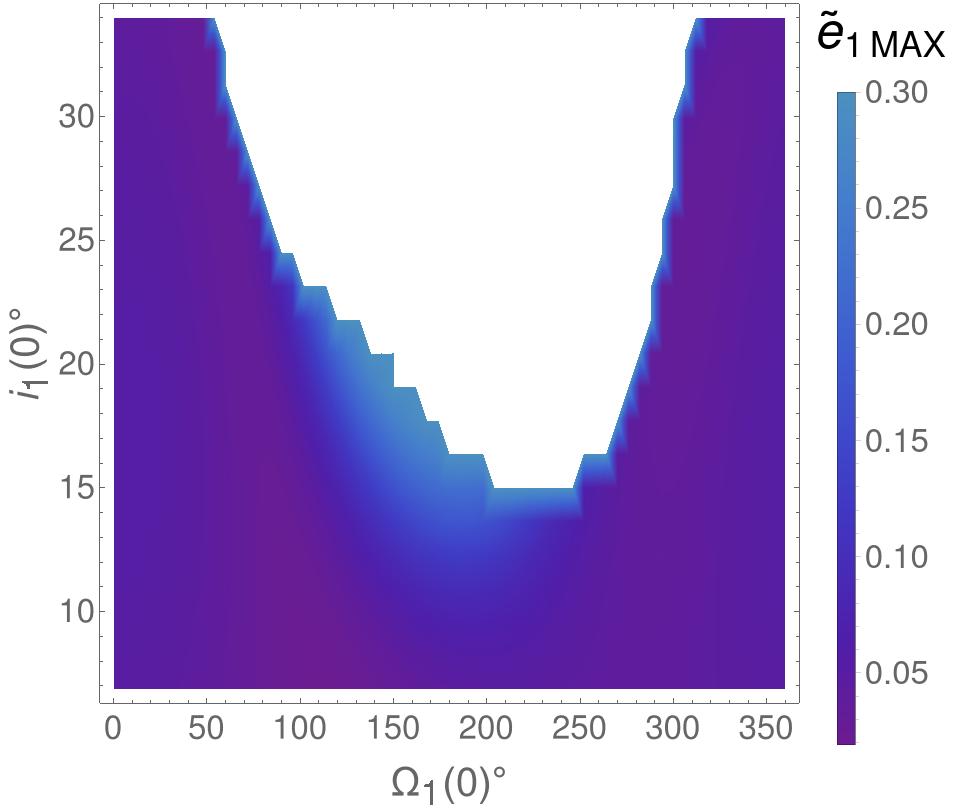}
\vskip -.05truecm
\end{minipage}
\vskip -.05truecm
\caption{Same as in Figure~\ref{fig.graph_astro_REL_S5_L6}, in the
  case with $\mathcal{N}_S=6$ and $\mathcal{N}_L=4$.}
\label{fig.graph_astro_REL_S6_L4}
\end{figure}

Moreover, also for this new model we can reproduce the same kind of
numerical exploration described in Section~\ref{sec:risult_astro}.  In
particular, we can compute the values of the numerical indicators
$\mathfrak{R}_{1\,{\rm MAX}}\,$, $\mathfrak{R}_{2\,{\rm MAX}}$ and
$\t{\e}_{1\,{\rm MAX}}$ corresponding to each pair
$\big(\i_1(0),\Omega_1(0)\big)\,$ of the $21\times 60$ points
definining the regular grid which covers $I_\i\times
I_\Omega=[6.865^\circ, 34^\circ]\times[0^\circ,360^\circ]$. In the
following, we analyze the color-grid plots for a few different values
of the parameter ruling the truncation in the trigonometric degree,
namely $\mathcal{N}_S\,$, and in the square root of the action, i.e.,
$\mathcal{N}_L\,$.  The color-grid plots for the maximal value reached
by $\mathfrak{R}_1\,$ and $\t{\e}_1$ are reported
in
Figures~\ref{fig.graph_astro_REL_S5_L6}--\ref{fig.graph_astro_REL_S6_L4}.

Let us recall that $\mathfrak{R}_{2\,{\rm MAX}}$ is an evaluation of
the maximal excursion of the inclination of \ups$\b$. For the sake of
brevity, its plots are omitted and they are not included in
Figures~\ref{fig.graph_astro_REL_S5_L6}--\ref{fig.graph_astro_REL_S6_L4},
because in our numerical explorations the ranges of values experienced
by $\mathfrak{R}_{2\,{\rm MAX}}$ are so narrow that their analysis
does not look so significant. Therefore, it is better to focus on the
plots of $\mathfrak{R}_{1\,{\rm MAX}}$ and $\t{\e}_{1\,{\rm MAX}}\,$;
let us recall that both of them refer to the eccentricity of \ups$\b$.
By comparing
Figures~\ref{fig.graph_astro_REL_S5_L6}--\ref{fig.graph_astro_REL_S6_L4},
one can appreciate a well known phenomenon concerning the constructive
algorithms {\it \`a la} Birkhoff: the greater the number of normalization steps (i.e., $\mathcal{N}_L-2$), the smaller the
domain of applicability (see, e.g.,~\cite{gio2003} for the discussion
about the determination of the optimal step).

By comparing
Figures~\ref{fig.graph_astro_REL_S6_L5}--\ref{fig.graph_astro_REL_S6_L4}
also with Figure~\ref{fig.Color_qper_e1_REL}, we observe that
in the cases with $\mathcal{N}_L=4,\,5$ our computational procedure is
able to reconstruct with a good accuracy the $U$--shaped border of the
stability domain. Note that the horizontal strip at
the bottom of the plots\footnote{This means that we are considering
  initial values of the inclination $\i_1(0)$ that are close to that
 of \ups$\c$.} corresponds to orbital motions which look stable, since the eccentricity of \ups$\b$ does not reach large
values (with the eventual exception of the narrow green areas that in
Figure~\ref{fig.Color_qper_e1_REL} are expected to correspond to
resonant regions). This highlights a main difference with the
non-relativistic model discussed in Section~\ref{sec:risult_astro},
because in that case there is an interval of values of $\Omega_1(0)$
centered about $180^\circ$ for which none of the initial inclinations
$\i_1(0)\in[6.865^\circ, 34^\circ]$ corresponds to a stable
orbital configuration (see Figure~\ref{fig.Color_qper_e1}). The
reliability of our simplified $2$~DOF Hamiltonian model (which is
defined in formula~\eqref{Ham.2DOF_REL} and takes into account also GR
corrections) is enforced also by the fact that it is able to capture
also this phenomenon.

\section*{Acknowledgments}
We are very grateful to Prof.~Christos Efthymiopoulos for his
encouragements, suggestions and the critical reading of the
manuscript.  This work has been partially supported by the MIUR-PRIN
20178CJA2B ``New Frontiers of Celestial Mechanics: theory and
Applications'' and by the MIUR Excellence Department Project
MatMod@TOV awarded to the Department of Mathematics, University of
Rome ``Tor Vergata'' .

\bibliographystyle{abbrv}
\bibliography{bibliography}

\begin{thebibliography}{10}

\bibitem{butetal1999}
R.~P. Butler, G.~W. Marcy, D.~A. Fischer, T.~M. Brown, A.~R. Contos, S.~G.
  Korzennik, P.~Nisenson, and R.~W. Noyes.
\newblock Evidence for multiple companions to $\upsilon$ {A}ndromedae.
\newblock {\em The Astrophysical Journal}, 526(2):916, dec 1999.

\bibitem{car2022}
C.~Caracciolo.
\newblock Normal form for lower dimensional elliptic tori in {H}amiltonian
  systems.
\newblock {\em Mathematics in Engineering}, 4(6):1--40, 2022.

\bibitem{caretal2022}
C.~Caracciolo, U.~Locatelli, M.~Sansottera, and M.~Volpi.
\newblock Librational {KAM} tori in the secular dynamics of the $\upsilon$
  {A}ndromed{\ae} planetary system.
\newblock {\em Monthly Notices of the Royal Astronomical Society},
  510(2):2147--2166, 2022.

\bibitem{curetal2011}
S.~Curiel, J.~Cant{\'o}, L.~Georgiev, C.~Ch{\'a}vez, and A.~Poveda.
\newblock A fourth planet orbiting $\upsilon$ {A}ndromedae.
\newblock {\em Astronomy \& Astrophysics}, 525:A78, 2011.

\bibitem{deietal2015}
R.~Deitrick, R.~Barnes, B.~McArthur, T.~R. Quinn, R.~Luger, A.~Antonsen, and
  G.~F. Benedict.
\newblock The three-dimensional architecture of the $\upsilon$ andromedae
  planetary system.
\newblock {\em The Astrophysical Journal}, 798(1):46, 2015.

\bibitem{gio2003}
A.~Giorgilli.
\newblock {\em Notes on exponential stability of {H}amiltonian systems}.
\newblock Pubblicazioni della Classe di Scienze, Scuola Normale Superiore,
  Pisa. Centro di Ricerca Matematica ``Ennio De Giorgi", 2003.

\bibitem{gio2022}
A.~Giorgilli.
\newblock {\em Notes on {H}amiltonian {D}ynamical {S}ystems}, volume 102.
\newblock Cambridge University Press, 2022.

\bibitem{GioDFGS-1989}
A.~Giorgilli, A.~Delshams, E.~Fontich, L.~Galgani, and C.~Sim\'o.
\newblock Effective stability for a {H}amiltonian system near an elliptic
  equilibrium point, with an application to the restricted three--body problem.
\newblock {\em J. Differential Equations}, 77:167--198, 1989.

\bibitem{giolocsan2014}
A.~Giorgilli, U.~Locatelli, and M.~Sansottera.
\newblock On the convergence of an algorithm constructing the normal form for
  lower dimensional elliptic tori in planetary systems.
\newblock {\em Celestial Mechanics and Dynamical Astronomy}, 119:397--424,
  2014.

\bibitem{giolocsan2017}
A.~Giorgilli, U.~Locatelli, and M.~Sansottera.
\newblock Secular dynamics of a planar model of the
  {S}un-{J}upiter-{S}aturn-{U}ranus system; effective stability in the light of
  {K}olmogorov and {N}ekhoroshev theories.
\newblock {\em Regular and Chaotic Dynamics}, 22:54--77, 2017.

\bibitem{gro1967}
W.~Gr{\"o}bner.
\newblock {\em Die Lie-reihen und ihre Anwendungen}, volume~3.
\newblock Deutscher Verlag der Wissenschaften, 1967.

\bibitem{hoaetal2022}
N.~H. Hoang, F.~Mogavero, and J.~Laskar.
\newblock Long-term instability of the inner {S}olar {S}ystem: numerical
  experiments.
\newblock {\em Monthly Notices of the Royal Astronomical Society},
  514(1):1342--1350, 2022.

\bibitem{las1999}
J.~Laskar.
\newblock Introduction to {F}requency {M}ap {A}nalysis.
\newblock In C.~Sim{\'o}, editor, {\em Hamiltonian Systems with Three or More
  Degrees of Freedom}, pages 134--150. Springer Netherlands, Dordrecht, 1999.

\bibitem{las2005}
J.~Laskar.
\newblock Frequency map analysis and quasiperiodic decompositions.
\newblock In E.~Lega, D.~Benest, and C.~Froeschl{\'e}, editors, {\em
  Hamiltonian systems and Fourier analysis: new prospects for gravitational
  dynamics}. Cambridge Scientific Pub Ltd, 2005.

\bibitem{lasgas2009}
J.~Laskar and M.~Gastineau.
\newblock Existence of collisional trajectories of {M}ercury, {M}ars and
  {V}enus with the {E}arth.
\newblock {\em Nature}, 459(7248):817--819, 2009.

\bibitem{lasrob2001}
J.~Laskar and P.~Robutel.
\newblock High order symplectic integrators for perturbed {H}amiltonian
  systems.
\newblock {\em Celestial Mechanics and Dynamical Astronomy}, 80(1):39--62,
  2001.

\bibitem{locetal2022}
U.~Locatelli, C.~Caracciolo, M.~Sansottera, and M.~Volpi.
\newblock Invariant {KAM} tori: from theory to applications to exoplanetary
  systems.
\newblock {\em I-Celmech training school, Springer PROMS}, 2022.

\bibitem{locetal2022a}
U.~Locatelli, C.~Caracciolo, M.~Sansottera, and M.~Volpi.
\newblock A numerical criterion evaluating the robustness of planetary
  architectures; applications to the $\upsilon$ {A}ndromed{\ae} system.
\newblock {\em Proceedings of the International Astronomical Union},
  15(S364):65–84, 2022.

\bibitem{locgio2000}
U.~Locatelli and A.~Giorgilli.
\newblock Invariant tori in the secular motions of the three-body planetary
  systems.
\newblock {\em Celestial Mechanics and Dynamical Astronomy}, 78(1):47--74,
  2000.

\bibitem{mas2023}
R.~Mastroianni.
\newblock Hamiltonian secular theory and {KAM} stability in exoplanetary
  systems with {3D} orbital architecture.
\newblock {\em PhD Thesis, Dep. of Mathematics ``Tullio-Levi Civita'', Univ. of
  Padua}, 2023.

\bibitem{mayque1995}
M.~Mayor and D.~Queloz.
\newblock A {J}upiter-mass companion to a solar-type star.
\newblock {\em Nature}, 378(6555):355--359, 1995.

\bibitem{mcaetal2010}
B.~E. McArthur, G.~F. Benedict, R.~Barnes, E.~Martioli, S.~Korzennik, E.~Nelan,
  and R.~P. Butler.
\newblock New observational constraints on the $\upsilon$ {A}ndromedae system
  with data from the {H}ubble {S}pace {T}elescope and {H}obby-{E}berly
  {T}elescope.
\newblock {\em The Astrophysical Journal}, 715(2):1203--1220, 2010.

\bibitem{miggoz2009}
C.~Migaszewski and K.~Go{\'z}dziewski.
\newblock Secular dynamics of a coplanar, non-resonant planetary system under
  the general relativity and quadrupole moment perturbations.
\newblock {\em Monthly Notices of the Royal Astronomical Society},
  392(1):2--18, 2009.

\bibitem{moglas2022}
F.~Mogavero and J.~Laskar.
\newblock The origin of chaos in the {S}olar {S}ystem through computer algebra.
\newblock {\em Astronomy \& Astrophysics}, 662:L3, 2022.

\bibitem{mor2002}
A.~Morbidelli.
\newblock {\em Modern celestial mechanics: aspects of solar system dynamics}.
\newblock 2002.

\bibitem{murder1999}
C.~D. Murray and S.~F. Dermott.
\newblock {\em Solar system dynamics}.
\newblock Cambridge university press, 1999.

\bibitem{pisetal2017}
D.~Piskorz, B.~Benneke, N.~R. Crockett, A.~C. Lockwood, G.~A. Blake, T.~S.
  Barman, C.~F. Bender, J.~S. Carr, and J.~A. Johnson.
\newblock Detection of water vapor in the thermal spectrum of the
  non-transiting hot {J}upiter {U}psilon {A}ndromedae b.
\newblock {\em The Astronomical Journal}, 154(2):78, 2017.

\bibitem{voletal2019}
M.~Volpi, A.~Roisin, and A.-S. Libert.
\newblock The {3D} secular dynamics of radial-velocity-detected planetary
  systems.
\newblock {\em Astronomy \& Astrophysics}, 626:A74, 2019.

\end{thebibliography}

\end{document}